%% file: main.tex
\documentclass[conference]{IEEEtran}
\usepackage{tikz}
\usepackage{amsmath}
\usepackage{amssymb}
\usepackage{url}
\usepackage{breakurl}           % break too-long urls in refs

\usepackage{balance}
\usepackage{array}
\usepackage{subcaption}
\usepackage{bbding} % letter
\usepackage[ruled,linesnumbered]{algorithm2e}
\usepackage{hyperref} %\path
\usepackage{booktabs}
\usepackage{graphicx}
\usepackage{multirow}
\usepackage{xcolor}
\usepackage{colortbl}  % optional, enhances table shading

\definecolor{lightgray}{gray}{0.9}
\usepackage{array}

\usepackage{float} 
\usepackage{placeins}
\usepackage{amsthm}

\usepackage{tcolorbox}
\usepackage{verbatim}

\providecommand\wenhao[1]{\textcolor{red}{\{\textbf{wenhao:} {\em#1}\}}}

\providecommand\xss[1]{\textcolor{red}{\{\textbf{xss:} {\em#1}\}}}

\newcounter{packednmbr}

\newenvironment{packeditemize}{
\begin{list}{$\bullet$}{
\setlength{\labelwidth}{8pt}
\setlength{\itemsep}{0pt}
\setlength{\leftmargin}{\labelwidth}
\addtolength{\leftmargin}{\labelsep}
\setlength{\parindent}{0pt}
\setlength{\listparindent}{\parindent}
\setlength{\parsep}{1pt}
\setlength{\topsep}{1pt}}}{\end{list}}

\newcommand\sysname{\textsc{CryptPEFT}\xspace}
\newcommand\ms{\textbf{\texttt{MS}}\xspace}
\newcommand\muu{\textbf{\texttt{MU}}\xspace}
\newcommand\atten{{\texttt{LinAtten}}\xspace}

\providecommand\hl[1]{{\textcolor{blue}{#1}}}
\renewcommand{\hl}[1]{#1}

\newcommand{\para}[1]{\vspace{3pt}\noindent\textbf{\textit{{#1. }}}}

\providecommand\ignore[1]{{}}
\usepackage[normalem]{ulem} % for uwave

\usepackage{minibox}
\newcounter{observcntr}

\usepackage{minibox}
\usepackage{tcolorbox} % 提供更好的盒子外观

\usepackage{amssymb}

\usepackage{makecell}
\usepackage{array}
\newcolumntype{L}[1]{>{\raggedright\let\newline\\\arraybackslash\hspace{0pt}}m{#1}}
\newcolumntype{C}[1]{>{\centering\let\newline\\\arraybackslash\hspace{0pt}}m{#1}}
\newcolumntype{R}[1]{>{\raggedleft\let\newline\\\arraybackslash\hspace{0pt}}m{#1}}

\newcounter{principlecntr}

\newcommand{\question}[1]{
    \par\vspace{6pt}\noindent
        \vspace{2pt}
        \begin{minipage}[t]{\columnwidth}
        \textbf{#1}
        \end{minipage}
    \vspace{2pt}
}

\DeclareRobustCommand*{\authorrefmark}[1]{\raisebox{0pt}[0pt][0pt]{\textsuperscript{\footnotesize\ensuremath{\ifcase#1\or *\or \dagger\or \ddagger\or%
    \mathsection\or \mathparagraph\or \|\or **\or \dagger\dagger%
    \or \ddagger\ddagger \else\textsuperscript{\expandafter\romannumeral#1}\fi}}}}
\usepackage{bbding} % corresponding letter

\usepackage{CJK}

\definecolor{darkred}{rgb}{0.545,0,0}
\usepackage{hyperref}
\hypersetup{
    colorlinks = true,
    allcolors = darkred,
}

\usepackage[available,functional,reproduced]{ndssbadges}

\begin{document}

%\begin{CJK}{UTF8}{gbsn}

% \title{\Large \bf \sysname: Parameter-Efficient Fine-Tuning Architecture for \\Private Neural Network Inference}

%v1
%\title{\sysname: Parameter-Efficient Fine-Tuning for Privacy-Preserving Neural Network Inference}

% \sysname: Privacy-Preserving Neural Network Inference for Parameter-Efficient Fine-Tuning

% 
\title{\sysname: Efficient and Private Neural  \\Network Inference via Parameter-Efficient Fine-Tuning\thanks{Corresponding authors: Wenhao Wang (\href{mailto:wangwenhao@iie.ac.cn}{wangwenhao@iie.ac.cn}) and Zihao Wang (\href{mailto:zihao.wang@ntu.edu.sg}{zihao.wang@ntu.edu.sg}).}}
% \syaname: Accelerating Private 
%v2
%\title{\sysname: Accelerating Encrypted Transformer Inference via Parameter-Efficient Fine-Tuning}

% 面向隐私推理加速的神经网络PEFT架构
\date{}
%for single author (just remove % characters)
%\author{
%{\rm Your N.\ Here}\\
%Your Institution
%\and
%{\rm Second Name}\\
%Second Institution
% copy the following lines to add more authors
% \and
% {\rm Name}\\
%Name Institution
%} % end author
\author{
{\rm  Saisai Xia\authorrefmark{1}\authorrefmark{2}, Wenhao Wang\authorrefmark{1}\authorrefmark{2}\textsuperscript{\Envelope}, Zihao Wang\authorrefmark{3}\textsuperscript{\Envelope}}, Yuhui Zhang\authorrefmark{1}\authorrefmark{2}, Yier Jin\authorrefmark{4}, Dan Meng\authorrefmark{1}\authorrefmark{2}, Rui Hou\authorrefmark{1}\authorrefmark{2}\\ \\
\authorrefmark{1}State Key Laboratory of Cyberspace Security Defense, Institute of Information Engineering, CAS \\
\authorrefmark{2}School of Cyber Security, University of Chinese Academy of Sciences \\
\authorrefmark{3}Nanyang Technological University \\
\authorrefmark{4}University of Science and Technology of China \\
}

\IEEEoverridecommandlockouts
\makeatletter\def\@IEEEpubidpullup{6.5\baselineskip}\makeatother
\IEEEpubid{\parbox{\columnwidth}{
		Network and Distributed System Security (NDSS) Symposium 2026\\
		23-27 February 2026, San Diego, CA, USA\\
		ISBN 979-8-9919276-8-0\\
		https://dx.doi.org/10.14722/ndss.2026.231102\\
		www.ndss-symposium.org
}
\hspace{\columnsep}\makebox[\columnwidth]{}}

\maketitle

\begin{abstract}
Publicly available large pretrained models (i.e., backbones) and lightweight adapters for parameter-efficient fine-tuning (PEFT) have become standard components in modern machine learning pipelines. However, preserving the privacy of both user inputs and fine-tuned adapters---often trained on sensitive data---during inference remains a significant challenge. Applying cryptographic techniques, such as multi-party computation (MPC), to PEFT settings still incurs substantial encrypted computation across both the backbone and adapter, mainly due to the inherent two-way communication between them.
To address this limitation, we propose \sysname, the first PEFT solution specifically designed for private inference scenarios. \sysname introduces a novel \emph{one-way communication (OWC)} architecture that confines encrypted computation solely to the adapter, significantly reducing both computational and communication overhead. To maintain strong model utility under this constraint, we explore the design space of OWC-compatible adapters and employ an automated architecture search algorithm to optimize the trade-off between private inference efficiency and model utility.
We evaluated \sysname using Vision Transformer backbones across widely used image classification datasets. Our
results show that \sysname significantly outperforms existing baselines, delivering speedups ranging from $20.62\times$ to $291.48\times$ in simulated wide-area network (WAN) and local-area network (LAN) settings.
On CIFAR-100, \sysname attains 85.47\% accuracy with just 2.26 seconds of inference latency. These findings demonstrate that \sysname offers an efficient and privacy-preserving solution for modern PEFT-based inference.

\end{abstract}

\input{tex/intro}
\input{tex/background}
\input{tex/design}
\input{tex/workflow}
\input{tex/search}
\input{tex/security}
\input{tex/evaluation}
\input{tex/discussion}
\input{tex/related}

\input{tex/conclusion}
\input{tex/ack}

\balance
\bibliographystyle{plain}
\bibliography{ref}

%\input{tex/workflow}

%\newpage
\appendices
%\setcounter{page}{1}
% for ae submission, we need the AE appendix to be a separated file
\input{tex/ndss_ae_appendix_template_v1}

%\end{CJK}
\end{document}

%% file: tex/intro.tex
\section{Introduction}

Pre-trained models have revolutionized machine learning, serving as the foundation for a variety of downstream applications~\cite{DBLP:conf/icml/ChenK0H20, DBLP:conf/nips/ChenKSNH20, vit}. By leveraging large-scale datasets and substantial computational resources, these models often learn representations that rival or exceed human-level performance. Subsequently, fine-tuning~\cite{DBLP:journals/pieee/ZhuangQDXZZXH21} re-purposes a pre-trained network for a specific task without necessitating training from scratch. For example, a model initially designed for general image classification can be adapted to detect tumors in medical scans~\cite{litjens2017survey, al2022etecadx, mahoro2024breast}, automatically tag images on social media~\cite{jin2022automatic, vu2020privacy}, or recognize traffic signs in autonomous vehicles~\cite{chen2022vision, prakash2021multi}. Such flexibility facilitates the specialization of foundational models and yields highly personalized services.

Despite these advantages, deploying fine-tuned models in user-facing contexts raises pressing privacy concerns. From the service provider's perspective, there is an incentive to maintain confidentiality of proprietary model parameters, while users often hesitate to disclose potentially sensitive input data. As a result, privacy-preserving inference\footnote{We use the terms ``privacy-preserving inference'' and ``private inference'' interchangeably throughout this paper.} based on secure multi-party computation (MPC)~\cite{knott2021crypten} has emerged as a promising solution to protect both the intellectual property of service providers and the confidentiality of user queries. However, existing secure computation protocols are often burdened by substantial communication and computational overhead~\cite{DBLP:conf/uss/MishraLSZP20, DBLP:conf/sp/MohasselZ17, DBLP:conf/ccs/RatheeR0CGR020, DBLP:conf/sp/TanKTW21}.
To alleviate these overheads, prior work has primarily focused on optimizing the resource-intensive low-level operations within MPC protocols, leading to the development of MPC-friendly approximations intended to expedite the inference process~\cite{li2022mpcformer, park2024powerformer, pang2024bolt, zimerman2024power, ao2024autofhe, zhang2023sal}. While such optimizations can reduce overhead, their effectiveness diminishes as models increase in complexity and size. When applied to extremely large architectures, these approaches may become impractical or infeasible.

In contrast, our study focuses on the development of MPC-friendly architectures tailored for fine-tuning, thereby addressing the fundamental challenges posed by the increasing size of modern models. By rethinking model design at the architectural level, we aim to ensure that privacy-preserving inference can scale effectively with minimal overhead. This emerging research paradigm has recently attracted attention in pioneering works~\cite{kundu2023learning, kundu2023making, zeng2023mpcvit, MPC-Minimized}. Some of these studies emphasize constructing compact models through techniques such as compressing or distilling large and powerful networks~\cite{kundu2023learning, kundu2023making, zeng2023mpcvit}. While these methods achieve improved efficiency, they often involve trade-offs that compromise model utility, which can be untenable in many practical applications. More recently, Rathee et al.~\cite{MPC-Minimized} proposed an alternative approach that retains the foundational architecture of the model, fine-tuning only the final layers to indirectly reduce the overall model size for privacy-preserving inference. While this method mitigates some utility losses, it remains hindered by persistent efficiency bottlenecks (\autoref{subsec:Effectiveness}).

\para{Our solution}
To address this challenge, we draw inspiration from Parameter-Efficient Fine-Tuning (PEFT) methodologies~\cite{hu2021lora, chen2022vision, chen2022adaptformer, zhang2022neural, he2023sensitivity}, which introduce a small set of trainable parameters---commonly referred to as \emph{adapters}---into a pre-trained \emph{backbone model} while keeping most of the original weights frozen. In standard (i.e., unencrypted plaintext) fine-tuning settings, PEFT has demonstrated both efficiency and robustness, often outperforming approaches that fine-tune only the final layers~\cite{chen2022adaptformer}. Given these advantages, it is natural to consider extending PEFT to private inference scenarios, where one might hope to perform the bulk of computation in plaintext on the public backbone, while limiting MPC-based encrypted computation to the adapter components---thus significantly improving overall inference efficiency.
Unfortunately, our study reveals that \textit{existing PEFT methods are fundamentally incompatible with private inference due to architectural limitations}. Specifically, conventional adapter designs require their (encrypted) outputs to be decrypted before being fed back into the backbone, creating a potential leakage point (c.f., \autoref{subsec:motivations}).
These limitations highlight the need for novel PEFT architectures that can operate entirely on encrypted data, without sacrificing privacy or efficiency.

In this paper, we propose \sysname, the first PEFT-based solution specifically designed for private inference. Our key insight is that conventional PEFT techniques inherently introduce a \emph{two-way communication (TWC)} mechanism between the backbone and the adapter, necessitating multiple encrypt-decrypt cycles. In contrast, \sysname enforces a \emph{one-way communication (OWC)} policy, ensuring that data flows unidirectionally from the backbone to the adapter without being fed back. By eliminating intermediate decryption, \sysname allows all computations to remain fully encrypted, thereby preserving strict privacy guarantees.

A natural concern arising from the removal of feedback loops, as mandated by the OWC policy, is whether such a restriction compromises the model's expressive power. To answer this, we begin with an empirical study to evaluate the impact of OWC on model performance across various downstream tasks. The results show that existing adapter architectures often suffer from reduced utility under OWC constraints. Based on this observation, we analyze how adapter structure and placement affect both utility and efficiency in private inference scenarios (\autoref{subsec:understanding}). Guided by these insights, we design a new adapter architecture specifically tailored for OWC-compliant settings. A central component of this design is \atten, a lightweight attention mechanism that replaces the traditional Softmax with MPC-friendly operations such as addition and multiplication. \atten significantly reduces the cost of private inference while maintaining high utility. To further enhance efficiency, \sysname adopts the Low-Rank Adaptation (LoRA) framework~\cite{hu2021lora}, which reduces computational and communicational overhead through low-rank decomposition. 

Finally, \sysname integrates a Neural Architecture Search (NAS) mechanism~\cite{DBLP:journals/jmlr/ElskenMH19} to automatically identify adapter configurations that optimize the trade-off between efficiency and utility for a given downstream task. \hl{Unlike conventional NAS in plaintext setting, which typically assumes computational cost scales proportionally with model size, private inference introduces fundamentally different cost dynamics. To enable cost-aware NAS, we profiled all key operations—including linear layers, matrix multiplications, normalization, and activation functions—under our target network environment, and derived cost estimation models based on the profiling results. Guided by these cost models, our NAS strategy is tailored to identify architectures that satisfy a target utility while minimizing private inference cost (\autoref{subsec:grid}).}

\ignore{\sysname adopts a model and protocol co-design approach to achieve an optimal balance between model utility and efficiency in privacy-preserving inference. 
On the model side, \sysname leverages the low-rank adaptation (LoRA) framework~\cite{hu2021lora} to reduce computational complexity through low-rank decomposition. While conventional PEFT often modifies layers throughout the network, \sysname focuses only on deeper layers to enhance efficiency while preserving expressiveness. Furthermore, \sysname integrates a neural architecture search (NAS) mechanism~\cite{DBLP:journals/jmlr/ElskenMH19} to identify architectures that optimize efficiency-utility tradeoffs. Notably, while the OWC paradigm reinforces privacy guarantees, the combination of LoRA and selective layer modifications reduces the search space, thereby lowering training overhead and further improving overall efficiency-utility tradeoffs. 
On the protocol side, we design mixed-degree polynomial approximations for the adapter, leveraging the insight that different adapters exhibit varying sensitivity to approximation errors. Additionally, we introduce an efficient private attention protocol that selectively focuses on the most significant components of the input, further enhancing model efficiency while maintaining model utility.}

\para{Evaluations}
In this study, we primarily focus on vision-related tasks, aligning with the majority of existing research in the field of private inference. Nonetheless, we view language models as a natural extension of our framework and plan to explore them in future work. Specifically, we built \sysname using public vision backbones (i.e., ViT-B), and evaluated its performance on widely-used image datasets, including CIFAR-10~\cite{cifar10}, CIFAR-100~\cite{CIFAR100}, Food-101~\cite{food101}, SVHN~\cite{SVHN} and Flowers-102~\cite{flowers102}. For each dataset, we identified the most suitable adapters through our NAS-based search procedure. 

We developed an end-to-end inference system for \sysname using the MPC framework \textsc{CrypTen}~\cite{crypten-github} and evaluated its private inference efficiency. Our experimental results show that \sysname achieves significant reductions in inference time compared to the baseline methods. Specifically, in a simulated wide-area network (WAN) environment characterized by a 400 Mbps bandwidth and 4 ms latency, it achieves an average speedup of $250.39\times$ over the traditional PEFT baseline method, and an average speedup of $20.62\times$ compared to the baseline that simply fine-tunes only the last layer.
%it reduces inference time by an average of \(226.41\times\) compared to the traditional PEFT baseline method and by an average of \(18.64\times\) relative to the baseline that simply fine-tunes only the last layer. %These improvements come with an average accuracy loss of less than 3\%.
\sysname also improves the model accuracy by 0.83\% on average over fine-tuning the last layer.
Taking the CIFAR-100 dataset as an example, \sysname 
%significantly outperforms the state-of-the-art method (i.e., MPCViT~\cite{zeng2023mpcvit}, 
achieves 
%a maximum accuracy of 77.46\%) in terms of model accuracy, with 
a model accuracy of 85.47\% and an inference time of just 2.26 seconds.

\para{Contributions} Our key contributions are outlined below:

% \item \textit{New observation}. We propose a novel one-way communication principle for PEFT frameworks, where only a small portion of computations need to be performed on ciphertext, reducing the overhead of private inference.
%我们提出了一个新颖的one-way communication架构设计原则，在使用ViTs进行安全推理时，仅有一小部分计算在密文上完成，降低隐私推理对计算性能的影响。
\noindent$\bullet$\textit{~New observations}. We propose \sysname, the first PEFT approach tailored to private inference scenarios. \sysname employs a one-way communication (OWC) strategy, completely removing the need for decryption within the network, guaranteeing strict end-to-end privacy. 
%Unlike existing PEFT methods that require partial decryption of intermediate representations, 

%Additionally, we propose to utilize LoRA on deeper layers to reduce communication and computational overhead, and incorporate NAS to automatically discover architectures, which maximizes efficiency while preserving accuracy.

\noindent$\bullet$\textit{~New designs}. 
Building on insights into adapter behavior under OWC constraints, we propose a redesigned adapter architecture and a NAS-guided search strategy to jointly enhance the utility and efficiency of private inference in \sysname.
    
\noindent$\bullet$\textit{~Extensive empirical studies}.
% We present evaluations on multiple vision and language tasks, demonstrating that our proposed method reduces computational overhead while maintaining high performance. Notably, our system achieves state-of-the-art efficiency in several private inference tasks. 
We implement \sysname on a standard vision backbone model, and evaluate it on widely used image datasets, including CIFAR-10, CIFAR-100, Food-101, SVHN and Flowers-102. 
%\hl{Experimental results demonstrate that \sysname reduces inference time by up to 20-fold compared with the state-of-the-art method~\cite{MPC-Minimized}.} 
\sysname surpasses existing solutions in both classification accuracy and inference efficiency, highlighting its potential for real-world, privacy-preserving deployment.
The source code for our research is publicly available at {\url{https://github.com/Saisai-Xia/CryptPEFT}}. %\footnote{\url{https://anonymous.4open.science/r/CryptPEFT-8405}}
%我们提供了在多个视觉任务上的实验评估，展示了所提方法在降低计算开销的同时仍能保持良好的模型性能。并能够在密文推理任务上达到目前state-of-the-art的效率。

%\para{Roadmap}
%The rest of the paper is organized as follows: \autoref{sec:back} provides the background of our research; \autoref{sec:Preliminaries} details the threat model and the motivation of our research; \autoref{sec:method} elaborates the \sysname framework and the theoretic analysis on its efficacy; \autoref{sec:eval} reports the evaluation results of \sysname; \autoref{sec:related} discusses prior related studies; \autoref{sec:discussion} discusses the potential future directions, and \autoref{sec:conclusion} concludes the paper.

% \wenhao{motivating CryptPEFT focus on inference, maybe turn down the emphasis on transformer storage and fine-tuning}

%% file: tex/background.tex
\section{Background}
\label{sec:back}
%Transformer架构在视觉和语言领域都达到了state-of-the-art的性能。
%The Transformer architecture has achieved state-of-the-art performance in both vision and language domains.
%This paper mainly focuses on the vision domain, with an evaluation of its application to the language domain in \autoref{subsec:llm}. In this section, we briefly review ViT, PEFT, and private inference techniques.
%本文主要以视觉领域介绍架构设计，我们的评估表明类似的设计同样适用于语言领域。本章介绍ViT架构、PEFT架构以及private inference相关的基础知识。

\subsection{Vision Transformer (ViT)}
\label{subsec:transformer_background}

Vision Transformer (ViT) is a neural network architecture that applies the Transformer model---originally developed for natural language processing---to image recognition tasks by treating images as sequences of patch embeddings. It has demonstrated competitive or superior performance compared to convolutional neural networks (CNNs), especially when trained on large-scale datasets.
A typical ViT consists of the Patch Embedding layer, Position Embedding layer, Transformer Encoder layer, and Classifier~\cite{vit}.  Given an image \( \mathbf{x} \in \mathbb{R}^{H \times W \times C} \) with $C$ channels and resolution $(H, W)$, the Patch Embedding layer divides $\mathbf{x}$ into multiple patches and flattens them into $\mathbf{x}_p \in \mathbb{R}^{N \times (P^2 \cdot C)}$,  where $(P, P)$ denotes the resolution of each image patch, and $N = \frac{H \times W}{P^2}$ represents the number of image tokens. Each row vector in $\mathbf{x}_p$ is then mapped to a 
$D$-dimensional space and concatenated with a learnable \texttt{[CLS]} token for further processing. The Position Embedding layer adds a set of learnable positional encodings to the output of the Patch Embedding layer to retain spatial information, resulting in a sequence of image tokens.

The Transformer Encoder layer consists of multiple identical Transformer layers. Each Transformer layer includes a Multi-Head Self-Attention (MHSA) layer and a Multi-Layer Perceptron (MLP) layer. For example, the image tokens $\mathbf{x}_{j-1}$ output by the $(j-1)$-th Transformer layer are first transformed into three vectors: $Q$, $K$, and $V$. These vectors are then used in the attention calculation within the MHSA layer of the $j$-th Transformer layer:
%ViT由~\cite{dosovitskiy2020image}首次引入视觉领域，典型的ViT由Patch Embedding Layer、Position Embedding Layer、Transformer Encoder Layers、Classification Head Layer构成，给定一个通道为C、分辨率为(H, W)的图像\( \mathbf{x} \in \mathbb{R}^{H \times W \times C} \)，Patch Embedding Layer将\( \mathbf{x} \)分割成多个patches并展平为$x_p \in \mathbb{R}^{N \times (P^2 \cdot C)}$，再将$x_p$的每个行向量的维度映射为D维，最后与可训练的\texttt{[CLS]}token进行拼接用于后续处理，其中(P, P)表示每个图像patch的分辨率，$N = \frac{H \times W}{P^2}$，表示共有N个image tokens。Position Embedding Layer负责将Patch Embedding Layer的输出与一组可学习的位置编码进行相加，从而保留位置信息，并最终形成一系列image tokens。Transformer Encoder Layers由多个相同的Transformer Encoder Block组成，每个Transformer Encoder Block主要包含一个Multi-Head Self-Attention Layer (MHSA) 和一个MLP Layer，以第j-1层Transformer Encoder Layer输出的image tokens $x_{j-1}$为例，其输入到第j层Transformer Encoder Block的MHSA之前，会被映射成三组向量，即Q、K、V，这三组向量进行注意力计算：
\[
\mathbf{x}_{j}^{'} = \text{Attention}(Q, K, V) = \text{Softmax}(QK^T / \sqrt{d_k}) V,
\]
where $d_k$ represents the feature dimension of each head in the MHSA. The output $\mathbf{x}_{j}^{'}$ is then passed through LayerNorm and the MLP. The computation can be formalized as:
%其中$d_k$为MHSA中的每个head的特征维度，$\mathbf{x}_{j}^{'}$继续输入至LayerNorm和MLP，计算流程可以形式化为：
\[
\mathbf{x}_j = \text{MLP}(\text{LN}(\mathbf{x}_{j}^{'})) + \mathbf{x}_{j}^{'}.
\]
For the subsequent Transformer layers, the computation of $\mathbf{x}_{j}$ continues as described above. The final Transformer layer outputs a set of image tokens, from which the previously appended \texttt{[CLS]} token is extracted. This \texttt{[CLS]} token, having captured global information, is then passed to the Classifier to complete the classification task. For more details on ViT, please refer to~\cite{vit}.
%对于后续的Transformer Encoder Block，$\mathbf{x}_{j}$继续执行上述流程，最后一层Transformer Encoder Block会输出image tokens，从中提取之前拼接的\texttt{[CLS]}token，将捕获了全局信息的\texttt{[CLS]}token输入至Classification Head Layer，从而完成分类任务。上述ViT的计算流程可以表示为图~\ref{}

%\subsection{ViT for Privacy-Preserving Inference Acceleration}

\subsection{Parameter-Efficient Fine-Tuning (PEFT)}
ViTs leverage global self-attention mechanisms across image patches, which enables strong representation learning but also results in a higher parameter count compared to traditional convolutional neural networks (CNNs). {Fine-tuning ViTs is particularly computationally expensive, as it typically involves updating all model parameters.} When adapting a pre-trained model to multiple downstream tasks, conventional fine-tuning necessitates storing separate parameter sets for each task, incurring significant storage and compute overhead.
Parameter-Efficient Fine-Tuning (PEFT) offers a practical alternative by introducing a small set of trainable parameters, such as adapters or low-rank matrices, while keeping the backbone frozen.
In this section, we detail AdaptFormer~\cite{chen2022adaptformer} and Low-Rank Adaptation (LoRA)~\cite{hu2021lora}, two representative PEFT methods that serve as baselines in our evaluations.

% In recent years, ViT adapters and Low-Rank Adaptation (LoRA) have become prominent in neural network architecture design. By leveraging Parameter-Efficient Fine-Tuning (PEFT) techniques during training and inference, these approaches effectively address the challenges of computational costs, storage demands, and transfer learning in traditional neural networks.

% \para{ViT-Adapter}
% Recently, ViT has become a mainstream architecture in vision tasks, particularly due to its enhanced performance in image classification. However, its high computational complexity and memory requirements remain major challenges in practical applications. ViT-Adapter addresses this issue through PEFT. It enables a pre-trained ViT to perform downstream tasks by tuning only task-specific adapters, eliminating the need for full-parameter fine-tuning. These adapters typically consist of low-rank matrices applied to specific layers, significantly reducing both the parameter count and computational cost. This approach enhances parameter efficiency in multi-task and transfer learning, while avoiding the high costs associated with training large-scale models.

\para{AdaptFormer}
%Recently, ViT has become a mainstream architecture in vision tasks, particularly due to its enhanced utility in image classification. However, its high computational complexity and memory requirements remain major challenges in practical applications. 
% ViT-Adapter addresses this issue by introducing adapters, allowing efficient fine-tuning of pre-trained Transformer models without adding a large number of parameters~\cite{chen2022vision}. Instead of full-parameter fine-tuning, ViT-Adapter inserts lightweight adapter networks between Transformer layers to adjust the model's parameters. 
AdaptFormer~\cite{chen2022adaptformer} enables ViTs to perform downstream tasks by tuning lightweight task-specific adapters, rather than the entire model. These adapters, often implemented as low-rank matrices inserted into selected layers, significantly reduce the number of trainable parameters, improving efficiency in multi-task and transfer learning scenarios.

%AdaptFormer addresses this issue through PEFT~\cite{chen2022adaptformer}. It enables a pre-trained ViT to perform downstream tasks by tuning only task-specific adapters, eliminating the need for full-parameter fine-tuning. 
%These adapters typically consist of low-rank matrices applied to specific layers, significantly reducing both parameter count and computational cost. This approach improves parameter efficiency in multi-task and transfer learning while avoiding the high costs of training large-scale models.%\xss{replace it with AdaptFormer}
%视觉Transformer（ViT）自提出以来，由于其基于自注意力机制（Self-Attention）而显著提升了图像分类任务的性能，逐渐成为视觉任务中的主流架构。然而，ViT在实际应用中的主要瓶颈之一是其庞大的计算复杂度和内存消耗。ViT-Adapter通过引入适配器模块（Adapter Module），在不增加大规模额外参数的情况下，实现了对预训练Transformer模型的有效微调。

%ViT-Adapter的核心思想是通过在Transformer的各层之间插入轻量级的适配器网络来调整原有模型的参数，而不是直接对Transformer本身进行全参数微调。适配器模块通常由一组低秩矩阵构成，这些矩阵仅对特定层进行调整，从而在参数量和计算量上大幅降低。这种策略使得ViT-Adapter能够在多任务学习和迁移学习中表现出更好的参数效率，同时避免了训练大规模参数所带来的高昂开销。

\ignore{
Compared to traditional Convolutional Neural Networks (CNNs), ViT-Adapter offers several advantages:
\begin{packeditemize}
    \item \textit{Enhanced modeling capability}. ViT captures long-range dependencies in images through global self-attention mechanisms, enabling it to handle complex visual tasks, especially with large-scale input images. 
    \item \textit{Improved transfer learning utility}. By fine-tuning with lightweight adapters, ViT-Adapter effectively leverages pre-trained Transformer models, enhancing utility in transfer learning scenarios with limited data.
    \item \textit{Parameter efficiency}. The adapters introduced by ViT-Adapter significantly reduce the number of parameters updated during fine-tuning, leading to lower computational overhead while maintaining high utility.
\end{packeditemize}
}

%相较于传统CNN，ViT-Adapter在以下方面展示了显著优势：
%\begin{packeditemize}
%    \item 更强的建模能力：ViT通过全局自注意力机制捕获图像中的长期依赖关系，而CNN则更侧重局部特征的提取。ViT能够更好地处理复杂的视觉任务，尤其是大尺度输入图像。
%    \item 迁移学习性能：ViT-Adapter通过轻量级的适配器模块进行微调，使得在有限数据的情况下能够充分利用预训练的Transformer模型，提升了迁移学习的效果。
%    \item 参数高效性：ViT-Adapter引入的适配器模块大幅减少了微调时需要更新的参数数量，相较于传统CNN模型，能够在保持较高性能的同时，显著降低训练的计算开销。
%\end{packeditemize}

\para{LoRA}
%低秩适配器（LoRA）是一种旨在通过低秩分解降低大规模预训练模型微调开销的技术。LoRA的核心思想是通过对预训练模型的部分权重矩阵进行低秩分解，从而减少微调时需要优化的参数量。LoRA将原有的权重矩阵分解为低秩矩阵的乘积，这样在训练过程中，仅调整低秩矩阵而不改动原始模型的权重。LoRA通常与大规模预训练语言模型（如GPT、BERT等）结合使用，也可以应用于视觉Transformer等架构。
Low-Rank Adaptation (LoRA)~\cite{hu2021lora} is a technique that reduces fine-tuning overhead for large pre-trained models by applying low-rank decomposition. It decomposes the weight matrices of the pre-trained model into low-rank matrices, which reduces the number of parameters requiring optimization during fine-tuning. 
%This allows for adjustments without modifying the original model's weights.
LoRA is commonly used with large language models like GPT and BERT, and can also be applied to architectures such as ViT.
{LoRA's key advantage is its ability to adapt to new tasks or datasets with minimal parameter changes, avoiding the need to retrain the entire model.} 

%This reduces computational cost and makes fine-tuning large models feasible in resource-limited environments. 

%Additionally, LoRA helps mitigate overfitting, especially when data is scarce. 
\ignore{Specifically, LoRA provides the following benefits:
%LoRA的主要优势在于其能够在不对整个模型进行重新训练的情况下，通过少量参数的调整来适应新的任务或数据集。这种方式不仅降低了计算成本，也使得在资源受限的环境中微调大型预训练模型成为可能。此外，LoRA还能有效避免传统微调方法中常见的过拟合问题，尤其是在数据较少的情况下。
%相较于传统架构，LoRA具有以下优势：
\begin{packeditemize}
    \item \textit{Parameter efficiency}. LoRA's low-rank decomposition reduces the number of parameters needing adjustment during fine-tuning, significantly lowering computational and storage costs compared to traditional full-parameter fine-tuning.
    \item \textit{Generalization}. By training only a small number of low-rank adapters, LoRA improves generalization, particularly for small datasets, making it especially effective in transfer learning.
    \item \textit{Compatibility and flexibility}. LoRA is compatible with various deep learning models, such as Transformers and ViTs, and requires minimal changes to the original architecture, offering strong adaptability.
    %参数高效性：LoRA的低秩分解使得在微调过程中需要调整的参数数量大幅减少，相比于传统的全参数微调方法，LoRA能够显著降低计算和存储开销。
    %\item 泛化能力：由于LoRA仅通过少量的低秩适配器进行训练，它在处理小数据集时比传统架构更具泛化能力，尤其在迁移学习中表现突出。
    %\item 兼容性和灵活性：LoRA可以与各种深度学习模型（包括Transformer、ViT等）兼容，且无需对原始网络架构进行大规模修改，适应性强。
\end{packeditemize}
}

% In summary, PEFT architectures, like ViT-Adapter and LoRA, reduce the number of parameters requiring optimization by using techniques like low-rank decomposition and adapters. This lowers computational complexity and storage needs, making fine-tuning large pre-trained models feasible in resource-limited settings. Unlike traditional CNNs, which necessitate fine-tuning the entire network, PEFT architectures enable more efficient fine-tuning through targeted parameter adjustments, effectively mitigating overfitting. Additionally, PEFT leverages the superior representational power of Transformers, enabling applications in complex tasks such as image classification and natural language processing.
%总结来看，ViT-Adapter和LoRA等新型架构在以下几个方面展现了显著的优势：
%计算和存储效率：ViT-Adapter和LoRA通过引入低秩分解和适配器模块等技术，显著减少了训练过程中需要优化的参数量，从而降低了计算复杂度和存储需求。这对于大规模预训练模型的微调尤其重要，使得在资源受限的环境下，仍能高效进行模型更新。

%迁移学习和多任务学习的优势：传统CNN和其他经典架构在迁移学习中的性能有限，因为它们通常需要对整个网络进行微调。而ViT-Adapter和LoRA通过精细化的参数调整，使得在新的任务或数据集上进行微调时更加高效，且可以避免过拟合。

%更强的建模能力和灵活性：ViT基于自注意力机制，能够捕获长程依赖关系，具有更强的表示能力。通过ViT-Adapter和LoRA等方法，这些优势得以在保持参数效率的同时进一步优化，尤其适用于复杂任务如图像识别、自然语言处理等。

\subsection{MPC-based Private Inference}
\label{subsec:mpcbackground}

{Private inference aims to protect sensitive data during model inference by employing cryptographic techniques such as multi-party computation (MPC) and homomorphic encryption (HE).} In this work, we build our inference system using the privacy-preserving machine learning framework \textsc{CrypTen}~\cite{crypten2020}, which implements secure computation protocols based on secret sharing. This section provides essential background on secret sharing based MPC protocols.
%Our proposed system can also be adapted for use with HE.

\para{Secret sharing}
Arithmetic secret sharing involves dividing a scalar value \( x \in \mathbb{Z}/Q\mathbb{Z} \), where \( \mathbb{Z}/Q\mathbb{Z} \) represents a finite ring of integers modulo \( Q \), across a set of parties \( \mathcal{P} = \{ p_1, p_2, \dots, p_n \} \). The sharing of \( x \) is represented by \( [x] = \{ [x]_p \}_{p \in \mathcal{P}} \), where each \( [x]_p \in \mathbb{Z}/Q\mathbb{Z} \) denotes the share held by party \( p \). The shares are constructed such that their sum (modulo \( Q \)) reconstructs the original value, i.e.,
\[
x = \sum_{p \in \mathcal{P}} [x]_p \pmod{Q}.
\]
To generate the shares for a value \( x \), the parties first create a \textit{zero-share} vector consisting of \( |\mathcal{P}| \) random numbers, ensuring their sum equals zero~\cite{cramer2005share}. One party, typically the one with access to the secret \( x \), adds the value \( x \) to its share and then discards the original value. The sum of all shares will reconstruct \( x \) without revealing the secret to any single party.

Binary secret sharing is a specific case of arithmetic secret sharing that operates in the binary field \( \mathbb{Z}/2\mathbb{Z} \), where arithmetic is performed modulo 2. In this setting, each party \( p \in \mathcal{P} \) holds a share \( \langle x \rangle_p \in \{ 0, 1 \} \), such that the reconstruction condition is given by:
\[
x = \bigoplus_{p \in \mathcal{P}} \langle x \rangle_p,
\]
where \( \bigoplus \) denotes the bitwise XOR operation. Binary secret sharing is often used in applications when binary operations are efficient.

%Arithmetic secret sharing shares a scalar value $x \in \mathbb{Z}/Q\mathbb{Z}$, where $\mathbb{Z}/Q\mathbb{Z}$ denotes a ring with $Q$ elements, across parties $p \in \mathcal{P}$. We denote the sharing of $x$ by $[x] = \{[x]_p\}_{p \in \mathcal{P}}$, where $[x]_p \in \mathbb{Z}/Q\mathbb{Z}$ indicates party $p$'s share of $x$. The shares are constructed such that their sum reconstructs the original value $x$, that is, $x = \sum_{p \in \mathcal{P}}{[x]_p \text{ mod } Q}$. To share a value $x$, the parties generate a pseudorandom zero-share with $|\mathcal{P}|$ random numbers that sum to $0$~\cite{cramer2005share}. The party that possesses the value $x$ adds $x$ to their share and discards $x$. 
%Binary secret sharing is a special case of arithmetic secret sharing that operates within the binary field $\mathbb{Z}/2\mathbb{Z}$. Each party $p \in \mathcal{P}$ holds a share $\langle x \rangle_p$, such that $x = \bigoplus_{p \in \mathcal{P}}{\langle x \rangle_p}$ is satisfied.

\para{Conversions}
The conversion from an arithmetic secret share \( [x] \) to a binary secret share \( \langle x \rangle_p \) is performed by first converting each party's arithmetic share \( [x]_p \) into binary shares. Each party creates a binary secret share, \( \langle [x]_p \rangle \), which represents the bits of their share \( [x]_p \). These binary shares are then combined by summing them to produce the binary secret share \( \langle x \rangle \), which represents the original value \( x \) in binary form.
Specifically, each party \( p \in \mathcal{P} \) generates a binary secret share \( \langle [x]_p \rangle \) of their share \( [x]_p \), and the parties then compute the total binary share \( \langle x \rangle \) by summing the individual binary shares:
\[
\langle x \rangle = \sum_{p \in \mathcal{P}} \langle [x]_p \rangle.
\]
This process typically uses a carry-lookahead adder to handle the binary addition efficiently, and it can be completed in \( \mathcal{O}(\log_2 |\mathcal{P}| \cdot \log_2 Q) \) communication rounds, where \( |\mathcal{P}| \) is the number of parties and \( Q \) is the modulus in the original arithmetic secret sharing scheme~\cite{catrina2010improved,damgaard2006unconditionally}. 

The conversion from $\langle x \rangle_p$ to $[x]$ is done by reconstructing the arithmetic share from the individual bits of the binary share \( \langle x \rangle \). Specifically, the arithmetic secret share is computed as:
\[
[x] = \sum_{b=1}^{B} 2^b \cdot [\langle x \rangle^{(b)}],
\]
where \( \langle x \rangle^{(b)} \) denotes the \( b \)-th bit of the binary secret share \( \langle x \rangle \), and \( B \) is the total number of bits needed to represent the binary share. The sum of the weighted bits reconstructs the original arithmetic value \( x \).

%{\color{red}The conversion from $[x]$ to $\langle x \rangle_p$ is implemented by having the parties create a binary secret share of their $[x]_p$ shares, and summing the resulting binary shares. Specifically, the parties create a binary secret share, $\langle [x]_p \rangle$, of all the bits in $[x]_p$. Subsequently, the parties compute $\langle x \rangle = \sum_{p \in \mathcal{P}}{\langle [x]_p \rangle}$ using a carry-lookahead adder in $\text{log}_2{(|\mathcal{P}|)\text{log}_2{(Q)}}$ communication rounds~\cite{catrina2010improved,damgaard2006unconditionally}. The conversion from $\langle x \rangle_p$ to $[x]$ is achieved by computing $[x] = \sum_{b=1}^{B}{2^b[\langle x \rangle^{(b)}]}$, where $\langle x \rangle^{(b)}$ denotes the $b$-th bit of the binary share $\langle x \rangle$ and $B$ is the total number of bits in the shared secret $\langle x \rangle$.}

\para{Private linear functions}
To perform private addition of two secret-shared values \( [z] = [x] + [y] \), each party \( p \in \mathcal{P} \) adds their respective shares of \( x \) and \( y \). That is, each party computes:
\[
[z]_p = [x]_p + [y]_p,
\]
where \( [x]_p \) and \( [y]_p \) are the individual shares of \( x \) and \( y \) held by party \( p \), and \( [z]_p \) is the corresponding share of the sum \( z = x + y \). 
Private multiplication between secret-shared values \( [x] \) and \( [y] \) is more complex and is typically implemented using pre-computed Beaver triples~\cite{beaver1992efficient}. A Beaver triple consists of three secret-shared values \( ([a], [b], [c]) \), where \( c = a \cdot b \).  Specifically, the parties compute $[\epsilon] = [x] - [a]$ and $[\delta] = [y] - [b]$, and decrypt $\epsilon$ and $\delta$ without information leakage due to the masking. They compute the result $[x][y] = [c]+\epsilon[b]+[a]\delta+\epsilon\delta$, using trivial implementations of addition and multiplication of secret shares with public values.
Private addition and multiplication can be used to implement a variety of linear functions in neural networks, such as dot products, outer products, matrix products, and convolutions.

%Private addition of two secret shared values, $[z] = [x] + [y]$, is implemented by having each party $p$ sum their shares of $[x]$ and $[y]$: each party $p \in \mathcal{P}$ computes $[z]_p = [x]_p + [y]_p$.
%Private multiplication is implemented using random Beaver triples~\cite{beaver1992efficient}, $([a], [b], [c])$ with $c=ab$. The parties compute $[\epsilon] = [x] - [a]$ and $[\delta] = [y] - [b]$, and decrypt $\epsilon$ and $\delta$ without information leakage due to the masking. They compute the result $[x][y] = [c]+\epsilon[b]+[a]\delta+\epsilon\delta$, using trivial implementations of addition and multiplication of secret shares with public values.
%All the linear functions in neural networks, such as dot products, outer products, matrix products, and convolutions, can be implemented as combinations of private addition and multiplication.

\para{Private non-linear functions}
Non-linear functions commonly used in neural networks, such as ReLU, Sigmoid, and Softmax, are typically implemented using approximations that combine private additions, multiplications, and comparisons. 
Private comparisons are implemented using a function that evaluates whether a secret-shared value \( [z] \) is less than 0 (i.e., evaluates \( [z < 0] \)) by following a series of steps: (1) First, the secret-shared value \( [z] \) is converted to its binary secret share \( \langle z \rangle \). This conversion splits \( [z] \) into individual bits, where each bit is shared among the parties. (2) The sign of \( z \) can be determined by evaluating the most significant bit (MSB) of the binary representation of \( z \). The sign bit \( \langle b \rangle \) is computed by performing a bit shift operation:
\[
\langle b \rangle = \langle z \rangle \gg (L-1),
\]
where \( L \) is the number of bits used to represent the secret-shared value \( z \), and the right shift operation extracts the sign bit. If \( z \) is negative, the MSB will be 1; otherwise, it will be 0. (3) Finally, the resulting binary sign bit \( \langle b \rangle \) is converted back to an arithmetic secret share \( [b] \). This allows the parties to compute the comparison securely, which can then be used for further operations such as determining the output of the ReLU function.

%Non-linear functions in neural networks, such as ReLU, Sigmoid and Softmax are mostly implemented using approximations with private additions, multiplications and comparisons. Additionally, private comparisons are implemented using a function that evaluates $[z < 0]$ by: (1) converting $[z]$ to a binary secret-share $\langle z \rangle$; (2) computing its sign bit, $\langle b \rangle = \langle z \rangle \gg (L-1)$; and (3) converting the resulting bit to an arithmetic sharing $[b]$. 

%% file: tex/design.tex
\section{\sysname Overview}
\label{sec:Preliminaries}

\subsection{Parameter-Efficient Fine-Tuning based Private Inference}
%\hl{重写第一部分，重点体现场景（为何需要模型提供商的adapter），协作计算（数据提供方参与计算的合理性）、威胁模型，隐私目标：公开信息可以泄露，除了能从public数据、输入、输出推测的数据外，不泄露adapter的信息；这里应该可以formal proof（或者在design的部分简单写一段关于安全性的简要分析，但由于比较trivial，所以不做formal proof，或者把formal proof放在附录）}
{Secure computation protocols are well-suited for efficiently handling linear operations (e.g., addition) and low-degree polynomial operations (e.g., multiplication). However, they incur substantial overhead when applied to operations such as comparison, division, or complex nonlinear computations. As a result, neural network operators that are efficient in plaintext settings often become prohibitively expensive under private inference. Prior work on accelerating private inference has largely focused on operator-level approximations or low-level protocol optimizations. In comparison, we aim to investigate the architectural compatibility between neural network designs and the constraints imposed by secure computation. Specifically, we focus on Parameter-Efficient Fine-Tuning (PEFT) methods, which have become standard practice in modern machine learning workflows.}

\para{System model}
%\subsection{Threat Model}
\label{subsec:threatmodel}
\hl{In our settings, we consider a common incentive mechanism in which the model service provider (\ms) offers secure model inference as a paid service, charging clients per query or based on usage. The model is fine-tuned on proprietary data for specific downstream tasks, while leveraging a publicly available backbone model.}
%
%In our setting, we consider a model service provider (\ms) that fine-tunes an adapter model using its proprietary training data while leveraging a publicly available backbone model. 
These proprietary data---such as medical diagnostic images---are highly valuable and specifically tailored for certain downstream tasks. Meanwhile, a model user (\muu) has access to the same public backbone but lacks access to the provider's private training data. To leverage the fine-tuned model's enhanced capabilities for a specific task, the \muu submits its processed input to the \ms and receives the corresponding inference results.%\wenhao{incentive model}

A key challenge in this interaction is preserving the privacy of both parties. The \muu seeks to protect the confidentiality of its input, ensuring that no sensitive information is leaked during inference. Simultaneously, the \ms aims to safeguard the proprietary adapter model from unauthorized access or reconstruction. Notably, since the \muu can observe both the input it provides and the output it receives, some information about the adapter is inherently exposed, potentially making the model susceptible to model inversion attacks. However, we argue that in our setting, the information naturally inferable from the input-output pairs is necessary. Our method ensures that beyond this unavoidable exposure, the \muu gains no additional knowledge about the adapter's private parameters. As such, we consider the defenses against model inversion attacks orthogonal to our work.
%, as our primary focus is on enforcing strict access control to the adapter beyond what is inherently revealed through inference.
%一个典型的隐私推理场景是模型服务商（Model Server， MS）和模型使用方（Model User, MU）都不想泄露各自的隐私，对MS而言，使用大规模私有数据训练或微调模型，模型参数和私有数据需要避免泄露，MU想使用该模型进行推理，不得不将自己的隐私数据发送给模型服务商，由MS完成推理过程并返回推理结果，对MU而言，自己的隐私数据不想暴露给MS，MU希望在本地部署模型并完成推理过程以避免隐私泄露，在明文推理上，MU和MS的需求是矛盾的，而2.3节提到的由密码学支持的隐私推理技术则可以实现上述需求，下面我们给出一个典型的隐私推理过程：
%图~\ref{fig:PEFT_infer}展示了一个使用PEFT技术的MS提供隐私推理的场景，
%We consider a common scenario for private neural network inference involving two parties: the model service provider (\ms) and the model user (\muu). The \ms is responsible for training or fine-tuning models on extensive private datasets, raising concerns about potential leakage of its model parameters and proprietary training data. Meanwhile, the \muu wishes to perform inference using the trained model without revealing its own sensitive inputs to the \ms.
Our research aims to develop efficient private inference techniques that protect the sensitive information of both parties.
In this scenario, the \muu and the \ms collaboratively run the following steps:

\begin{packeditemize}
    \item \textit{\textbf{\muu-side pre-processing.}} The \muu initially performs local computations on the plaintext backbone model using its input data. It then encrypts the resulting intermediate outputs and sends this encrypted data to the \ms, who possesses a specialized adapter.
    \item \textit{\textbf{\ms-side computation.}} Upon receiving the encrypted intermediate data, the \ms processes it using the adapter, ensuring that all computations remain in encrypted form. This involves performing private linear and nonlinear operations collaboratively with the \muu, utilizing secure computation protocols outlined in \autoref{subsec:mpcbackground}.
    \item \textit{\textbf{Result retrieval.}} The \ms sends the encrypted output to the \muu, who decrypts it to obtain the final inference result.
\end{packeditemize}

\ignore{We begin by outlining a typical private inference process as follows:
\begin{packeditemize}
    \item MS使用隐私推理引擎（Private Inference Engine， PIE）处理私有模型（Secret Model， SModel）后得到crypto\_SModel
    \item MU使用PIE处理隐私数据（Secret Input， SInput）后，得到crypto\_SInput，并发送给MS
    \item MS在密文上执行隐私推理，得到密文上的隐私输出crypto\_SOutput，并发送给MU
    \item MU使用PIE处理crypto\_SOutput后，得到明文上的推理结果SOutput
\end{packeditemize}
注意，上述
}

\para{Threat model}
To formalize the privacy guarantees, we adopt the semi-honest (i.e., honest-but-curious) adversarial model.
%we assume that both the \ms and the \muu follow a semi-honest (honest-but-curious) model. 
This implies that both parties adhere to the prescribed protocol but may attempt to extract additional information from the data they receive or observe during the interaction. 
This setup is practical since the server is driven by financial incentives to adhere to protocols and deliver high-quality services, while the client is encouraged to comply with the protocol to access those services. As a result, this semi-honest model is widely used in existing research. \hl{However, our approach can be extended to support malicious adversaries that may arbitrarily deviate from the protocol. Specifically, the protocols can be strengthened by incorporating homomorphic message authentication codes (MACs)~\cite{damgaard2012multiparty}, a standard technique in authenticated secret sharing, which enable each party to verify the correctness of computations performed by others without revealing private data.}
%Although the parties do not violate the protocol's steps, they may analyze the exchanged data to infer details not explicitly disclosed.
%may attempt to glean additional information about each other's private data. Although they do not deviate from the protocol steps, they may scrutinize any data received or observed to infer hidden details not explicitly disclosed by the protocol.
Under this threat model, two key privacy objectives must be met:

\begin{packeditemize}
    \item \textit{\textbf{Confidentiality of \muu's input.}} The \ms should not learn any information about the private input data of \muu. This guarantee is especially crucial when the input data includes proprietary or sensitive information, such as personal health records or confidential business data. 

    \item \textit{\textbf{Confidentiality of \ms's private adapter.}} The \muu obtains the output corresponding to their own input but cannot obtain any additional information of the \ms's private adapter. Any compromise in this regard could undermine the \ms's intellectual property and competitive edge.
\end{packeditemize}

%\wenhao{discuss malicious security}
% discussion section?

%隐私推理过程中，我们假设双方是半诚实的，即双方遵循协议规定，但会尝试从协议的中间结果中获取额外信息

%，在私有神经网络推理~\cite{pang2024bolt,srinivasan2019delphi,hao2022iron,reagen2021cheetah}的相关研究中，半诚实假设是一个常用的假设，采用这一假设的原因有两个，一是MS为MU提供推理服务，二者在一定程度上需要互相信任，二是半诚实假设更切合实际
% 这里要明确涉及的参与方包括：模型提供方M，它基于私有数据（例如医疗、法律等行业数据）微调了一个大模型对外提供服务；数据提供方D，提供隐私输入，经过模型处理，希望得到隐私的结果。
% 计算流程：
% \begin{packeditemize}
%     \item D将其输入I进行处理F(I)后此处的处理可以依赖于公开的backbone，因此F接受一个隐含的输入backbone: B，以密文的形式E(F(I))提供给M。
%     \item M基于E(F(I))结合自己的隐私adapter：A，和public B进行计算，由于在M本地计算，因此adapter和B都是明文，并将最终的计算结果R返回给D。
% \end{packeditemize}
\ignore{
针对上述隐私推理，我们的隐私目标如下：
\begin{packeditemize}
    \item MS无法从crypto\_SInput中恢复出MU的明文的任何信息;
    \item MU可以从crypto\_SOutput恢复出推理结果，但无法获取更多关于SModel的信息。
\end{packeditemize}
}

\subsection{One-way Communication Policy}
\label{subsec:motivations}

\ignore{
\begin{figure}
    \centering
    \includegraphics[width=0.8\columnwidth]{figure/PEFT_infer_1.pdf}
    \caption{PEFT\_infer}
    \label{fig:PEFT_infer}
\end{figure}
}

\begin{figure}
    \centering
    \begin{subfigure}{\linewidth}
        \centering
        \includegraphics[width=\textwidth]{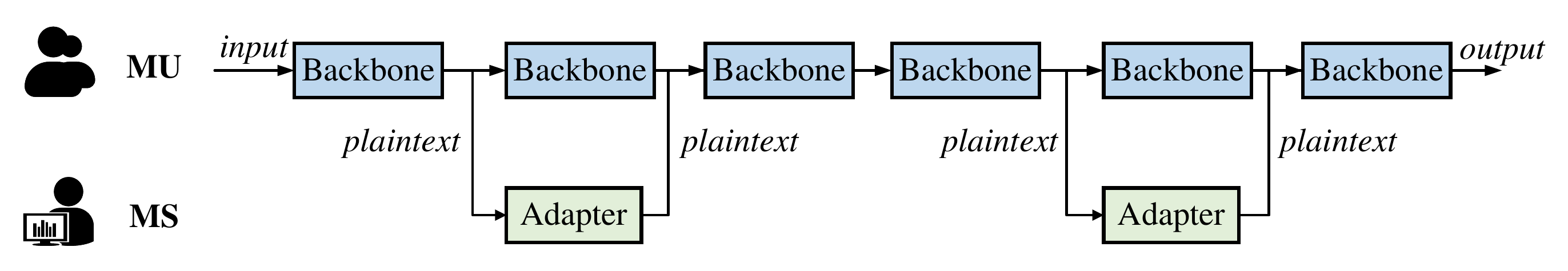}
        \caption{Plaintext inference.}
        \label{fig:plaintext}
    \end{subfigure}
    \begin{subfigure}{\linewidth}
        \centering
        \includegraphics[width=\textwidth]{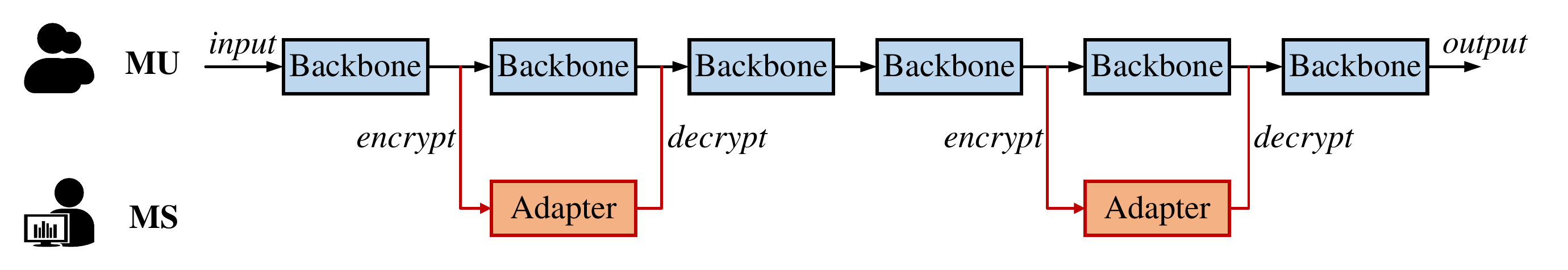}
        \caption{Encrypt-then-decrypt inference (\textit{insecure}).}
        \label{fig:encrypt-then-decrypt}
    \end{subfigure}
    \begin{subfigure}{\linewidth}
        \centering
        \includegraphics[width=\textwidth]{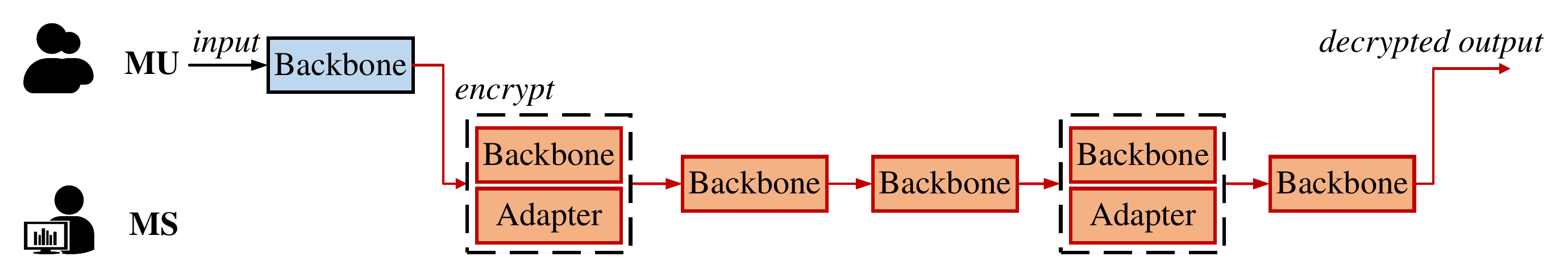}
        \caption{Private inference (\textit{secure but slow}).}
        \label{fig:slow}
    \end{subfigure}
    \caption{Inference workflows in different scenarios. Orange-shaded components indicate computations performed under secure computation protocols.}
    \label{fig:motivations}
\end{figure}

\para{Motivations}
In this paper, we focus on PEFT methods such as AdaptFormer~\cite{chen2022adaptformer} and LoRA~\cite{hu2021lora}, which introduce a small number of trainable parameters while keeping the majority of pre-trained weights frozen. {This approach has been shown to significantly reduce training time, memory consumption, and fine-tuning overhead.} As illustrated in \autoref{fig:plaintext}, inference in plaintext settings typically involves separate computation on the backbone and the adapter, followed by merging intermediate representations before proceeding through subsequent layers. In such environments, PEFT consistently outperforms simpler strategies like fine-tuning only the final layers.

However, in the private inference scenario, existing PEFT techniques typically require bidirectional communication and partial decryption between the backbone and adapter, introducing potential leakage points. 
%require partial decryption or two-way communication, each of which introduces potential leakage points. 
%Although MPC protocols generally protect data through cryptographic means, any disclosure of intermediate activations or parameters---even partial---can compromise these privacy guarantees~\cite{DBLP:conf/ccs/PasquiniAB21}. Such vulnerabilities are especially pronounced when adapters and backbones exchange intermediate features, particularly if decryption is necessary to combine them. Traditional architectures that rely on back-and-forth communication between model components therefore undermine the strict privacy guarantees provided by MPC.
Specifically, a naive (and inherently \textit{insecure}) approach to applying PEFT in this setting is illustrated in \autoref{fig:encrypt-then-decrypt}: the \muu uses the public backbone to compute intermediate results in plaintext, encrypts these results, and sends them to the \ms, which holds the adapter. The \ms then processes it with the adapter using secure computation protocols, and returns the partial output to the \muu so that the \muu can decrypt them and continue the inference process. While this design minimizes the computational load by retaining plaintext operations in the backbone, it directly violates the privacy objectives set forth in \autoref{subsec:threatmodel}. As a result, conventional PEFT architectures that depend on iterative interactions between the backbone and adapter are fundamentally incompatible with the strict end-to-end privacy guarantees expected in private inference. To fully meet those objectives, most computations---including those performed by the backbone---must be performed over encrypted data (\autoref{fig:slow}). This requirement renders the naive design impractically slow and resource-intensive, highlighting the core challenge of applying PEFT in private inference scenarios. 

%The limitations outlined above highlight the need for novel design paradigms that seamlessly integrate MPC protocols and modern fine-tuning strategies from the ground up. Rather than merely retrofitting existing designs to the private inference scenario, a more cohesive framework is necessary—one that inherently minimizes data leakage risks while maintaining high inference accuracy and efficiency. 
% By emphasizing one-way data flow, low-rank parameter adaptations, and selective layer updates, researchers can build models that align more closely with MPC requirements. Ultimately, such MPC-friendly architectures will be critical for enabling large-scale, private inference without incurring prohibitive computational overhead.

% \subsection{Observations}

%由于transformer模型快速增长的存储开销和训练成本，我们考虑MS使用PEFT技术，基于冻结公开的backbone模型，通过训练和微调adapter实现面向特定下游任务的高效模型。在该场景下，MS和MU都持有公开的Backbone及其参数，此外MS拥有adapter作为其私有模型。
% Due to the increasing storage overhead and training costs of transformer models, we consider the use of PEFT by \ms. The \ms leverages a publicly available backbone model, freezing its parameters, and then trains and fine-tunes an adapter to efficiently tailor the model for specific downstream tasks. In this scenario, both the \ms and the \muu possess the public backbone and its parameters, while the \ms holds the adapter as its proprietary model.
%As shown in Figure~\ref{fig:PEFT_infer}, 

%如图1所示，MU和MS联合进行隐私推理的过程如下：
\ignore{
\begin{packeditemize}
    \item MS使用PIE处理SModel，得到crypto\_SModel
    \item MU使用SInput在本地的Backbone上执行部分明文推理，得到中间推理结果，需要使用Adapter时，MU使用PIE处理中间推理结果，并通过网络发送给MS
    \item MS在密文上执行隐私推理，得到密文上的隐私输出crypto\_SOutput，并把发送给MU
    \item MU使用PIE处理crypto\_SOutput后，得到明文上的推理结果SOutput，使用SOutput在本地的Backbone上执行后续的明文推理，需要使用Adapter时，重复上述过程。
\end{packeditemize}
}

\para{Solution}
%\label{subsec:intuitions}
A critical challenge in private inference lies in ensuring that neither the \ms nor the \muu obtains unnecessary information beyond what is essential for producing the final result. Traditional fine-tuning pipelines often rely on iterative or bidirectional data exchange, a process we refer to as \textit{two-way communication} (TWC).
To tackle the challenges above, we introduce a key insight for reconciling PEFT with private inference: \emph{one-way communication (OWC)}. This approach involves transferring data \emph{only once and in a single direction}, eliminating the need for decryption or the back-and-forth exchange of intermediate representations. 

%Below, we explain how OWC addresses information leakage, simplifies the process, and seamlessly integrates into MPC environments.
 
In TWC, each step typically involves decrypting partial outputs and exchanging them between components (e.g., the adapter and backbone). This intermediate decryption poses significant risks, as it exposes internal features that malicious or curious parties could exploit to extract private information.
In contrast, OWC eliminates this vulnerability by enforcing that intermediate activations never return to earlier layers or to the other party in plaintext. Instead, all data processing—including that performed by the adapter—happens in a forward-only manner. Specifically, any intermediate representation remains in encrypted form as it passes through the model. Critically, \emph{no} step in the pipeline requires decrypting these activations until the final output is ready for the legitimate recipient. This design intrinsically aligns with MPC's principle of keeping all intermediate computations locked behind cryptographic methods, thereby eliminating leakage risks.

%% file: tex/workflow.tex
\subsection{\sysname Workflow}
\label{workflow}

\begin{figure*}
    \centering
    \includegraphics[width=0.85\textwidth]{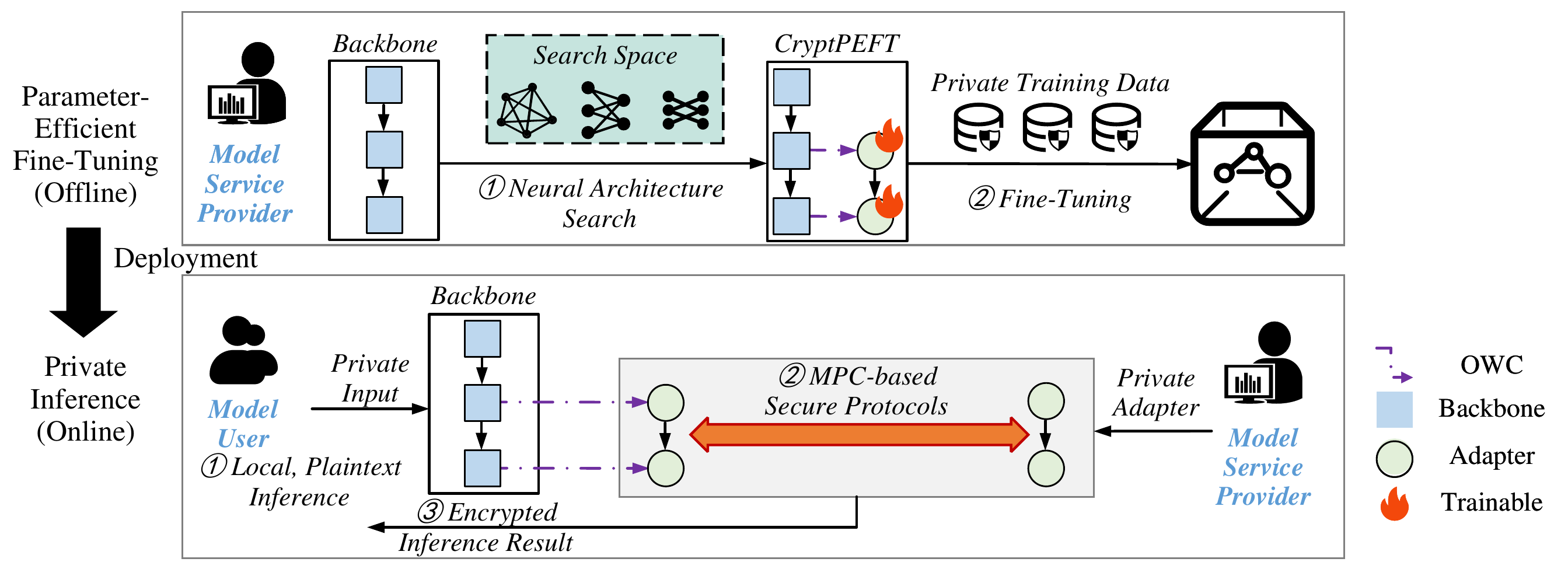}
    \caption{Workflow of \sysname.}
    \label{fig:CryptPEFT_infer}
\end{figure*}

In this paper, we propose \sysname, the first PEFT architecture specifically designed for private PEFT inference scenarios.
%To address the challenges of adapting PEFT to private inference, we propose \sysname. 
\autoref{fig:CryptPEFT_infer} provides an overview of \sysname's workflow, illustrating the secure collaboration between the \muu and the \ms while ensuring compliance with the OWC policy.

Specifically, the \muu holds a public backbone network, which it uses to perform local inference on the input data in plaintext. When specialized adapter-based processing, owned by the \ms, is required, the \muu encrypts the intermediate results using an MPC-based secure protocol. These encrypted intermediate results are then transmitted to the \ms for secure processing using the adapter. Notably, the \muu does not need to wait for the \ms's response to continue its local, plaintext computations. The \ms focuses on receiving ciphertext from the \muu, performing private inference using the adapter, and returning the encrypted prediction results. Upon receiving these encrypted outputs, the \muu decrypts them to produce the final inference results. Throughout this workflow, only the final prediction result is transmitted from the \ms to the \muu in ciphertext form—a necessary step—while all other intermediate results flow unidirectionally from the \muu to the \ms in encrypted form.

A potential concern with removing feedback loops is whether it diminishes the overall expressive power of the model. Traditional architectures often rely on residual or skip connections, effectively implementing partial ``feedback'' of intermediate activations. In practice, many deep learning models already exhibit strong utility even when most layers remain unaltered. Coupled with PEFT, our experience has shown that carefully designed adapter structures and the strategic placement of adapters in deeper layers (while maintaining a unidirectional flow) can maintain robust model utility. This approach preserves the core objective of fine-tuning, i.e., adapting pre-trained models to new tasks or data distributions, while adhering to the OWC policy essential for efficient private inference. The details for the design of OWC-compliant adapters will be presented in \autoref{sec:method}.

Ultimately, OWC introduces a novel design paradigm for integrating cryptographic techniques with modern PEFT architectures. Instead of retrofitting existing models—often burdened by bidirectional data flows—OWC serves as a guiding principle that informs model construction to inherently protect intermediate computations. As elaborated in the following sections, OWC is foundational to building secure and parameter-efficient pipelines, enabling the adaptation of large pre-trained models to new tasks under strict privacy constraints without incurring excessive overhead.

%% file: tex/search.tex
\section{\sysname Design Details}
\label{sec:method}

In \autoref{subsec:motivations}, we established that enforcing OWC policy is essential for enabling efficient PEFT-based private inference.
%Building on this insight, we introduce the first core design principle of \sysname:
However, traditional adapter architectures experience a notable degradation in accuracy when subject to OWC constraints. To overcome this challenge, we begin with key empirical observations and design principles in \autoref{subsec:understanding}, which serve as the foundation for our approach. Based on these principles, we describe the architecture and placement strategies for the adapters in \autoref{subsubsec:adapter_arch} and \autoref{subsubsec:organizational_strategy}, respectively. In \autoref{subsec:grid}, we introduce an algorithm that automates the search for optimal adapter configurations. %Lastly, \autoref{subsubsec:approximation} outlines the approximation techniques employed to enhance the efficiency of private inference under secure computation constraints.

% 列举实验的发现，以及如何基于这些发现提出结构设计的一些原则，最后是怎么search（grid search也可以）出最终的结构的。
%在本节中，我们将主要介绍如何设计一个有效的CryptPEFT，传统的PEFT在设计时需要考虑如何设计PEFT中使用的Adapter、Adapter的组织方式（包括放置的位置和与Backbone的连接方式），最后还需要全局性的进行超参搜索以保证成本和性能的折中，CryptPEFT也不例外，设计CryptPEFT还需要考虑在密文上计算的高效性[引几篇密文隐私推理加速的文章，MPC和FHE]，即Efficient Approximation of Adapter

%In this section, we present the design of an effective CryptPEFT. Like traditional PEFT, this involves defining the adapter structure, as well as its placement and connection to the backbone. We also introduce a method to search for optimized hyperparameters to balance the cost of private inference with model performance. Additionally, we discuss an approach for efficiently approximating the adapter to optimize performance during private inference.

% \subsection{Methodology Overview}
% 在本节，我们从总体上介绍CryptPEFT的结构以及如何应用在隐私推理场景，CryptPEFT的设计细节和策略我们会在第4节讨论。

% 我们延用3.2节的隐私推理场景，将传统的PEFT结构替换成CryptPEFT，如图~\ref{fig:CryptPEFT_infer}所示，MU持有公开的Backbone，可以先在本地使用明文进行推理，一旦需要使用MS提供的Adapter，MU使用PIE处理中间结果得到密文，将密文发送给MS，而无需等待MS的响应即可继续执行后续的明文推理，MS只负责接收从MU发来的密文并执行隐私推理，直到最后，隐私推理结束，MS把最终的推理结果（密文形式）发送给MU，MU使用PIE处理后得到明文上的推理结果。这一过程中，除了最终的推理结果需要MS发送给MU外(这是不可避免的)，其余中间结果都是MU向MS以密文形式发送，满足OWC。

\subsection{Understanding OWC-Compliant Adapters}
\label{subsec:understanding}

% Besides privacy benefits, OWC also has important implications for protocol design. Typical two-party MPC protocols can become cumbersome when intermediate states must be re-encrypted and re-decrypted at every layer. Each additional encryption-decryption cycle introduces overhead in terms of both computation and communication. By contrast, OWC naturally streamlines this procedure: partial features are passed forward (still in encrypted form) and do not require ``round-trips'' that would inflate the overall latency. Consequently, adopting OWC reduces the total number of cryptographic operations, thereby increasing efficiency.\wenhao{update this paragraph}

%OWC is especially synergistic with low-rank adapter techniques (e.g., LoRA~\cite{hu2021lora}) and selective fine-tuning, since these approaches already prioritize minimal modifications to the model. Rather than scattering small adapters throughout the network—which would necessitate repeated two-way communication—one can place a small number of trainable parameters at deeper layers and pass data forward through these adapters only once. This strategy inherently complements OWC by localizing the encrypted computations to a sequence of carefully orchestrated operations, significantly reducing both the leakage points and the computational overhead.

In compliance with the OWC policy, data flows unidirectionally from the backbone to the adapter. This fundamental constraint necessitates a careful reconsideration of the adapter architecture to maintain both high accuracy and private inference efficiency. We first examine the impact of several critical factors in adapter design with empirical studies, ultimately leading to the formulation of the key design constraints. Specifically, we aim to address the following questions.

\question{(Q1): \textit{Are traditional adapters effective for \sysname?}}

To evaluate how the OWC constraint affects model performance, we conducted a series of experiments using traditional PEFT methods such as LoRA and AdaptFormer, both originally designed under the Two-Way Communication (TWC) paradigm. In these experiments, we held all training hyperparameters and downstream tasks (i.e., dataset for specific tasks) unchanged, modifying only the communication policy---from TWC to OWC. As shown in \autoref{tab:TWC_vs_OWC}, this shift leads to considerable fluctuations in model utility. In most downstream tasks, we observe a clear performance drop. Specifically, on the SVHN dataset---a street-view digit recognition task with only 10 labels and a data distribution that deviates significantly from standard object classification---the model's utility decreases by an average of 28.63\%.

\begin{table}
\centering
\caption{Classification accuracy (\%) of traditional adapters across different datasets under the TWC and OWC settings.}
\label{tab:TWC_vs_OWC}
\begin{tabular}{ccc | cc}
\toprule
\multirow{2}{*}{\textbf{Datasets}} &   \multicolumn{2}{c}{\textbf{LoRA}} & \multicolumn{2}{c}{\textbf{AdaptFormer}}  \\
\cmidrule(l){2-5}
& TWC & OWC & TWC & OWC \\
\midrule
CIFAR-10 & 97.31 & 95.84 {\cellcolor{lightgray}\color{red}(↓1.47)} & 97.34 & 95.91 {\cellcolor{lightgray}\color{red}(↓1.43)} \\
CIFAR-100 & 87.41 & 83.73 {\cellcolor{lightgray}\color{red}(↓3.68)} & 87.23 & 83.72 {\cellcolor{lightgray}\color{red}(↓3.51)} \\
%Caltech-101 & 94.93 & 95.79 {\cellcolor{lightgray}\color{teal}(↑0.86)} & 94.53 & 95.91 {\cellcolor{lightgray}\color{teal}(↑1.38)} \\
Food-101 & 83.95 & 80.13 {\cellcolor{lightgray}\color{red}(↓3.82)} & 83.91 & 80.02 {\cellcolor{lightgray}\color{red}(↓3.89)} \\
SVHN & 91.81 & 63.17 {\cellcolor{lightgray}\color{red}(↓28.64)} & 91.72 & 63.11 {\cellcolor{lightgray}\color{red}(↓28.61)} \\
Flowers-102 & 84.37 & 80.76 {\cellcolor{lightgray}\color{red}(↓3.61)} & 84.42 & 80.89 {\cellcolor{lightgray}\color{red}(↓3.53)} \\
\midrule
\end{tabular}
\end{table}

\ignore{
\begin{table}[htbp]
\centering
\caption{Comparison of LoRA and AdaptFormer on different datasets under TWC and OWC settings}
\label{tab:TWC_vs_OWC}
\resizebox{\columnwidth}{!}{%
\begin{tabular}{llccc}
\toprule
\textbf{Method} &       & \textbf{CIFAR-10} & \textbf{CIFAR-100} & \textbf{Caltech-101} \\
\midrule
\multirow{2}{*}{LoRA} 
& \cellcolor{lightgray}TWC & \cellcolor{lightgray}97.31 & \cellcolor{lightgray}87.41 & \cellcolor{lightgray}94.93 \\
& OWC & 95.84 {\color{red}(↓1.47)} & 83.73 {\color{red}(↓3.68)} & 95.79 {\color{green}(↑0.86)} \\
\midrule
\multirow{2}{*}{AdaptFormer} 
& \cellcolor{lightgray}TWC & \cellcolor{lightgray}97.34 & \cellcolor{lightgray}87.23 & \cellcolor{lightgray}94.53 \\
& OWC & 95.91 {\color{red}(↓1.43)} & 83.72 {\color{red}(↓3.51)} & 95.91 {\color{green}(↑1.38)} \\
\midrule
\end{tabular}
}
\vspace{0.5em}
\resizebox{\columnwidth}{!}{%
\begin{tabular}{llccc}
\toprule
\textbf{Method} &       & \textbf{Food-101} & \textbf{SVHN} & \textbf{Flowers-102} \\
\midrule
\multirow{2}{*}{LoRA} 
& \cellcolor{lightgray}TWC & \cellcolor{lightgray}83.95 & \cellcolor{lightgray}91.81 & \cellcolor{lightgray}84.37 \\
& OWC & 80.13 {\color{red}(↓3.82)} & 63.17 {\color{red}(↓28.64)} & 80.76 {\color{red}(↓3.61)} \\
\midrule
\multirow{2}{*}{AdaptFormer} 
& \cellcolor{lightgray}TWC & \cellcolor{lightgray}83.91 & \cellcolor{lightgray}91.72 & \cellcolor{lightgray}84.42 \\
& OWC & 80.02 {\color{red}(↓3.89)} & 63.11 {\color{red}(↓28.61)} & 80.89 {\color{red}(↓3.53)} \\
\bottomrule
\end{tabular}
}
\end{table}
}

This degradation highlights a core limitation of conventional adapter architectures under the OWC constraint: their \textit{inability to effectively capture inter-token dependencies}. In TWC-based PEFT methods, task-specific features extracted by the adapter are re-integrated into the backbone, allowing the backbone's attention mechanisms to recompute and propagate information across tokens. However, OWC restricts such bidirectional interaction, preventing the features extracted by adapters from propagating across the tokens---including the \texttt{[CLS]} token, which is crucial for downstream prediction. As a result, the model loses a key pathway for token-level information exchange, which is especially detrimental for tasks with complex or non-standard input distributions. To address this challenge, our adapter design must incorporate an internal attention block within the adapter that explicitly models inter-token relationships. This allows the adapter to refine the representation of the \texttt{[CLS]} token and recover utility.

%In traditional PEFT frameworks, features extracted for downstream tasks are integrated into the backbone, allowing attention re-computation on the backbone. However, the OWC principle requires these features to remain separate from the backbone, necessitating attention re-computation on the adapter.
 
%\hl{@saisai: explain why attention is necessary} This highlights the need for including the attention mechanism in the adapter architecture tailored under the OWC constraint.\wenhao{两个现象：(1)普遍会降低utility，这可能是由于无法capture inter-token dependencies;（2）有些数据集会下降特别多，说明适配不同的下游任务时的generability会比较差。}

%为了验证上述问题，我们做了一系列直观的对比实验，对于一个传统的PEFT方法，例如LoRA和AdaptFormer，其依赖TWC原则，在控制变量的情况下，我们仅将TWC原则改为OWC原则，使用相同的训练超参和相同的下游任务，实验结果如\autoref{tab:TWC_vs_OWC}，我们观察到，从TWC转换到OWC时，存在严重的utility波动问题，在大部分的下游任务上，模型utility都大幅下降，特别是当任务风格出现较大变化时，以SVHN为例，这是一个街景门牌号数字识别任务，只有10个标签，和普通的物品分类任务有较大区别，模型的utility平均下降28.63\%。在我们实验的个别下游任务中，模型utility有一定的上升，我们猜想这是受益于OWC原则，在我们后面针对\sysname的实验上也证实了这一猜想。一系列实验表明，传统的adapter缺少注意力机制，在OWC原则下，无法主动学习到各个Token之间的关系，在很多下游任务中表现不佳，我们需要设计一个适配OWC的adapter。

\question{(Q2): \textit{Are existing attention mechanisms efficient for \sysname?}}

\begin{figure}
    \centering
    \includegraphics[width=\columnwidth]{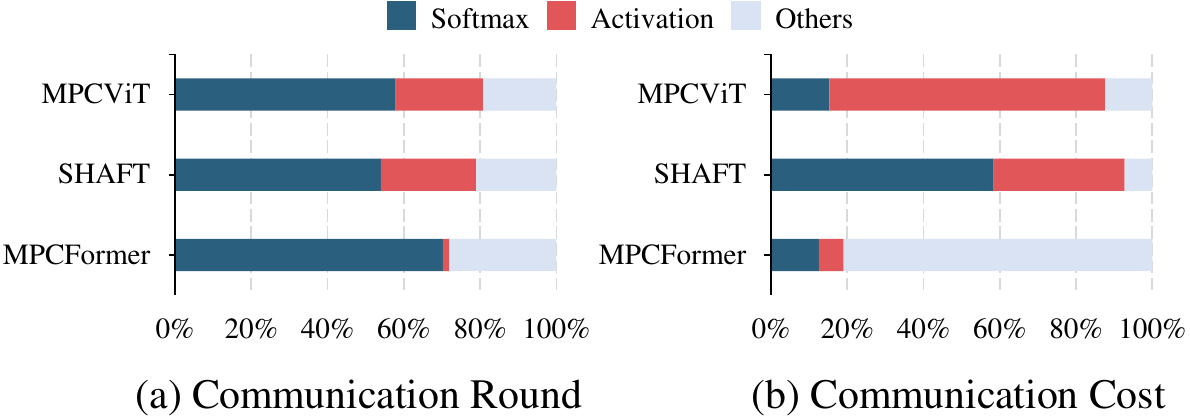}
    \caption{Communication overhead breakdown: Softmax and activations are dominant contributors in existing methods.}
    \label{fig:Q2_comm_Rounds_Cost}
\end{figure}

%\para{Observation 3: Traditional attention schemes are costly for private inference (not only attention, but also Lora)}
%In standard Transformer models, Attention and MLP are widely employed to capture long-range dependencies within input sequences (\autoref{subsec:transformer_background}). 
In standard Transformer architectures, attention and MLP layers play a central role in modeling long-range dependencies within input sequences (\autoref{subsec:transformer_background}).
However, in private inference settings, the Softmax operation in attention and the GELU activation in MLP introduce significant computational and communication overhead. To address this, recent works have proposed MPC-friendly alternatives. MPCFormer~\cite{li2022mpcformer} approximates both Softmax and GELU using low-degree polynomials. MPCViT~\cite{zeng2023mpcvit} adopts a hybrid attention mechanism—combining ReLU Softmax Attention~\cite{RSAttention} and Scaling Attention~\cite{ScaleAttention}—and replaces GELU with the more MPC-efficient ReLU. SHAFT~\cite{ndss/KeiC25_SHAFT} further improves communication efficiency by introducing a constant-round Softmax protocol and an MPC-tailored GELU approximation.
Since MPC-based private inference is highly sensitive to communication---both in rounds and bandwidth---we evaluate the communication overhead of these methods {within an adapter block} using CrypTen~\cite{crypten2020}.
As shown in \autoref{fig:Q2_comm_Rounds_Cost}, the Softmax and activation functions in MPCViT and SHAFT account for over 80\% of total communication overhead with the private inference of an adapter block. While MPCFormer reduces bandwidth usage, it still incurs a high number of communication rounds.
These results highlight that directly applying existing attention and activation designs in adapter modules can severely impact private inference efficiency, underscoring the need for more communication-efficient alternatives.
%while the approaches in MPCViT and SHAFT achieve functional correctness, they incur high communication cost and round complexity. MPCFormer performs better in terms of communication cost but still suffers from a high number of communication rounds. 
%These findings suggest that directly adopting such designs in adapter modules would significantly degrade private inference efficiency, underscoring the need for more communication-friendly alternatives.
%针对上述问题，有一系列不同方向的研究，如MPCFormer~\cite{li2022mpcformer}使用高效的多项式近似Softmax和GELU。MPCViT~\cite{zeng2023mpcvit}混合使用ReLU Softmax Attention~\cite{RSAttention}和Scaling Attention~\cite{ScaleAttention}，并使用对MPC更加友好的ReLU替换MLP中的GELU。SHAFT(NDSS2025)~\cite{ndss/KeiC25_SHAFT}提出了常数轮的Softmax协议以及对MPC更加友好的GELU近似。
%基于MPC的隐私推理对通信十分敏感，具体来说，通信的延迟和带宽共同影响整体的通信效率，对MPC友好的结构需要具备低通信轮数和低通信量的特点。我们用CrypTen~\cite{crypten2020}复现上述方案，并分析了Softmax和activation function的通信轮数和通信量，实验结果如\autoref{fig:Q2_comm_Rounds_Cost}所示，MPCViT和SHAFT中提出的Softmax和activation function在communication round和communication cost方面均占比较高，MPCFormer中提出的Softmax和activation function在communication cost上占比不高，但在communication round方面占比仍然很高，因此，直接在adapter中使用这些方案，会降低隐私推理的效率。

%However, in practice, many deep learning models already exhibit strong utility even when most layers remain unaltered. Coupled with PEFT, our experience has shown that selectively positioning adapters in deeper layers (while maintaining a unidirectional flow) can achieve robust model utility. This approach preserves the essence of fine-tuning—adapting the model to new tasks or data distributions—while respecting the strict one-way flow of information critical for MPC.

%Ultimately, OWC offers a fresh design perspective on how to integrate cryptographic methods with modern fine-tuning strategies. Rather than retrofitting existing architectures—often riddled with bidirectional data transfers—OWC is a guiding principle that shapes model construction to inherently safeguard intermediate computations. As we discuss in subsequent sections, adopting OWC is the cornerstone for building a secure, parameter-efficient pipeline, ensuring that large pre-trained models can be adapted to new tasks under rigorous privacy constraints without incurring prohibitive computational costs.

\question{(Q3): \textit{How does the placement of adapters affect model utility?}
\label{ques:Q2}
}

Conventional PEFT methods typically insert adapters into every layer of the backbone, resulting in a number of adapters that scales with model depth. While this design improves model expressiveness, it substantially compromises the efficiency of private inference. In this work, we revisit the adapter placement strategy within the CryptPEFT framework, aiming to preserve utility while minimizing the number of adapters. A central question we explore is whether positioning adapters in shallower or deeper layers significantly affects model performance. This sparse adapter setting not only improves computational efficiency but also reduces the search space for neural architecture search (NAS).

\begin{figure}
    \centering
    \includegraphics[width=0.8\columnwidth]{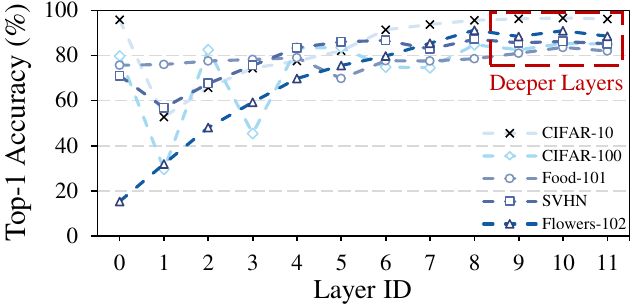}
    \caption{Impact of adapter placement on model utility.}
    \label{fig:Q3_adapter_pos}
\end{figure}

%\wenhao{用的什么adapter，是OWC还是TWC}\xss{OWC}
To investigate this,
%we conduct a series of experiments where only a single adapter is inserted into the backbone, and its position is varied across different layers within the network.
%\xss{To systematically investigate the impact of adapter placement, 
we conduct a series of experiments in which a single adapter---comprising a low-rank decomposition module, an attention mechanism, and a MLP module---is integrated into the backbone architecture under the constraints of OWC. The adapter is inserted at varying depths across different layers of the network to evaluate its influence on model utility while adhering to OWC.
%}
As shown in \autoref{fig:Q3_adapter_pos}, the results indicate that, across many downstream tasks, placing the adapter in deeper layers consistently yields superior performance—highlighted by the deep red dashed boxes in the figure. These findings suggest that, under tight resource or efficiency constraints, prioritizing adapter placement in deeper layers is a more effective strategy for preserving model utility.

%传统的PEFT方法强调对模型的每个基础层都应用adapter，adapter的数量会随着模型深度而增加，这显然不利于提升隐私推理效率，我们重新思考adapter在CryptPEFT中的组织方式，并尝试使用尽可能少的adapter达到可观的utility，当使用少量adapter时，adapter的放置位置表现得更加重要，我们的疑问：将adapter放置在较浅的位置或者放置在较深的位置是否会显著影响模型性能？为了消除上述疑问，我们进行了一系列实验，仅为整个Backbone使用一个adapter，通过改变adapter的位置，研究放置adapter的最佳位置。实验结果如\autoref{fig:Q2_adapter_pos}所示，我们观察到，在多个下游任务中，adapter的最佳位置是Deeper Layers——使用深红色虚线框示意。因此，在使用少量adapter时，我们需要优先将adapter放置在Deeper Layers

\para{Summary of design constraints}
In summary, to fully harness the utility benefits of PEFT within \sysname's workflow, \sysname enforces 4 key constraints and incorporates a neural architecture search (NAS) mechanism. 
%We detail these 4 critical components below:

% \begin{itemize}
%\noindent$\bullet$\textit{~Constraint 1: Ensuring One-Way Communication (OWC).}
%This constraint enforces a unidirectional flow of encrypted data, preventing any back-and-forth exchange that might require decryption mid-inference. By adhering to OWC, \sysname effectively reduces potential leakage points.

\noindent$\bullet$\textit{~Constraint 1: Ensuring One-Way Communication (OWC).}
This constraint enforces a unidirectional flow of encrypted data, preventing any back-and-forth exchange that might require decryption mid-inference. By adhering to OWC, \sysname effectively reduces potential leakage points.

\noindent$\bullet$\textit{~Constraint 2: Integrating attention mechanisms with the adapters.} 
Adapters augmented with attention mechanisms demonstrate a strong capacity to model inter-token dependencies and capture global contextual representations, thereby facilitating effective updates to the \texttt{[CLS]} token. When adhering to OWC, models incorporating such attention-based adapters are still capable of preserving a high level of utility.

%\noindent$\bullet$\textit{~Constraint 2: Focusing on LoRA.}
%LoRA~\cite{hu2021lora} applies a low-rank decomposition to efficiently capture task-specific knowledge within adapters. By confining trainable parameters to compact low-rank matrices, \sysname substantially lowers computational overhead in MPC settings.

\noindent$\bullet$\textit{~Constraint 3: Using MPC-friendly attention and activation functions.}
The fundamental operations supported by MPC are addition and multiplication, while other operations require the use of approximation algorithms or protocols, which inevitably increase the communication rounds and communication cost. Therefore, adapters should aim to minimize the use of complex approximation algorithms and protocols.

\noindent$\bullet$\textit{~Constraint 4: Focusing on deeper layers.}
Rather than inserting adapters throughout the entire backbone, \sysname targets deeper layers, which typically carry high-level features essential for final predictions. This selective adjustment yields a favorable trade-off between model accuracy and efficiency.

\noindent$\bullet$\textit{~Neural Architecture Search (NAS).}
To identify the optimal configurations of adapter structures and placements, \sysname employs a NAS approach. By exploring different architecture candidates under the above constraints, it systematically discovers a model architecture that maximizes model accuracy while minimizing private inference latency.

\begin{figure*}[h!]
    \centering
    \includegraphics[width=0.8\textwidth]{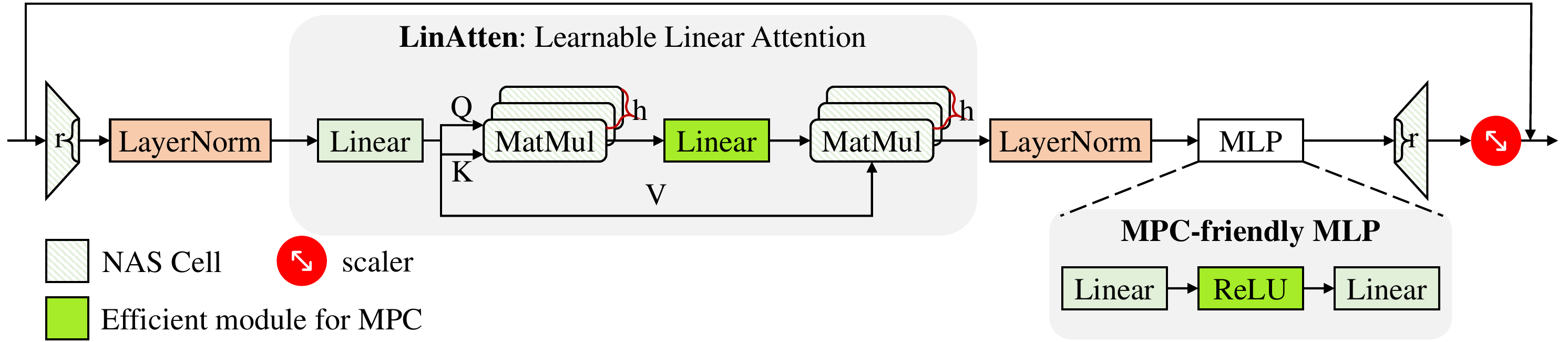}
    \caption{The proposed adapter structure in \sysname.}
    \label{fig:adapter}
\end{figure*}

\subsection{\sysname Adapter Architecture}
\label{subsubsec:adapter_arch}

\autoref{fig:adapter} shows the core design of our adapter. The fundamental goal is to \emph{update the \texttt{[CLS]} token exclusively within the adapter}, ensuring that no information is propagated back into the backbone. 
%This design is essential for preserving the OWC constraint and avoiding any bidirectional interaction.
Since the backbone's \texttt{[CLS]} token cannot be updated by an external module under OWC, our adapter {incorporates} an attention block to extract and refine features specifically for the \texttt{[CLS]} token.

\ignore{To mitigate the overhead of private SoftMax computation, we extend the concept of "Non-local Neural Networks"~\cite{ScaleAttention} and propose a novel attention mechanism: learnable linear attention (\atten). "Non-local Neural Networks"提出了一个通用的非局部操作来捕获长距离依赖，该操作的形式化定义为\[
\text{Non-local}(x) = \frac{1}{C(x)} \sum_{i} f(x_i, x) g(x_i)
\]
其中，$f(\cdot)$是相似度函数，$g(\cdot)$是一个线性变换，$\frac{1}{C(\cdot)}$是一个归一化操作，在\atten中，对$f(\cdot)$和$g(\cdot)$，我们沿用经典注意力机制中的点积操作和矩阵乘，但为了加速隐私推理，我们把$\frac{1}{C(\cdot)}$近似为线性变换，最终，\atten的形式化定义为\[
\text{\atten}(Q, K, V) = L(QK^T)V
\]
其中$L(\cdot)$表示线性变换，形式化为$L(x) = W_L x$.}
%------------ end of ignore

To reduce the cost of private SoftMax in existing attention mechanisms, we draw inspiration from \textit{non-local neural networks}~\cite{ScaleAttention} and propose a novel attention mechanism: \textit{learnable linear attention (\atten)}. Non-local neural networks introduce a general non-local operation to capture long-range dependencies, defined as:
\[
\text{Non-local}(x) = \frac{1}{C(x)} \sum_{i} f(x_i, x) g(x_i).
\]
Here, \( f(\cdot) \) is a similarity function, \( g(\cdot) \) is a linear transformation, and \( \frac{1}{C(\cdot)} \) is a normalization factor. In \atten, we adopt the standard dot-product operations from classical attention for both \( f(\cdot) \) and \( g(\cdot) \), i.e., matrix multiplication. Moreover, to enhance private inference efficiency, we replace the normalization term \( \frac{1}{C(\cdot)} \) with a learnable linear transformation.
As a result, \atten is formally defined as:
\[
\text{\atten}(Q, K, V) = L(QK^T)V,
\]
where \( L(\cdot) \) denotes a learnable linear transformation, parameterized as \( L(\mathbf{x}) = W_L \mathbf{x} \).
\atten is a highly efficient alternative that relies solely on linear operations, i.e., multiplication and addition, making it particularly well-suited for private inference while maintaining effective attention modeling. Our evaluation shows that \atten achieves a good utility-efficiency tradeoff (\autoref{tab:performance} and \autoref{tab:LAN_batch64} in \autoref{sec:eval}).

Although \atten eliminates expensive operations such as SoftMax, directly applying it at the full feature dimensionality can remain computationally burdensome in private inference settings due to the scale of the involved matrix operations. To alleviate this issue, we adopt a low-rank adaptation strategy~\cite{hu2021lora} to strike a balance between model expressiveness and computational efficiency. By constraining the learnable parameters to low-rank matrices, \sysname significantly reduces both computational and communication overhead in secure MPC environments. Specifically, adapter weight matrices are decomposed into low-rank components parameterized by a tunable rank $r$, enabling fine-grained control over the model's parameter count and the private inference efficiency.

To further optimize the efficiency of the MLP module under MPC, we replace the GELU activation with ReLU, a more computation-friendly alternative in secure MPC environments, thereby yielding an MPC-friendly MLP design.

%To enhance the efficiency of the MLP module under MPC, we substitute the GELU activation function with ReLU, which is more suitable for secure computation, thus constructing an MPC-friendly MLP.

%LoRA~\cite{hu2021lora} introduces a low-rank decomposition that enables adapters to capture task-specific information efficiently while significantly reducing the number of trainable parameters.

% The adapter's transformation is given by:
% \begin{equation}
% \text{Adapter}(\mathbf{x}_{j}) = \text{scaler} \cdot \text{Up}\bigl(\text{Encode}(\text{Down}(\mathbf{x}_{j}))\bigr),
% \end{equation}
% where:
% \begin{equation}
%     \text{Down}(\mathbf{x}) = W_{\text{down}} \cdot \mathbf{x} + B_{\text{down}}, \quad
%     \text{Up}(\mathbf{x}) = W_{\text{up}} \cdot \mathbf{x} + B_{\text{up}},
% \end{equation}
% and
% {\footnotesize
% \begin{equation}
% \text{Encode}(\mathbf{x}) = \text{MLP}\Bigl(\text{LN}\bigl(\text{MHSA}(\text{LN}(\mathbf{x})) + \mathbf{x}\bigr)\Bigr) 
% \;+\; \bigl(\text{MHSA}(\text{LN}(\mathbf{x})) + \mathbf{x}\bigr).
% \end{equation}
% }

%\wenhao{linear attention}

%We further introduce a \textit{scaler} factor---a multiplicative constant---to balance the contribution of the adapter's output. While public scenarios often fix this scaler at small values (e.g., 0.1), our more constrained OWC architecture may require alternative settings for optimal accuracy-efficiency tradeoffs. \autoref{subsec:grid} demonstrates the impact of varying this parameter on the utility of the model.
We also introduce a \textit{scaler} factor—a tunable multiplicative constant—to modulate the contribution of the adapter output. While this factor is often fixed to small values (e.g., 0.1) in plaintext settings, our OWC-constrained architecture may benefit from alternative configurations to better navigate the accuracy-efficiency tradeoff. The effects of varying this \textit{scaler} factor are empirically explored in \autoref{subsec:grid}.

Finally, the number of attention heads 
$h$ in \atten affects both the model's representational capacity and the cost of core operations such as matrix multiplication. We allow $h$ to be adjusted based on the requirements of downstream tasks, retaining only the minimal number of heads necessary to sustain competitive accuracy while minimizing overhead.

%The number of attention heads \(h\) in \atten affects both model capacity and the computational overhead of functions such as MatMul. We allow \(h\) to vary according to downstream tasks, retaining only as many heads as necessary to achieve competitive accuracy while minimizing computation cost.

\subsection{Organizational Strategy of the Adapter}
\label{subsubsec:organizational_strategy}

\begin{figure}
    \centering
    \includegraphics[width=\columnwidth]{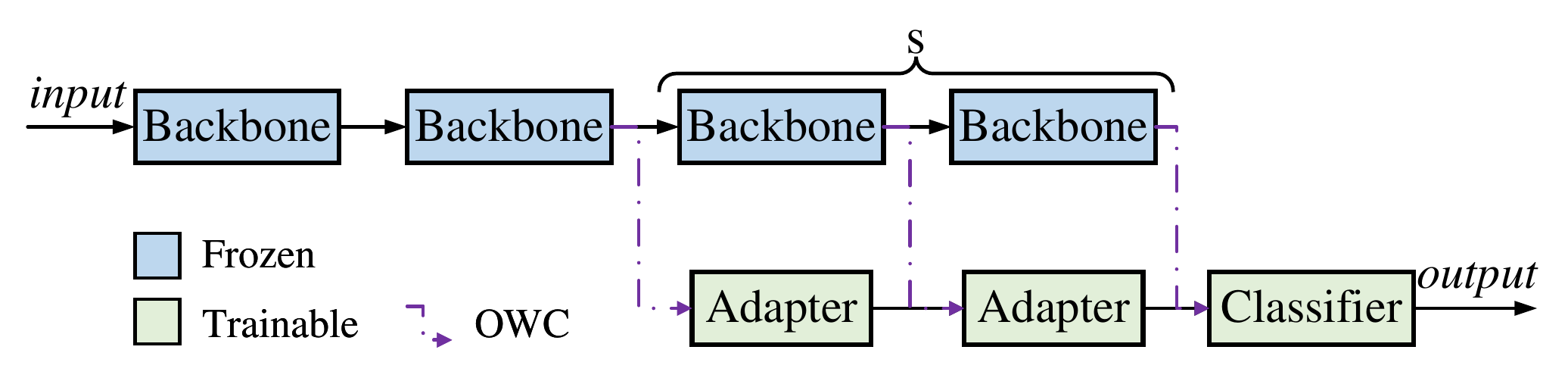}
    \caption{Adapter placement: trainable adapters and classifier are selectively inserted into deeper layers of a frozen backbone. }
    \label{fig:CryptPEFT_design}
\end{figure}

As shown in \autoref{fig:CryptPEFT_design}, the backbone remains frozen, and we selectively insert adapters and a classifier in the deeper layers. Let \(s\) denote the number of final layers chosen for parameter-efficient fine-tuning. By restricting adapter placement to the last \(s\) Transformer layers, we can capture high-level semantic features with fewer trainable parameters---an essential consideration for preserving low-overhead private inference.

%Moreover, each adapter block includes a \textit{residual connection}, ensuring stable gradient flow. 

%The frozen backbone remains untouched, while trainable adapters and a new classifier are selectively inserted in deeper layers.

%Empirically, adjusting \(s\) allows us to balance model utility (when \(s\) is large) against private inference speed (when \(s\) is small). \hl{We provide guidelines for selecting \(s\) in our experiments (\autoref{})}, highlighting that even a small scope of transfer learning can retain most of the task-critical knowledge.

\subsection{Automated Search for Adapters}
\label{subsec:grid}

%\hl{In a typical NAS workflow, three components are essential: the search space, the performance-cost estimator, and the search strategy. The search strategy repeatedly samples candidate architectures from the search space, evaluates each candidate using the estimator, and uses the results as feedback to guide future sampling. In our work, we customize all three components to meet the specific requirements of PEFT-based private inference.}

\hl{In a typical NAS workflow, four components are essential: the search space, the performance-cost estimator, the sampling controller, and the search strategy. The search strategy iteratively selects candidate architectures from the search space via the sampling controller, evaluates them using the estimator, and leverages the results as reward to optimize the controller. In our work, we tailor all these components to address the specific requirements of PEFT-based private inference.}

%throughout the search process.
%对scaler固定值的选取，我们进行了一系列实验，我们保证其余参数相同，仅改变scaler，在不同下游任务下，观察不同scaler对模型utility的影响，实验结果如\autoref{fig:ablation_adapter}所示，平均来看，scaler在设置为0.5时，模型的utility表现最佳，因此在整个搜索过程中，我们固定scaler的值为0.5。
\begin{figure}[t]
    \centering
    \includegraphics[width=0.9\columnwidth]{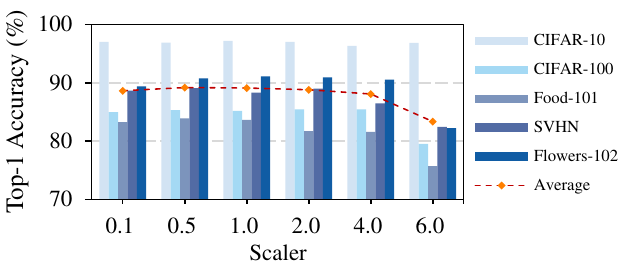}
    \caption{Impact of $scaler$ on model utility.}
    \label{fig:ablation_adapter}
\end{figure}

%\hl{Our search space is explicitly designed around the adapter structure described in \autoref{subsubsec:adapter_arch}. Each adapter contains two NAS cells representing distinct design choices: (1) a \textit{learnable linear attention (\atten)} with a configurable number of attention heads, and (2) a low-rank compression module characterized by its rank. In addition to the internal structure of each adapter, we also treat the total number of adapters as an independent search dimension.} 
\hl{Our search space is explicitly designed around the adapter structure described in \autoref{subsubsec:adapter_arch}. In this paper, we focus on micro-searching~\cite{zoph2018learning}, i.e., searching for NAS cells that can be repeatedly stacked to construct the network, which has proven effective and significantly reduces the search space. Specifically, each adapter contains two NAS cells representing distinct design choices: (1) the \atten module with a configurable number of attention heads, and (2) the low-rank compression module characterized by its rank. In addition to the internal structure of each adapter, we also treat the total number of adapters as a separate search dimension.}
In practice, we fix the \textit{scaler} parameter based on preliminary tuning to simplify the overall search space. To identify an appropriate fixed value, we conducted a series of controlled experiments in which all other parameters were held constant while varying only \textit{scaler}. We evaluated the model utility across multiple downstream tasks, and as shown in \autoref{fig:ablation_adapter}, a value of 0.5 consistently yielded the best average performance. As such, we fix \textit{scaler} at 0.5 for all subsequent evaluations. 
\hl{Accordingly, the full search space is defined by three key hyperparameters: 
$h$ (the number of heads), $r$ (the rank for low-rank compression), and $s$ (the number of adapters), which jointly affect both model utility and the cost of private inference.}

%\hl{Unlike plaintext NAS, which primarily optimizes for model utility under the assumption that cost scales proportionally with parameter count, private inference introduces fundamentally different constraints. In this setting, the cost of each candidate architecture is no longer solely determined by its parameter size, but instead depends on three distinct factors: communication volume, the number of communication rounds, and local computation time. These costs depend on network characteristics such as bandwidth and latency, as well as whether operations can be batched into a single communication round. To enable accurate cost estimation, we profiled all key operations—including linear layers, matrix multiplications, normalization, and activation functions—under our target network environment.}
\hl{Unlike conventional NAS in plaintext setting, which typically assumes computational cost scales proportionally with model size, private inference introduces fundamentally different cost dynamics. Here, the cost of each candidate architecture depends not only on parameter size, but also on communication cost, the number of communication rounds, and local computation time. These factors are influenced by system characteristics such as bandwidth, latency, and the ability to batch operations into a single communication round. To enable cost-aware NAS, we profiled all key operations, including linear layers, matrix multiplications, normalization, and activation functions, under the target network environment, and derived cost estimation models based on the profiling results.} \hl{Specifically, the communication cost is modeled as $(0.001153\,h + 0.000187\,r + 0.000578)\,s + 0.005692$, while the number of communication rounds is calculated as $26\,s + 3$.\footnote{\hl{Each adapter involves 26 communication rounds: 9 for linear operations, 2 for matrix multiplication, 6 for normalization, and 9 for activation. The linear layer contributes 1 round, and the normalization layer contributes 2 rounds.}}
\ignore{\begin{align*}
\text{comm\_cost} &= (0.001153\,h + 0.000187\,r + 0.000578)\,s \\ & \quad \quad + 0.005692, \\
\text{comm\_rounds} &= 26\,s + 3.
\end{align*}
Here, the communication rounds are computed as follows: 1 round for a linear layer, 2 rounds for a normalization layer, and 26 rounds per adapter—comprising 9 rounds for linear operations, 2 for matrix multiplication, 6 for normalization, and 9 for activation. }
Under a fixed network environment, both communication cost and the number of communication rounds contribute linearly to the overall communication time, which can therefore be modeled as $(c_1 h + c_2 r+ c_3)s+c_4$. For example, in a typical WAN setting (400 Mbps bandwidth, 4 ms latency), the communication time is:
\begin{align*}
\textit{comm\_time} &= (0.02117\,h + 0.00344\,r + 0.35828)\,s + 0.15541,
\end{align*}
with a coefficient of determination $R^2 = 0.9975$.
%We observe that the hyperparameters $h$, $r$, and $s$ can almost perfectly explain the variation in communication time with a coefficient of determination $R^2 = 0.9975$.
%For example, under a typical WAN configuration (400 Mbps bandwidth and 4 ms latency), the communication time can be modeled as (with a coefficient of determination $R^2 = 0.9975$):
%Under a fixed network environment, both communication cost and communication round affect the latency in linear factors. Therefore, 
%We observed a near-perfect linear correlation between communication time and the hyperparameters (Pearson correlation coefficient $\rho = 0.9955$). Taking a typical WAN setting (400 Mbps bandwidth, 4 ms latency) as an example (\wenhao{replace with r, h, s}):
%Taking the typical WAN environment with 400 Mbps bandwidth and 4 ms latency as an example, the communication time can be modeled as (with a coefficient of determination $R^2 = 0.9975$):
Similarly, the computation time is modeled as (with $R^2 = 0.9873$):
\begin{align*}
\textit{comp\_time} &= (0.01711\,h + 0.00121\,r + 0.12311)\,s + 0.16581.
\end{align*}
Consequently, the total private inference latency can be modeled as the sum of communication time and computation time.}

%\xss{The comp\_time here actually includes the local plaintext execution time, so, should we revise the description?}
\ignore{\xss{相似的，我们也构建了针对LAN的cost model, 
the communication time is modeled as ($R^2 = 0.9969$):
\begin{align*}
\text{comm\_time} &= (0.010666\,h + 0.001562\,r + 0.058425)\,s + 0.06299.
\end{align*}
Similarly, the computation time is modeled as ($R^2 = 0.9924$):
\begin{align*}
\text{comp\_time} &= (0.0168\,h + 0.0014\,r + 0.0647)\,s + 0.1793.
\end{align*}
}}

\ignore{
\begin{align*}
\text{inference\_time}
&= \text{comm\_time} + \text{comp\_time} \notag \\
&= 1.3311\,\text{comm\_time} + 0.1934 \notag \\
&= (0.028259\,H + 0.004585\,R + 0.014173)\,S + 0.8681.
\end{align*}
}

\ignore{
{\small
\begin{algorithm}[t]
\caption{NAS-based search algorithm.}
\label{alg:gridsearch}
\KwIn{$\textbf{\textit{H}} = \{h_1, \dots, h_n\}$, $\textbf{\textit{R}} = \{r_1, \dots, r_m\}$, $\textbf{\textit{S}} = \{s_1, \dots, s_p\}$, $param\_limit$}
\KwOut{$best\_structure = (h^*, r^*, s^*)$, $max\_utility$}

$max\_utility \leftarrow 0.0$\;
$best\_structure \leftarrow \emptyset$\;

\For{each $h$ in $\textbf{\textit{H}}$}{
    \For{each $r$ in $\textbf{\textit{R}}$}{
        \For{each $s$ in $\textbf{\textit{S}}$}{
            $param\_count \leftarrow \texttt{GetParameters}(h, r, s)$\;
            \If{$param\_count > param\_limit$}{
             \textbf{continue} \tcp*{prune}
            }
            $utility \leftarrow \texttt{GetUtility}(h, r, s)$\;
            \If{$current\_utility > max\_utility$}{
                $max\_utility \leftarrow current\_utility$\;
                $best\_structure \leftarrow (h, r, s)$\;
            }
        }
    }
}
\Return $best\_structure, max\_utility$\;
\end{algorithm}
}}

{\scriptsize
\begin{algorithm}[t]
\caption{NAS-based search algorithm.}
\label{alg:gridsearch}
\KwIn{Targets $\mathcal{U}_{\text{target}}, \mathcal{L}_{\text{target}}, \mathcal{T}_{\text{target}}$; Latency model $\text{Latency}(*,*,*)$; Search spaces $\mathcal{H}, \mathcal{R}, \mathcal{S}$}
\KwOut{Best config $(h^\ast, r^\ast, s^\ast)$; Best utility $U^\ast$}

$(s,\Delta) \gets (1,1)$\;
$(h^\ast, r^\ast, s^\ast, \mathcal{U}^\ast) \gets (\bot, \bot, \bot, -\infty)$\;
\While{$s \leq max(\mathcal{S})$}{
    $ \tau \gets 0$\;
    \While{True}{
        $(h,r) \gets \mathcal{C}(\theta)$\;
            
        \If{$\text{Latency}(h, r, s) > \text{Latency}(h_{\text{init}},r_{\text{init}},s{+}\Delta)$ or $\text{Latency}(h, r, s) > \mathcal{L}_{\text{target}}$}{ $(Reward,\tau) \gets (1/\text{Latency}(h, r, s), \tau + 1)$\;}
        \Else{
            $ \mathcal{U} \gets \text{Eval}(h,r,s)$\;
            $ Reward \gets \mathcal{U} + 1/\text{Latency}(h, r, s)$\;
            $ \tau \gets \tau+1$\;
            \If{$\mathcal{U} > \mathcal{U}^\ast$}{$(h^\ast,r^\ast,s^\ast,\mathcal{U}^\ast) \gets (h,r,s,\mathcal{U})$\;

            $ \tau \gets 0$\;
            }
            \lIf{$\mathcal{U^\ast} \geq \mathcal{U}_{\text{target}}$}{\Return $(h^\ast,r^\ast,s^\ast,U^\ast)$}   
        }
        $ \theta \gets \mathcal{OPT}(\mathcal{C}, \theta,Reward)$\;
        \lIf{$\tau \geq \mathcal{T}_{\text{target}}$}{\textbf{break}}
    }

    % \ForEach{$r \in \mathcal{R} \setminus \{r_{\text{init}}\}$}{
    %     \lIf{$\text{Latency}(h^\ast, r, s) > \text{Latency}(h_{\text{init}},r_{\text{init}},s{+}\Delta)$}{\textbf{break}}
    %     \If{$\text{Latency}(h^\ast, r, s) \leq \mathcal{L}_{\text{target}}$}{
    %         $\mathcal{U} \gets \text{Eval}(h^\ast,r,s)$\;
    %         \lIf{$\mathcal{U} > \mathcal{U}^\ast$}{$(r^\ast,s^\ast,\mathcal{U}^\ast) \gets (r,s,\mathcal{U})$}
    %         \lIf{$\mathcal{U^\ast} \geq \mathcal{U}_{\text{target}}$}{\Return $(h^\ast,r^\ast,s^\ast,U^\ast)$}
            
    %     }
    % }
    $s \gets s + 1$;
    % \uIf{$\mathcal{U}^\ast \geq \mathcal{U}_{\text{target}}$}{\Return $(h^\ast, r^\ast, s)$}
    % \uElseIf{$\text{Latency}(h^\ast, r^\ast, s) > \text{Latency}(h_{\text{init}},r_{\text{init}},s{+}1)$}{$s \gets s + 1$}
    % \Else{$s \gets s + 1$}
}
\Return $(h^\ast,r^\ast,s^\ast,U^\ast)$;
\end{algorithm}
}

%\wenhao{here, from our cost model, we find that the cost is mostly controlled by $s$, then give our search strategy}
Guided by the cost models, our NAS objective is to identify architectures that achieve a target utility while minimizing the latency of private inference. 
Since latency is primarily influenced by the number of adapters ($s$), our customized NAS strategy (outlined in \autoref{alg:gridsearch})
prioritizes minimizing \(s\), while tuning \(h\) and \(r\) to improve utility within the latency constraint. {The value of $s$ is increased only when necessary, i.e., when the desired utility cannot be met by adjusting \(h\) and \(r\), or when increasing \(h\) or \(r\) incurs a higher cost than adding an additional adapter.}
Specifically, we first use a sampling controller to sample values for $h$ and $r$ (line 6), and estimate the latency using our latency model (line 7). If the sampled adapter meets the latency constraint, we evaluate the utility (line 11) and use both utility and latency to compute the reward (line 12), which guides the controller toward configurations with higher utility and lower latency (line 20). If repeated samples fail to reach the target utility, the current search round is terminated (line 21), and the number of adapters is incremented before restarting the search (line 23). The process returns either the adapter configuration that meets the target utility (line 18), or the one closest to it (line 25).
{Overall, these adaptations make our NAS particularly well-suited for private inference, where both latency and communication cost are critical constraints.}

%\wenhao{add here}Specifically, 我们首先利用 sampling controller 对 $h$ 和 $r$ 进行采样 (line 6)，并通过 Latency model 评估当前配置下的隐私推理延迟 (line 7)。若该适配器（adapter）满足延迟约束，则进一步计算模型的 utility (line 11)，并将 utility 与隐私推理延迟共同用于构建奖励函数(line 12)，以引导 sampling controller 朝着更低延迟和更高 utility 的方向优化采样(line 20)。若连续多次采样仍未达到目标 utility，则终止当前采样轮次(line 21)，增加适配器数量后重新开始搜索(line 23)。最终，若搜索过程中能够达到目标 utility，则直接返回对应的候选适配器配置(line 18)；否则，返回最接近目标 utility 的适配器配置(line 25)。

%. The algorithm explores user-defined candidate sets for the rank \(r\), number of attention heads \(h\), and scope of transfer learning \(s\).

%With this strategy, our NAS converges in approximately 4.62, 4.71, 7.78, 7.29, and 0.58 hours on CIFAR-10, CIFAR-100, Food-101, SVHN, and Flowers-102, respectively.

\ignore{
While the design constraints outlined above reduce the dimensionality of our search space, identifying the optimal combination of ($r$, $scaler$, $h$, $s$) for a specific downstream task remains a nontrivial challenge. To address this, we adopt a NAS-based search strategy, as outlined in \autoref{alg:gridsearch}. The algorithm explores user-defined candidate sets for the rank \(r\), number of attention heads \(h\), and scope of transfer learning \(s\). }

% \wenhao{需要看一下这些是否有需要留下的内容}
% \para{策略三的有效性}
% 这里我们进行一系列实验说明是否有必要为Adapter寻找更适配的scaler，我们使用4.2节给出的模型结构，设置s=4，Backbone为ViT-B-16，使用的数据集为CIFAR-100，在一定范围内改变scaler的值，其余参数保持默认，实验结果如图~\ref{}，其中r表示使用LoRA技术时设置的秩(rank)，结果表明CryptPEFT中适配Adapter的scaler与传统PEFT有较大出入，有必要为Adapter寻找更适配的scaler。

% \para{划分迁移学习范围的重要性}
% 在CryptPEFT中，我们对Backbone进行迁移学习，其中一个目标是尽可能高的模型性能，另一个目标是可训练参数量尽可能小以避免过长的推理时间，大的迁移学习范围也许对模型性能有会有提升，但也会带来过高的推理延迟，因此模型性能和推理延迟的折中是设计CryptPEFT需要考虑的问题，划分迁移学习范围能够实现这一目的。

% 我们使用4.2节给出的模型结构，设置scaler=4.0，r=120，Backbone为ViT-B-16，使用的数据集为CIFAR-100，在一定范围内改变s的值，其余参数保持默认，实验结果如图~\ref{},结果表明，划分一个小的迁移学习范围，可以在保持99\%以上性能的情况下，减少超80\%的参数，因此划分一个合理的迁移学习范围非常重要

% 接下来我们开始介绍如何搜索CryptPEFT的结构，给定下游任务，根据模型性能和隐私推理时间的折中给定参数量范围（以小于Backbone参数的5\%为例）我们的搜索目标是找到准确率最高的CryptPEFT结构，根据我们给出的Adapter的设计策略和组织策略，对模型准确率有影响的参数有r、scaler、h、s，从图~\ref{}[寻找适配的scaler的实验那张图]来看，scaler对模型准确率的影响趋势比较独立，因此我们把scaler从搜索空间去除，在整个搜索过程中，使用固定的值scaler=4.0，我们的搜索过程如算法~\ref{alg:gridsearch}所示：

\ignore{In practice, we fix the \textit{scaler} parameter based on preliminary tuning to simplify the overall search space. To identify an appropriate fixed value, we conducted a series of controlled experiments in which all other parameters were held constant while varying only \textit{scaler}. We evaluated the model utility across multiple downstream tasks, and as shown in \autoref{fig:ablation_adapter}, a value of 0.5 consistently yielded the best average performance. As such, we fix \textit{scaler} at 0.5 for all subsequent evaluations. The algorithm then exhaustively evaluates all remaining parameter combinations while optimizing for two objectives:
\begin{enumerate}
    \item \textit{\textbf{Private inference latency ($\mathcal{L}_{\text{target}}$):}} Must not exceed a specified threshold to ensure private inference efficiency.
    \item \textit{\textbf{Model utility ($\mathcal{U}_{\text{target}}$):}} We select the configuration achieving the highest accuracy within the allowable parameter range.
\end{enumerate}
}

\ignore{
\subsection{Efficient Approximations for Private Inference}\xss{delete this sec}
\label{subsubsec:approximation}

Privacy-preserving neural network inference often relies on polynomial or piecewise-linear approximations of nonlinear components (e.g., GELU, Softmax) to reduce the complexity of cryptographic operations~\cite{zimerman2024power}. \hl{Our adapter likewise inherits these nonlinearities}, but it can seamlessly incorporate any existing approximation technique developed for Transformers, since the adapter's MHSA and MLP modules closely mirror those in standard Transformer blocks. For example, we follow Power-Softmax~\cite{zimerman2024power} to approximate Softmax and replace GELU with a ReLU-based variant for faster, secure computations. This approach preserves the adapter's expressiveness while reducing the overhead in MPC protocols.

%\wenhao{how we search for approximations}

%\wenhao{private attention protocol}
}

%% file: tex/security.tex
\section{Security Analysis}
\label{subsec:security}
The security of \sysname is established under the Universal Composability (UC) framework~\cite{canetti2001universally}, which guarantees that the protocol remains secure even when composed with arbitrary other protocol instances. Specifically, the UC framework relies on a simulation-based definition, where a protocol is considered secure if for every real-world adversary \( A \), there exists a polynomial-time simulator \( S \) such that no environment \( \mathcal{Z} \) can distinguish whether it is interacting with the real protocol execution or with the ideal functionality emulated by \( S \). In \sysname, all computation is performed using \textsc{CrypTen}~\cite{crypten-github}, which provides a suite of cryptographic primitives including arithmetic and binary secret sharing, as well as secure conversions between them (\autoref{subsec:mpcbackground}). These primitives have been rigorously proven secure under standard simulation-based techniques, and their security holds under the UC composition theorem~\cite[Appendix B]{knott2021crypten}. Since all higher-level operations in \sysname are constructed as compositions of these UC-secure primitives, the overall protocol inherits their composability guarantees. Consequently, \sysname is UC-secure against semi-honest adversaries, and its security holds even in complex, concurrent environments where multiple protocol instances may be executed simultaneously.

%% file: tex/evaluation.tex
\section{Evaluations}
\label{sec:eval}

%\wenhao{需要考虑一下我们有没有值得写的与implementation相关的内容，如果有，可以添加半页纸到1页纸：包括关于自动搜索策略的实现，以及密文推理系统的实现}

\subsection{Experiment Methodology}
\label{subsec:eval_setup}
%我们的测试平台硬件环境为Intel Xeon Silver 4310 CPU @ 2.10GHz、64G内存、NVIDIA GeForce RTX 4090 GPU。我们使用GPU用于搜索和微调\sysname。对于隐私推理的效率评估，考虑到目前大多数隐私推理框架仅针对CPU平台进行实现，为了公平起见，我们仅使用CPU进行隐私推理，即使用CPU明文计算backbone，并使用CPU对adapter进行隐私推理，特别地，我们仅使用CPU的单线程实现。
%其中，明文计算我们使用pytorch version 2.3.1框架，隐私推理则使用CrypTen框架。我们使用linux下的Traffic Control命令行工具，模拟网络带宽为局域网(10 Gbps,  1 ms)、广域网（500Mbps、50 ms）下的推理时延。

%%v_1
% \begin{table*}
% \centering
% \caption{Search space for the utility-first and efficiency-first configurations, where \texttt{param\_limit} represents the upper bound ratio of learnable parameters.}
% {\small
% \begin{tabular}{ccccc}
% \toprule
% \textbf{Configurations} & \textbf{$H$} & \textbf{$R$} & \textbf{$S$} & \textbf{\path{param_limit}} \\
% \midrule
% {Utility-first} & {\{1, 2, 4, 6, 10, 12\}} & {\{60, 120, 180, 240, 300\}} & {\{2, 4, 6\}} & {5\%} \\
% {Efficiency-first} & {\{1, 2\}} & {\{60, 120\}} & {\{1, 2\}} & {2\%} \\
% \bottomrule
% \end{tabular}
% \label{tab:searchspace}
% }
% \end{table*}

%%v_2

% \begin{table}
% \centering
% \caption{The searched adapter structures by NAS.}
% {\small
% \begin{tabular}{@{}cccc@{}}
% \toprule
% \textbf{Configurations}          & \textbf{CIFAR-100} & \textbf{Food-101}  & \textbf{SVHN}      \\ \midrule
% Utility-first & {\{12, 300, 4\}} & {\{4, 300, 4\}} & {\{12, 180, 6\}} \\
% Efficiency-first  & {\{1, 60, 2\}} & \{{2, 120, 2\}} & {\{2, 60, 2\}} \\ \bottomrule
% \end{tabular}
% \label{tab:best_adapter_setting}
% }
% \end{table}

\para{Hardware and software configurations}
Our experimental platform consists of an Intel Xeon Silver 4310 CPU, 64 GB of RAM, and two NVIDIA GeForce RTX 4090 GPUs. The GPU was employed for automated architecture search and fine-tuning. Since most existing private inference frameworks are optimized for CPU-based implementation, our primary evaluations were conducted on the CPU.
We used PyTorch version 2.3.1 for plaintext computations and \textsc{CrypTen}~\cite{crypten-github} for private inference.
Specifically, plaintext computations for the backbone and private inference for the adapter were executed on the CPU with \textsc{CrypTen}'s default configurations.
%in a single-threaded configuration.
Additionally, we examined the benefits of GPU acceleration in the context of batched task processing, utilizing the GPU to evaluate its impact on private inference efficiency.

%The GPU is used for searching and fine-tuning \sysname and for the evaluation of GPU acceleration.
%To evaluate the efficiency of private inference, we focus on CPU platforms, as most existing frameworks are designed for CPU-based implementations. For fairness, all private inference tasks are performed exclusively on the CPU. 

To emulate different network conditions, we utilized the Traffic Control (TC) tool on Linux. Following the setup in BumbleBee~\cite{lu2023bumblebee} and SHAFT~\cite{ndss/KeiC25_SHAFT}, we reproduced inference latency scenarios under two representative network environments: a wide-area network (WAN) with 400 Mbps bandwidth and 4 ms latency, and a local-area network (LAN) with 1 Gbps bandwidth and 0.5 ms latency. We evaluated the impact of varying bandwidth on private inference efficiency in \autoref{subsec:Effectiveness}.
%In addition, 

\para{Evaluation metrics}
To evaluate the effectiveness of \sysname, we focus on two key aspects: model utility, measured by classification accuracy, and private inference efficiency, evaluated in terms of communication time and total time during private inference, averaged over 10 runs. %\wenhao{we should use batch=1 by default, and evaluate the effect of different batch sizes }
%为了全面评估CryptPEFT在隐私推理加速中的有效性，我们主要从两个方面进行评估：模型性能和推理效率。具体来说，我们通过测量CryptPEFT在不同任务上的准确率来确定其性能。此外，我们使用端到端的隐私推理系统评估CryptPEFT所需要的推理时间和通信量，使用端到端隐私推理的总时间作为其推理效率。通过同时分析这些指标，我们能够从性能和效率之间的平衡角度出发，验证该方法在实际应用中的实用性和可扩展性。

\ignore{
\wenhao{remove this part here; put a similar part into sec. B, show 2 best configurations which optimize for utility and efficiency respectively} % 与现在这两个的区别在于，我们找到曲线中的2个sweet points（例如我们的目标不是达到最大的utility，而是以小的代价尽可能地达到高的utility），最好能找到2个
Based on these considerations, we define two configurations for optimizing the structure of \sysname:
% to accommodate varying priorities for optimized structures: 
(1) \textit{utility-first} configuration, which prioritizes model accuracy and tolerates higher latency during private inference, and (2) \textit{efficiency-first} configuration, which emphasizes minimizing inference time, potentially at the expense of some utility loss. 
To support these configurations, we design distinct search spaces, as detailed in \autoref{tab:searchspace}. 
The utility-first configuration adopts a broader search space with more learnable parameters to maximize accuracy. In contrast, the efficiency-first configuration restricts the search space by eliminating components that significantly impact inference latency, such as excessively large low-rank matrices and \atten modules with a high number of heads, and by reducing the number of adapters, thereby improving inference efficiency while maintaining acceptable model utility.
}

%For the utility-first configuration, a broader search space is employed, incorporating a greater number of learnable parameters to prioritize model accuracy. In comparison, the efficiency-first configuration narrows the search space by excluding components that substantially impact inference efficiency, such as excessively large low-rank matrices, \atten modules with a high number of heads, and by reducing the number of adapters. This design aims to optimize inference efficiency while ensuring an acceptable level of model utility.

\ignore{
\begin{table}[t]
\centering
\caption{Search space for the utility-first and efficiency-first configurations, where \texttt{param\_limit} represents the upper bound ratio of learnable parameters.\wenhao{may remove}}
{
\begin{tabular}{c | c | c}
\toprule
\multicolumn{1}{c|}{\textbf{Search Dimensions}} & \multicolumn{1}{c|}{\textbf{Utility-first}} & \multicolumn{1}{c}{\textbf{Efficiency-first}} \\
\midrule
$H$ & \{1, 2, 4, 6, 10, 12\} & \{1, 2\} \\
$R$ & \{60, 120, 180, 240, 300\} & \{60, 120\} \\
$S$ & \{1, 2, 3, 4, 5, 6\} & \{1, 2\} \\
\path{param_limit} & 5\% & 2\% \\
\bottomrule
\end{tabular}
\label{tab:searchspace}
}
\end{table}
}

%我们为\sysname设置了两种需求，一是性能优先，即在保证隐私的前提下更加注重模型性能，对隐私推理延迟的容忍度更高，二是效率优先，即相比模型性能，更加注重隐私推理时间，针对这两种需求，我们给出两种不同的搜索空间，如表~\ref{tab:searchspace}所示，性能优先下，我们允许的搜索空间更广，支持的可学习参数更多，效率优先下，我们排除了显著影响推理效率的模块，如过大的低秩矩阵、head数过高的MHSA模块、并减少adapter的数量，目标是在保证性能的情况下，达到更优的效率。

\para{Baselines}
We selected two baseline approaches: 

\noindent$\bullet$\textit{~PEFT}---the traditional PEFT architecture, where we applied two typical methods—LoRA~\cite{hu2021lora} and AdaptFormer~\cite{chen2022adaptformer}—to the backbone by implementing their proposed adapters. 
%We then evaluate the utility of these methods. 
When evaluating inference efficiency, the encryption-then-decryption approach may lead to privacy leaks (\autoref{fig:encrypt-then-decrypt}). To align with the privacy guarantees of \sysname, we followed the inference process depicted in~\autoref{fig:slow} for the PEFT baselines.

\noindent$\bullet$\textit{~Simple Fine-tuning (SFT)}. A straightforward approach to achieving OWC involves fine-tuning only the final few layers of the backbone while keeping the majority of the layers frozen~\cite{MPC-Minimized}. In private inference, this strategy enables plaintext execution for the frozen layers, while the fine-tuned final layers are executed with secure computation protocols, thereby reducing computational and communication overhead.

%Inspired by~\cite{MPC-Minimized}, this approach proposes private inference only on the last few layers of the backbone model, while leaving. For model utility evaluation, the majority of the layers were frozen, and only the final layers were fine-tuned. For inference efficiency evaluation, 

%作为对比，我们选择两个baseline方案：(PEFT) 传统的PEFT架构，这里我们使用两种典型的工作，LoRA~\cite{hu2021lora}和AdaptFormer~\cite{chen2022adaptformer}，将他们提出的adapter应用到backbone上，评估这些方法的性能，评估推理效率时，加密再解密的方法会导致不安全(~\ref{fig:encrypt-then-decrypt})，为了和我们的\sysname的安全保证对齐，我们对PEFF方法使用图~\ref{fig:slow}所示的推理过程；第二种baseline方案是(Simple Fine-tuning)，该方法受~\cite{MPC-Minimized}启发，其提出的只对模型最后几层做隐私推理的方案满足OWC原则，评估性能时，冻结大部分层，微调模型的最后几层，评估推理效率时，大部分的层使用明文推理，最后几层运行在PIE环境下，对其进行隐私推理。

\ignore{
\begin{table}
\centering
\caption{Summary of datasets, features, and challenges.\wenhao{may remove}}
\label{tab:summary_datasets}
\begin{tabular}{>{\raggedright\arraybackslash}p{1.4cm} 
                >{\raggedright\arraybackslash}p{3cm} 
                >{\raggedright\arraybackslash}p{3.2cm}}
\toprule
\textbf{Datasets} & \textbf{Features} & \textbf{Challenges} \\
\midrule
CIFAR-10 & Low resolution, simple background & Distinguishing similar classes at low resolution \\
CIFAR-100 & Low resolution, fine-grained & Small inter-class differences \\
Food-101 & High resolution, multi-angle shots & High intra-class variation in food appearance \\
SVHN & Real street scenes, tilted/occluded digits & Digit deformation, background noise \\
Flowers-102 & Natural scenes, flower close-ups & Variations in pose, color, and complex structure \\
\bottomrule
\end{tabular}
\end{table}
}

\para{Datasets}
We used ViT-B-16 as the backbone model for evaluation and evaluated the effectiveness of \sysname on five representative datasets: CIFAR-10~\cite{cifar10}, CIFAR-100~\cite{CIFAR100}, Food-101~\cite{food101}, SVHN~\cite{SVHN} and Flowers-102~\cite{flowers102}.
Prior studies, such as MPCViT~\cite{zeng2023mpcvit} and AdaptFormer~\cite{chen2022adaptformer}, conducted evaluations on subsets of these datasets. \hl{TinyImageNet was intentionally excluded due to its strong visual and distributional similarity to ImageNet, the dataset used to pretrain our ViT-B-16 backbone. This overlap can lead to inflated performance, reducing its effectiveness for evaluating domain generalization. Instead, we included Flowers-102 to introduce greater style diversity. Its relatively small size and large number of classes also present a challenging low-resource setting, enabling a more comprehensive evaluation of model adaptability under realistic, distribution-shifted conditions relevant to private inference.}

%As detailed in \autoref{tab:summary_datasets}, these datasets cover a wide range of visual domains, enabling a comprehensive evaluation of \sysname's generalizability.

%namely CIFAR-10, CIFAR-100, SVHN, and Food-101. We extended this set by including Caltech-101 and Flowers-102 to enhance task diversity. 

%MPCViT~\cite{zeng2023mpcvit}、AdaptFormer~\cite{chen2022adaptformer}等主流工作都使用了上述部分数据集，如CIFAR-10、CIFAR-100、SVHN和Food-101，为了扩充下游任务的多样性，我们在之前工作使用的数据集基础上，又新增了Caltech-101和FLowers-102，详细信息如\autoref{tab:summary_datasets}所示，我们使用这些数据集验证\sysname的通用性。
%\wenhao{用一句话解释一下为何没有用医疗数据集，例如目前主流的工作都面向我们采用的数据集进行测试}

%To test the generalization ability of \sysname across different tasks, we 
%评估过程中，我们使用的Backbone为ViT-B-16，我们使用三种不同风格的数据集：CIFAR-100, Food-101和SVHN，以验证\sysname在不同风格任务上的泛化能力.

\subsection{Utility}
%and Efficiency Evaluation}
% \label{subsec:Effectiveness}

\begin{figure}
    \centering
    \includegraphics[width=0.9\columnwidth]{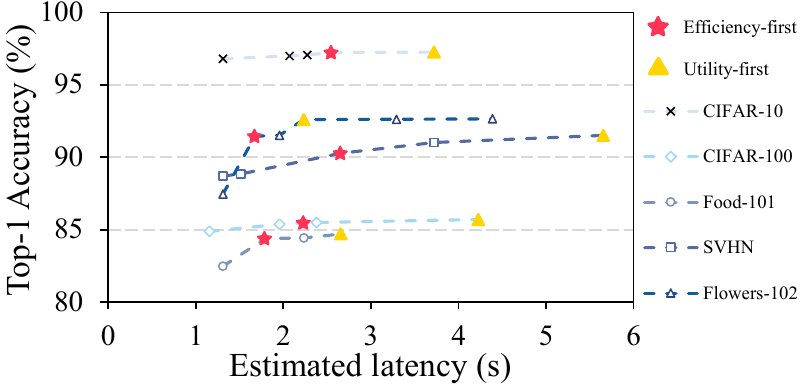}
    \caption{The relationship between model utility and estimated private inference latency targeting the WAN environment.}
    \label{fig:sweet_point}
\end{figure}

\para{NAS}
%\sysname针对各种下游任务执行搜索算法时，我们使用WAN下的cost model来计算每个候选adapter在WAN网络环境下的预估隐私推理延迟，通过设置目标隐私推理延迟实现搜索剪枝来加速基于强化学习的神经网络搜索过程，\autoref{fig:sweet_point}展示了不同目标隐私推理延迟下的搜索结果，我们根据这一结果选择了两种sweet point，即Efficiency-first和Utility-first，分别表示牺牲部分模型utility以追求隐私推理效率和牺牲部分隐私推理效率来实现最大化模型utility两种策略，以Efficiency-first为例，\sysname提出的搜索算法在CIFAR-10,CIFAR-100,Food-101,SVHN和Flowers-102上的执行时间分别为4.15，3.60，3.95，4.73和0.62 GPU小时，更详细的 structures of adapter 展示在\autoref{tab:best_adapter_setting}.
\hl{\sysname performs NAS by leveraging a cost model tailored to the target network environment to estimate the end-to-end private inference latency under each candidate adapter. Given a target utility, the search strategy selects the adapter with the lowest estimated latency that meets the utility constraint. We first use the WAN setting as a case study to examine the relationship between target utility and resulting private inference latency (\autoref{fig:sweet_point}). The results show that, across various downstream tasks, higher utility generally leads to increased latency. However, beyond a certain point, further utility gains are marginal despite significantly increased latency. Based on this observation, we define two trade-off strategies: \textit{efficiency-first}, which sacrifices some utility for lower latency, and \textit{utility-first}, which accepts higher latency to maximize utility.}

The adapter structures (in the format of \{$h$, $r$, $s$\}) discovered under both WAN and LAN settings are summarized in \autoref{tab:best_adapter_setting}. Our results indicate that the optimal adapter design varies across downstream tasks and depends on the performance objective. For example, under the utility-first setting targeting the WAN environment, the Food-101 dataset requires 12 heads in the \atten module, while only 4 are needed under the efficiency-first setting.
In the following evaluations, we primarily adopt the efficiency-first setting, as it offers strong utility with significantly improved efficiency. \hl{Under this setting, \sysname completed the search in 4.15, 1.77, 3.87, 4.73, and 0.62 hours on CIFAR-10, CIFAR-100, Food-101, SVHN, and Flowers-102, respectively.}

\begin{table}[t!]
\centering
\caption{Optimized adapter structures obtained by NAS.}
\label{tab:best_adapter_setting}
\resizebox{\columnwidth}{!}{%
\begin{tabular}{c|c|c|c|c}
\toprule
\multirow{2}{*}{\textbf{Datasets}}  & \multicolumn{2}{c}{\textbf{Utility-first}} & \multicolumn{2}{c}{\textbf{Efficiency-first}}\\ \cmidrule(l){2-5}
 & LAN & WAN & LAN & WAN  \\ \midrule
CIFAR-10 & {\{1, 180, 2\}} & {\{10, 180, 2\}} & {\{4, 120, 1\}} & {\{2, 120, 2\}} \\
CIFAR-100 & {\{1, 300, 2\}} & {\{2, 300, 2\}} & {\{1, 240, 1\}} & {\{1, 300, 1\}}  \\
Food-101  & {\{10, 300, 1\}} & {\{12, 300, 1\}}  & {\{6, 180, 1\}} & {\{4, 180, 1\}}  \\
SVHN  & {\{12, 120, 3\}} & {\{12, 180, 3\}} & {\{12, 60, 1\}} & {\{12, 300, 1\}}  \\
Flowers-102  & {\{1, 300, 1\}} & {\{1, 300, 1\}} & {\{1, 120, 1\}} & {\{1, 180, 1\}}  \\

\bottomrule
\end{tabular}
}
\end{table}
\ignore{
\begin{table*}
\centering
\caption{Comparison of fine-tuning methods on model utility in terms of classification accuracy. \sysname achieves classification accuracy comparable to the baseline methods, while reducing the number of learnable parameters.}
%\sysname improves the model accuracy by 0.35\% simple fine-tuning.
{\small
\begin{tabular}{@{}cccccc@{}}
\toprule
\multicolumn{2}{c}{\multirow{2}{*}{\textbf{Method}}}                                                 & \multirow{2}{*}{\begin{tabular}[c]{@{}c@{}}\textbf{Learnable}\\ \textbf{Params (M)}\end{tabular}} & \multicolumn{3}{c}{\textbf{Datasets}} \\ \cmidrule(l){4-6} 
\multicolumn{2}{c}{} & & \textbf{CIFAR-100}  & \textbf{Food-101} & \textbf{SVHN} \\ 
\midrule
\multirow{2}{*}{PEFT} & LoRA & 0.16 & \textbf{87.41\%} & 83.95\% & 91.81\% \\
& AdaptFormer & 1.19 & 87.23\% & 83.91\% & 91.72\% \\ \hline
\multicolumn{2}{c}{\multirow{2}{*}{Simple Fine-tuning}} & 7.09 & 86.24\% & 84.97\% & 87.35\% \\ 
\multicolumn{2}{c}{} & 14.17 & 86.57\% & 85.56\% & 90.10\% \\ \hline
\multirow{2}{*}{\sysname} & \begin{tabular}[c]{@{}c@{}}Utility-first\end{tabular} & 4.02 / 4.02 / 2.84 & 87.11\% / 86.73\% & \textbf{86.65\%} / 86.24\% & 93.67\% / \textbf{93.90\%} \\
& \begin{tabular}[c]{@{}c@{}}Efficiency-first\end{tabular} & 0.55 / 0.55 / 0.55 & 85.69\% / 85.37\% & 84.51\% / 83.58\% & 89.27\% / 88.86\% \\
\bottomrule
\end{tabular}
\label{tab:performance}
}
\end{table*}
}

%\para{Model utility}
\para{Utility comparison}
We began by evaluating the model utility under various configurations of \sysname. 
In the baseline approach, the neural architecture search component is absent, resulting in the use of the same adapter structure for all downstream tasks. In contrast, our approach customizes \sysname to align with the specific characteristics of each downstream task. 
% \hl{As outlined in \autoref{subsubsec:approximation}, non-linear functions are typically approximated using polynomials during private inference. \sysname incorporates an efficient approximation mechanism to address this. We report the utility before and after applying the approximation components in the format ``\textit{accuracy} / \textit{approximated accuracy}''.} \wenhao{delete this, if not needed}
As shown in \autoref{tab:performance}, the utility-first configuration of \sysname outperforms the traditional PEFT baseline by 1.45\%. This improvement is particularly significant considering that the PEFT baseline requires the majority of computations on both the backbone and adapter to be performed with secure protocols (\autoref{fig:slow}), which introduces substantial communication and computational overhead (\autoref{tab:LAN_batch64}). Furthermore, even the efficiency-first configuration of \sysname, which sacrifices some utility for inference efficiency, outperforms the baseline that simply fine-tuning the last layer by an average of 0.80\%.

%which achieves the highest secure efficiency in the baseline approach.
% 这句不加了吧 ok
%, demonstrate that \sysname incurs an accuracy drop of less than 2\% compared to traditional PEFT baseline in both utility-first and efficiency-first settings. Furthermore, \sysname outperforms the simple fine-tuning baseline by an average of 0.35\%. 
%In comparison, the state-of-the-art method, MPCViT~\cite{zeng2023mpcvit}, achieves a maximum accuracy of 77.8\%, which is significantly lower than that of \sysname. \xss{update，感觉这里不提MPCViT比较好}

We also report the average number of parameters involved in encrypted computation across the downstream tasks.  \sysname achieves a significant reduction, lowering the parameter count by 88.81\% in the efficiency-first setting for the WAN environment, compared to the last-layer fine-tuning baseline (0.80 million vs. 7.15 million). This reduction translates to lower computational and communicational overhead in private inference, which is further analyzed in the following.

%针对不同的\sysname的配置，我们首先进行性能评估，baseline方案中，缺少神经网络搜索部分，因此对不同的下游任务，都使用相同的模型结构，我们的方案中，不同的下游任务，都使用与其匹配的\sysname进行评估，实验结果如表\ref{tab:performance}所示，与baseline相比，\sysname的准确率在不同需求下（Performance priority和Efficiency priority），平均下降不足1\%，在Learnable Params(额外引入adapter的总参数)部分，我们的方案针对CIFAR-100/Food-101/SVHN给出各自对应的Learnable Params (形式为xx/xx/xx)，如4.2.4节分析，隐私推理过程中，非线性函数一般用多项式进行近似，\sysname同样实现了高效近似，这里我们给出了\sysname使用近似组件前后的性能，形式为(acc/acc\_approx)，可以看出，近似后的模型性能下降平均仅为0.5\%(或者说，近似后的模型保持了99\%以上的性能)

%体现近似函数的次数

%\wenhao{我们主要做ViT，但ViT是否应该也有不同的规模，针对不同的数据集、准确率，尽可能全面地提供评估数据}

%\subsection{Applications to the Language Domain}
%\label{subsec:llm}
%评估一个语言模型的效果
%\FloatBarrier % 确保图片不会跨越到下一小节

\subsection{Private inference efficiency}
\label{subsec:Effectiveness}

%\textsc{CrypTen}~\cite{knott2021crypten} is an MPC-based privacy-preserving machine learning framework built on PyTorch.
We implemented an end-to-end private inference system for \sysname using the privacy-preserving machine learning framework \textsc{CrypTen}~\cite{crypten-github}.
To ensure a fair comparison, we adopted the efficiency-first configuration and applied the same approximation techniques---specifically, the state-of-the-art Softmax and GELU approximations introduced by SHAFT---to all baseline methods, as described in \autoref{subsec:understanding}.
We evaluated the inference efficiency of \sysname and the baselines under both simulated LAN and WAN settings.
%Performance was measured using three key metrics: communication cost, communication time, and total inference time.

%\xss{现在表格里面只放了总推理时间，三个指标太宽了放不下}\wenhao{挑一个数据集展示breakdown}
%\hl{When evaluating the baselines, we used the state-of-the-art approximations of Softmax and GELU provided by SHAFT.}
%(\xss{需要额外提一下，评估baseline时，用到的softmax和gelu都采用了SHAFT在CrypTen上实现的SOTA方法}).
%For the evaluation, we adopted the efficiency-first configuration and applied the same approximation components across all methods, as detailed in \autoref{subsubsec:approximation}. We compared the efficiency of \sysname against baseline methods in both simulated LAN and WAN network environments.
%The inference efficiency was evaluated using 3 metrics: communication cost, communication time, and total time.

%The efficiency of inference depends not only on computational complexity but also on the communication rate and volume, as inference requires communication among the involved parties.

\begin{table*}[t]
\centering
\caption{Comparison of model utility (i.e., accuracy). \sysname achieves classification accuracy comparable to the baseline methods, while significantly reducing the number of parameters involved in encrypted computation.}
% (PI is used for simplification in the table) parameters
\label{tab:performance}
%\sysname improves the model accuracy by 0.35\% simple fine-tuning.
%\resizebox{\textwidth}{!}{%
\begin{tabular}{cc|ccccccc}
\toprule
\multicolumn{2}{c|}{\multirow{2}{*}{\textbf{Methods}}} & \multicolumn{1}{c}{\textbf{Avg. private}} & \multirow{2}{*}{\textbf{CIFAR-10}} & \multirow{2}{*}{\textbf{CIFAR-100}} & \multirow{2}{*}{\textbf{Food-101}} & \multirow{2}{*}{\textbf{SVHN}} & \multirow{2}{*}{\textbf{Flowers-102}} & \textbf{Avg.} \\
                & & \textbf{params. (M)} & & & & & & \textbf{utility} \\
\midrule
\multirow{2}{*}{\makecell{\textbf{PEFT}\\(baseline)}} & LoRA         & 86.01  & 97.31\% & \textbf{87.41\%} & 83.95\% & 91.81\% & 84.37\% & 88.97\% \\
 & AdaptFormer  & 87.05  & 97.34\% & 87.23\% & 83.91\% & 91.72\% & 84.42\% & 88.92\% \\
\midrule
\multirow{2}{*}{\makecell{\textbf{SFT}\\(baseline)}} & Last Layer   & 7.15  & 97.28\% & 86.24\% & 84.99\% & 87.35\% & 88.81\% & 88.93\% \\
 & Last 2 layers& 14.23 & \textbf{97.51\%} & 86.57\% & \textbf{85.56\%} & 90.10\% & 90.63\% & 90.07\% \\
\midrule
\multirow{4}{*}{\textbf{\sysname}} & {\cellcolor{lightgray}Utility-first (LAN) } & \cellcolor{lightgray}{1.28}  & \cellcolor{lightgray}97.29\% & \cellcolor{lightgray}85.63\% & \cellcolor{lightgray}84.84\% & \cellcolor{lightgray}\textbf{91.82\%} & \cellcolor{lightgray}\textbf{92.60\%} & \cellcolor{lightgray}\textbf{90.43}\% \\
& \cellcolor{lightgray}Efficiency-first (LAN)    & \cellcolor{lightgray}\textbf{0.46}  & \cellcolor{lightgray}97.19\% & \cellcolor{lightgray}85.37\% & \cellcolor{lightgray}84.56\% & \cellcolor{lightgray}90.03\% & \cellcolor{lightgray}{91.32\%} & \cellcolor{lightgray}89.69\% \\
& \cellcolor{lightgray}Utility-first (WAN) & \cellcolor{lightgray}1.53 & \cellcolor{lightgray}97.26\% & \cellcolor{lightgray}85.70\% & \cellcolor{lightgray}84.70\% & \cellcolor{lightgray}91.51\% & \cellcolor{lightgray}\textbf{92.60\%} & \cellcolor{lightgray}{90.35\%}  \\
& \cellcolor{lightgray}Efficiency-first (WAN)    & \cellcolor{lightgray}{0.80}  & \cellcolor{lightgray}97.23\% & \cellcolor{lightgray}85.47\% & \cellcolor{lightgray}84.38\% & \cellcolor{lightgray}90.27\% & \cellcolor{lightgray}{91.45\%} & \cellcolor{lightgray}89.76\% \\

\bottomrule
\end{tabular}

\end{table*}

\begin{table*}[t]
\centering
\caption{Comparisons of private inference latency (unit: second) within typical LAN and WAN environments. \hl{We also report the latency estimated by our cost model for comparison (\uwave{with wavy underline}).}}
\label{tab:LAN_batch64}
%\resizebox{\textwidth}{!}{%
\begin{tabular}{lcc|cccccccc}
\toprule
 & \multicolumn{2}{c|}{\multirow{1}{*}{\textbf{Methods}}} & \multicolumn{1}{c}{\textbf{CIFAR-10}} & \multicolumn{1}{c}{\textbf{CIFAR-100}} & \multicolumn{1}{c}{\textbf{Food-101}} & \multicolumn{1}{c}{\textbf{SVHN}} & \multicolumn{1}{c}{\textbf{Flowers-102}}\\ \midrule
\multirow{6}{*}{\rotatebox{90}{LAN}} & \multirow{2}{*}{\makecell{\textbf{PEFT}\\(baseline)}} & LoRA & 271.34 & 270.39 & 266.42 & 266.59 & 267.82 \\ & & AdaptFormer & 269.02 & 269.80 & 270.75 & 268.31 & 271.65 \\ \cmidrule(l){2-8}
& \multirow{2}{*}{\makecell{\textbf{SFT}\\(baseline)}} & Last layer & 23.30 & 23.56 & 23.44 & 23.17 & 23.72 \\
& & Last 2 layers & 46.89 & 46.96 & 44.95 & 45.28 & 47.24 \\
\cmidrule(l){2-8}
& \multirow{1}{*}{\textbf{\sysname}} & \cellcolor{lightgray}Efficiency-first & \cellcolor{lightgray}\textbf{0.81} (\uwave{0.83}) & \cellcolor{lightgray}\textbf{1.13} (\uwave{1.10}) &  \cellcolor{lightgray}\textbf{1.04} (\uwave{1.06}) & \cellcolor{lightgray}\textbf{0.90} (\uwave{0.87}) & \cellcolor{lightgray}\textbf{0.75} (\uwave{0.75}) \\ 
\midrule
\multirow{6}{*}{\rotatebox{90}{WAN}}
& \multirow{2}{*}{\makecell{\textbf{PEFT}\\(baseline)}} & LoRA & 548.27 & 545.91 &  547.08 & 546.43 & 545.56 \\ 
& & AdaptFormer & 549.95 & 551.08 & 550.64 & 549.66 & 548.96 \\ \cmidrule(l){2-8}
& \multirow{2}{*}{\makecell{\textbf{SFT}\\(baseline)}} & Last layer & 45.32 & 45.38 & 45.25 & 45.56 & 44.26 \\
& & Last 2 layers & 90.70 & 90.80 & 90.07 & 88.08 & 90.54 \\ \cmidrule(l){2-8}
& \multirow{1}{*}{\textbf{\sysname}} & \cellcolor{lightgray}Efficiency-first & \cellcolor{lightgray}\textbf{2.45} (\uwave{2.55}) & \cellcolor{lightgray}\textbf{2.26} (\uwave{2.24})&  \cellcolor{lightgray}\textbf{1.85} (\uwave{1.79}) & \cellcolor{lightgray}\textbf{2.78} (\uwave{2.66}) & \cellcolor{lightgray}\textbf{1.61} (\uwave{1.68}) \\ 

\bottomrule
\end{tabular}
%}
\end{table*}

\begin{table}[h!]
\centering
\caption{Performance breakdown analysis.}
\label{tab:breakdown}
{
\begin{tabular}{cc|ccc}
\toprule
& \textbf{Metrics} & \makecell{\textbf{SFT}\\(baseline)} & \textbf{\sysname} & \textbf{Improvements}\\
\midrule
% & Comm. (GB)        &  1.55           &    \cellcolor{lightgray}0.06    & \cellcolor{lightgray}$25.83\times$     \\
% & Comm. round       &  77        &       \cellcolor{lightgray}29   & \cellcolor{lightgray}$2.66\times$   \\
%\cmidrule(l){2-5}
% \midrule
\multirow{4}{*}{\rotatebox{90}{LAN}} &
Comm. (GB)        &  1.55           &    \cellcolor{lightgray}0.05    & \cellcolor{lightgray}$31.00\times$     \\
& Comm. round       &  77        &       \cellcolor{lightgray}29   & \cellcolor{lightgray}$2.66\times$   \\
& Comm. time (s)  &      14.40       &       \cellcolor{lightgray}0.55  & \cellcolor{lightgray}$26.18\times$     \\
& Total time (s)  &     23.56        &       \cellcolor{lightgray}1.13    & \cellcolor{lightgray}$20.85\times$  \\

\cmidrule(l){2-5}
\multirow{4}{*}{\rotatebox{90}{WAN}} &
Comm. (GB)        &  1.55           &    \cellcolor{lightgray}0.06    & \cellcolor{lightgray}$25.83\times$     \\
& Comm. round       &  77        &       \cellcolor{lightgray}29   & \cellcolor{lightgray}$2.66\times$   \\
& Comm. time (s)  &      34.42       &      \cellcolor{lightgray}1.51   & \cellcolor{lightgray}$22.79\times$    \\
& Total time (s) &      45.38       &      \cellcolor{lightgray}2.26   & \cellcolor{lightgray}$20.08\times$    \\
\bottomrule
\end{tabular}
}
\end{table}

%%%% v3
\ignore{
\begin{table*}[h!]
\centering
\caption{Comparison of fine-tuning methods on private inference efficiency within typical LAN and WAN environments.}
\resizebox{\textwidth}{!}{%
\begin{tabular}{@{}ccccccccccccc@{}}
\toprule
 & \multicolumn{2}{c}{\multirow{4}{*}{\textbf{Methods}}} & \multicolumn{9}{c}{\textbf{Datasets}} \\ \cmidrule(l){4-12} 
& \multicolumn{2}{c}{} & \multicolumn{3}{c}{\textbf{CIFAR-100}} & \multicolumn{3}{c}{\textbf{Food-101}} & \multicolumn{3}{c}{\textbf{SVHN}}\\ \cmidrule(l){4-12} 
& \multicolumn{1}{c}{} & & \begin{tabular}[c]{@{}c@{}}Comm\\ Cost (GB)\end{tabular} & \begin{tabular}[c]{@{}c@{}}Comm\\ Time (s)\end{tabular} & \begin{tabular}[c]{@{}c@{}}Total\\ Time (s)\end{tabular} & \begin{tabular}[c]{@{}c@{}}Comm\\ Cost (GB)\end{tabular} & \begin{tabular}[c]{@{}c@{}}Comm\\ Time (s)\end{tabular} & \begin{tabular}[c]{@{}c@{}}Total\\ Time (s)\end{tabular} & \begin{tabular}[c]{@{}c@{}}Comm\\ Cost (GB)\end{tabular} & \begin{tabular}[c]{@{}c@{}}Comm\\ Time (s)\end{tabular} & \begin{tabular}[c]{@{}c@{}}Total\\ Time (s)\end{tabular} \\ \midrule
\multirow{6}{*}{\rotatebox{90}{LAN}} & \multirow{2}{*}{PEFT} & LoRA & 4.70 & 6.69 & 46.29 & 4.70 & 6.68 & 46.46 & 4.70 & 5.91 & 46.33 \\ & & AdaptFormer & 4.77 & 6.39 & 46.70 & 4.77 & 6.83 & 46.94 & 4.77 & 6.04 & 43.90 \\ \cmidrule(l){2-12}
& \multirow{2}{*}{Simple Fine-tuning} & Last layer & 0.39 & 0.47 & 4.05 & 0.39 & 0.52 & 4.02 & 0.39 & 0.49 & 4.08 \\
& & Last 2 layers & 0.78 & 0.96 & 7.78 & 0.78 & 0.95 & 7.85 & 0.78 & 1.07 & 7.88 \\ \cmidrule(l){2-12}
& \multirow{1}{*}{\sysname} & Efficiency-first & \textbf{0.05} & \textbf{0.10} & \textbf{0.79} & \textbf{0.06} & \textbf{0.13} & \textbf{0.95} & \textbf{0.06} & \textbf{0.10} & \textbf{0.99} \\ 
\midrule
\multirow{6}{*}{\rotatebox{90}{WAN}}
& \multirow{2}{*}{PEFT} & LoRA  & 4.70 & 139.19 & 176.79 & 4.70 & 139.81 & 175.88 & 4.70 & 138.79 & 175.03 \\ 
& & AdaptFormer & 4.77 & 141.31 & 178.08 & 4.77 & 141.50 & 178.11 & 4.77 & 141.12 & 177.83 \\ \cmidrule(l){2-12}
& \multirow{2}{*}{Simple Fine-tuning} & Last layer & 0.39 & 11.77 & 15.59 & 0.39 & 11.78 & 15.38 & 0.39 & 11.63 & 15.26 \\
& & Last 2 layers & 0.78 & 23.10 & 29.59 & 0.78 & 22.93 & 29.33 & 0.78 & 23.22 & 30.70 \\ \cmidrule(l){2-12}
& \multirow{1}{*}{\sysname} & Efficiency-first & \textbf{0.05} & \textbf{1.76} & \textbf{2.52} & \textbf{0.06} & \textbf{2.13} & \textbf{2.96} & \textbf{0.06} & \textbf{2.14} & \textbf{2.93} \\ 

\bottomrule
\end{tabular}
}
\label{tab:LAN_batch64}
\end{table*}
}

%%%% v2
\ignore{
\begin{table*}
\centering
\caption{Comparison of fine-tuning methods on inference efficiency within the LAN environment.}
\resizebox{\textwidth}{!}{%
\begin{tabular}{@{}cccccccccccc@{}}
\toprule
\multicolumn{2}{c}{\multirow{4}{*}{\textbf{Method}}} & \multirow{4}{*}{\begin{tabular}[c]{@{}c@{}}\textbf{Learnable}\\ \textbf{Params (M)}\end{tabular}} & \multicolumn{9}{c}{\textbf{Datasets}}\\ \cmidrule(l){4-12} 
\multicolumn{3}{c}{} & \multicolumn{3}{c}{\textbf{CIFAR-100}} & \multicolumn{3}{c}{\textbf{Food-101}} & \multicolumn{3}{c}{\textbf{SVHN}}\\ \cmidrule(l){4-12} 
\multicolumn{2}{c}{} & & \begin{tabular}[c]{@{}c@{}}Comm\\ Cost (GB)\end{tabular} & \begin{tabular}[c]{@{}c@{}}Comm\\ Time (s)\end{tabular} & \begin{tabular}[c]{@{}c@{}}Total\\ Time (s)\end{tabular} & \begin{tabular}[c]{@{}c@{}}Comm\\ Cost (GB)\end{tabular} & \begin{tabular}[c]{@{}c@{}}Comm\\ Time (s)\end{tabular} & \begin{tabular}[c]{@{}c@{}}Total\\ Time (s)\end{tabular} & \begin{tabular}[c]{@{}c@{}}Comm\\ Cost (GB)\end{tabular} & \begin{tabular}[c]{@{}c@{}}Comm\\ Time (s)\end{tabular} & \begin{tabular}[c]{@{}c@{}}Total\\ Time (s)\end{tabular} \\ \midrule
\multirow{2}{*}{PEFT} & LoRA & 0.16  & 4.70 & 6.69 & 46.29 & 4.70 & 6.68 & 46.46 & 4.70 & 5.91 & 46.33 \\ & AdaptFormer & 1.19 & 4.77 & 6.39 & 46.70 & 4.77 & 6.83 & 46.94 & 4.77 & 6.04 & 43.90 \\ \hline
\multicolumn{2}{c}{\multirow{2}{*}{Simple Fine-tuning}} & 7.09                                                                             & 0.39 & 0.47 & 4.05 & 0.39 & 0.52 & 4.02 & 0.39 & 0.49 & 4.08 \\
\multicolumn{2}{c}{} & 14.17 & 0.78 & 0.96 & 7.78 & 0.78 & 0.95 & 7.85 & 0.78 & 1.07 & 7.88 \\ \hline
\multirow{1}{*}{\sysname} & Efficiency-first & 0.55 / 0.55 / 0.55 & \textbf{0.05} & \textbf{0.10} & \textbf{0.79} & \textbf{0.06} & \textbf{0.13} & \textbf{0.95} & \textbf{0.06} & \textbf{0.10} & \textbf{0.99} \\ 
\bottomrule
\end{tabular}
}
\label{tab:LAN_batch64}
\end{table*}
}

As shown in \autoref{tab:LAN_batch64}, \sysname significantly reduces private inference latency in both LAN and WAN environments. Taking CIFAR-100 as an example, \sysname achieves a $20.85\times$ speedup over the last-layer fine-tuning baseline and a $238.76\times$ speedup over AdaptFormer in the LAN setting. In the WAN setting, it yields a $20.08\times$ and $243.84\times$ speedup over the same baselines, respectively. Notably, \sysname demonstrates an inference time of 2.26 seconds with an accuracy of 85.47\% on CIFAR-100 under the WAN environment.

\ignore{
\begin{table*}
\centering
\caption{Comparison of fine-tuning methods on inference efficiency within the WAN environment.}
\resizebox{\textwidth}{!}{%
\begin{tabular}{@{}cccccccccccc@{}}
\toprule
\multicolumn{2}{c}{\multirow{4}{*}{\textbf{Method}}} & \multicolumn{9}{c}{\textbf{Datasets}}\\ \cmidrule(l){3-11} 
\multicolumn{2}{c}{} & \multicolumn{3}{c}{\textbf{CIFAR-100}} & \multicolumn{3}{c}{\textbf{Food-101}} & \multicolumn{3}{c}{\textbf{SVHN}}\\ \cmidrule(l){3-11} 
\multicolumn{1}{c}{} & & \begin{tabular}[c]{@{}c@{}}Comm\\ Cost (GB)\end{tabular} & \begin{tabular}[c]{@{}c@{}}Comm\\ Time (s)\end{tabular} & \begin{tabular}[c]{@{}c@{}}Total\\ Time (s)\end{tabular} & \begin{tabular}[c]{@{}c@{}}Comm\\ Cost (GB)\end{tabular} & \begin{tabular}[c]{@{}c@{}}Comm\\ Time (s)\end{tabular} & \begin{tabular}[c]{@{}c@{}}Total\\ Time (s)\end{tabular} & \begin{tabular}[c]{@{}c@{}}Comm\\ Cost (GB)\end{tabular} & \begin{tabular}[c]{@{}c@{}}Comm\\ Time (s)\end{tabular} & \begin{tabular}[c]{@{}c@{}}Total\\ Time (s)\end{tabular} \\ \midrule
\multirow{2}{*}{PEFT} & LoRA  & 4.70 & 139.19 & 176.79 & 4.70 & 139.81 & 175.88 & 4.70 & 138.79 & 175.03 \\ & AdaptFormer & 4.77 & 141.31 & 178.08 & 4.77 & 141.50 & 178.11 & 4.77 & 141.12 & 177.83 \\ \hline
\multirow{2}{*}{Simple Fine-tuning} & Last layer & 0.39 & 11.77 & 15.59 & 0.39 & 11.78 & 15.38 & 0.39 & 11.63 & 15.26 \\
& Last 2 layers & 0.78 & 23.10 & 29.59 & 0.78 & 22.93 & 29.33 & 0.78 & 23.22 & 30.70 \\ \hline
\multirow{1}{*}{\sysname} & Efficiency-first & \textbf{0.05} & \textbf{1.76} & \textbf{2.52} & \textbf{0.06} & \textbf{2.13} & \textbf{2.96} & \textbf{0.06} & \textbf{2.14} & \textbf{2.93} \\ 
\bottomrule
\end{tabular}
}
\label{tab:WAN_batch64}
\end{table*}
}

%WAN 64, v2
\ignore{
\begin{table*}
\centering
\caption{Comparison of fine-tuning methods on inference efficiency within the WAN environment.}
\resizebox{\textwidth}{!}{%
\begin{tabular}{@{}cccccccccccc@{}}
\toprule
\multicolumn{2}{c}{\multirow{4}{*}{\textbf{Method}}} & \multirow{4}{*}{\begin{tabular}[c]{@{}c@{}}\textbf{Learnable}\\ \textbf{Params (M)}\end{tabular}} & \multicolumn{9}{c}{\textbf{Datasets}}\\ \cmidrule(l){4-12} 
\multicolumn{3}{c}{} & \multicolumn{3}{c}{\textbf{CIFAR-100}} & \multicolumn{3}{c}{\textbf{Food-101}} & \multicolumn{3}{c}{\textbf{SVHN}}\\ \cmidrule(l){4-12} 
\multicolumn{2}{c}{} & & \begin{tabular}[c]{@{}c@{}}Comm\\ Cost (GB)\end{tabular} & \begin{tabular}[c]{@{}c@{}}Comm\\ Time (s)\end{tabular} & \begin{tabular}[c]{@{}c@{}}Total\\ Time (s)\end{tabular} & \begin{tabular}[c]{@{}c@{}}Comm\\ Cost (GB)\end{tabular} & \begin{tabular}[c]{@{}c@{}}Comm\\ Time (s)\end{tabular} & \begin{tabular}[c]{@{}c@{}}Total\\ Time (s)\end{tabular} & \begin{tabular}[c]{@{}c@{}}Comm\\ Cost (GB)\end{tabular} & \begin{tabular}[c]{@{}c@{}}Comm\\ Time (s)\end{tabular} & \begin{tabular}[c]{@{}c@{}}Total\\ Time (s)\end{tabular} \\ \midrule
\multirow{2}{*}{PEFT} & LoRA & 0.16  & 4.70 & 139.19 & 176.79 & 4.70 & 139.81 & 175.88 & 4.70 & 138.79 & 175.03 \\ & AdaptFormer & 1.19 & 4.77 & 141.31 & 178.08 & 4.77 & 141.50 & 178.11 & 4.77 & 141.12 & 177.83 \\ \hline
\multicolumn{2}{c}{\multirow{2}{*}{Simple Fine-tuning}} & 7.09                                                                             & 0.39 & 11.77 & 15.59 & 0.39 & 11.78 & 15.38 & 0.39 & 11.63 & 15.26 \\
\multicolumn{2}{c}{} & 14.17 & 0.78 & 23.10 & 29.59 & 0.78 & 22.93 & 29.33 & 0.78 & 23.22 & 30.70 \\ \hline
\multirow{1}{*}{\sysname} & Efficiency-first & 0.55 / 0.55 / 0.55 & \textbf{0.05} & \textbf{1.76} & \textbf{2.52} & \textbf{0.06} & \textbf{2.13} & \textbf{2.96} & \textbf{0.06} & \textbf{2.14} & \textbf{2.93} \\ 
\bottomrule
\end{tabular}
}
\label{tab:WAN_batch64}
\end{table*}
}

\para{Performance breakdown}
%\autoref{tab:breakdown}展示了\sysname在communication rounds、communication cost、communication time and total time上的performance breakdown，我们以CIFAR-100数据集为例，从baseline方案中选择最高效的一种，即SFT的Last layer作为对比。相比SFT，\sysname减少了28.57\%的communication rounds，减少了98.06\%的通信量。由于MPC方案中通信时间占主导地位，因此在LAN和WAN网络环境下，总推理时间方面\sysname相比baseline分别加速$22.87\times$、$21.51\times$。
\autoref{tab:breakdown} provides a detailed performance breakdown of \sysname on the CIFAR-100 dataset, including communication cost, communication round, communication time, and total inference time. For comparison, we used the most efficient baseline among the SFT configurations---fine-tuning only the last layer. Compared to this baseline, \sysname reduces the number of communication rounds by 62.37\% and the communication cost by an average of 96.45\% in both LAN and WAN environments. Since communication is the dominant contributor to latency in MPC-based inference, these reductions result in significant performance gains: \sysname achieves speedups of $20.85\times$ in LAN settings and $20.08\times$ in WAN settings.

\begin{table}
\centering
\caption{Comparison of utility and private inference latency (unit: second) between MPCViT and \sysname.}
\label{tab:mpcvit_vs_CryptPEFT}
{
\begin{tabular}{ll|c|ccc}
\toprule
 \multicolumn{2}{c|}{\textbf{Settings}}  &\multicolumn{1}{c|}{ \textbf{Metrics}} & \textbf{MPCViT} & \textbf{\sysname} & \textbf{Improvements} \\
\midrule
\multirow{4}{*}{\rotatebox{90}{CIFAR-10}} &
\multirow{2}{*}{\rotatebox{90}{LAN}}  &  Utility  &  94.27\%       &       \cellcolor{lightgray}97.19\%  & \cellcolor{lightgray}\color{teal}↑2.92\%    \\
& & Total time &  10.64   & \cellcolor{lightgray}0.81   & \cellcolor{lightgray}$13.14\times$   \\
\cmidrule(l){3-6}
& \multirow{2}{*}{\rotatebox{90}{WAN}}  & Utility  &  94.27\%        &       \cellcolor{lightgray}97.23\%   & \cellcolor{lightgray}\color{teal}↑2.96\%  \\
& &  Total time & 24.10   & \cellcolor{lightgray}2.45   & \cellcolor{lightgray}$9.84\times$   \\
\midrule
\multirow{4}{*}{\rotatebox{90}{CIFAR-100}} &
\multirow{2}{*}{\rotatebox{90}{LAN}}  &  Utility  &  77.46\%       &       \cellcolor{lightgray}85.37\%  & \cellcolor{lightgray}\color{teal}↑7.91\%     \\
& & Total time & 10.38   &  \cellcolor{lightgray}1.13   & \cellcolor{lightgray}$9.19\times$   \\
\cmidrule(l){3-6}
& \multirow{2}{*}{\rotatebox{90}{WAN}}  & Utility &   77.46\%        &       \cellcolor{lightgray}85.47\%    & \cellcolor{lightgray}\color{teal}↑8.01\%   \\
& & Total time &  22.97  & \cellcolor{lightgray}2.26   & \cellcolor{lightgray}$10.16\times$   \\
\bottomrule
\end{tabular}
}
\end{table}

%Using CIFAR-100 as an example, \sysname achieves a speedup of $6.19\times$ compared to simply fine-tuning the final layer, and a speedup of $70.67\times$ compared to the traditional PEFT method AdaptFormer. Taking the CIFAR-100 dataset as an example, in the WAN setting, \sysname achieves an amortized latency of 2.52 seconds, necessitates only 0.05 GB of communication, and attains a classification accuracy of 85.37\%. 

%Similarly, we conducted experiments with the batch size set to 64 in the WAN environment. As shown in \autoref{tab:WAN_batch64}, compared to the baseline approach, \sysname benefits from a smaller number of parameters, which leads to reduced communication overhead during private inference. As a result, it maintains higher inference efficiency even in low-bandwidth network environments. In contrast, the baseline approach, due to the large number of parameters involved in private inference, requires higher network bandwidth, resulting in lower inference efficiency under low-bandwidth conditions. In this environment, using CIFAR-100 as an example, \sysname achieves an acceleration of $6.19\times$ compared to fine-tuning only the final layer, and an acceleration of $70.67\times$ compared to the traditional PEFT method, AdaptFormer. Taking the CIFAR-100 dataset as an example, in the WAN setting, \sysname achieves an amortized latency of 2.52 seconds, necessitates only 0.05 GB of communication, and attains a classification accuracy of 85.37\%. 

\para{\hl{Reliability of the cost model}}
\hl{The cost model plays a critical role in our NAS framework.  To evaluate its accuracy, we compare the estimated end-to-end latency with actual measurements under both LAN and WAN settings (\autoref{tab:LAN_batch64}). The results show that the estimation error remains within a $\pm4.3\%$ margin, indicating that the model provides reliable latency predictions to guide the search process.}

\para{Comparison with MPCViT~\cite{zeng2023mpcvit}}
We also compared \sysname with MPCViT, the current state-of-the-art that leverages NAS and approximation techniques for efficient private inference for ViT. MPCViT provides pre-trained models for various downstream tasks, including CIFAR-10 and CIFAR-100, which align with our evaluation benchmarks. As shown in \autoref{tab:mpcvit_vs_CryptPEFT}, \sysname outperforms MPCViT on both datasets, achieving higher utility and a $9.19\times$ to $13.14\times$ speedup in private inference. These gains result from \sysname's PEFT-specific design, which leverages a public backbone and adopts the OWC policy to minimize encrypted computation, thereby enhancing both model utility and efficiency.

%在设计MPC友好的神经网络架构方向上，MPCViT，实现了utility和隐私推理效率的折中，是当前的sota方法，MPCViT开源了几个针对不同下游任务的模型，其中包括CIFAR-10和CIFAR-100，这是和我们方案相交的两个下游任务。因此，我们在CIFAR-10和CIFAR-100上对MPCViT和\sysname的Utility和隐私推理效率进行比较，评估结果如\autoref{tab:mpcvit_vs_CryptPEFT}所示，\sysname在改善utility的同时，对隐私推理实现了$7.71\times$ to over $10.89\times$的加速。

\ignore{
\begin{figure}
    \centering
    \includegraphics[width=0.87\columnwidth]{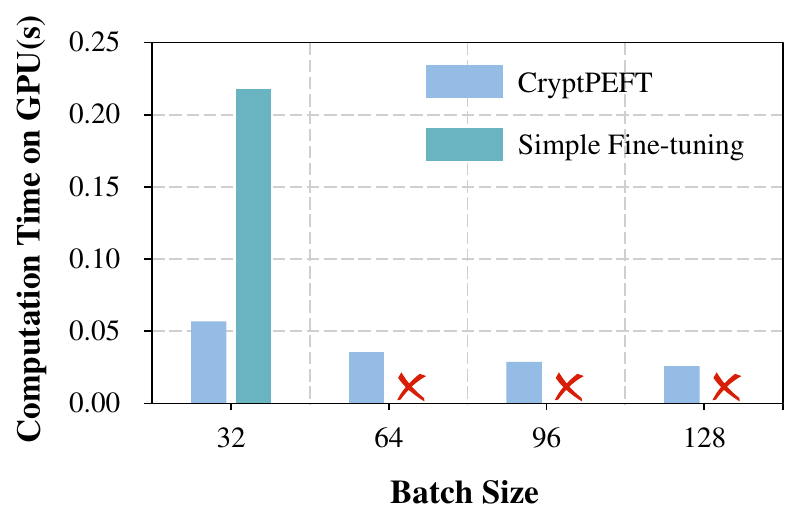}
    \caption{Amortized computation time on GPU with different batch sizes.}
    \label{fig:gpu_time}
\end{figure}
}
\ignore{
\para{GPU acceleration on private inference}
%额外的，我们测试了GPU对隐私推理的加速能力，在\autoref{subsec:eval_setup}描述的Hardware and software configurations下，GPU相比CPU，通信时间几乎一致，计算时间大约有10倍的加速，然而隐私推理对GPU的显存需求巨大，对模型参数量十分敏感，相比目前存在的方案，CryptPEFT因具有小的Learnable Params，有着天然的优势，图1展示了不同batch size下，\sysname 和 Simple Fine-tuning 方案在隐私推理过程中的GPU上计算时间，Simple Fine-tuning 引入7.09M的Leaenable Params，在大的batch size下无法推理成功，GPU显存不足，我们标记在了图中
We further evaluated the efficiency improvement of GPU acceleration for \sysname. The results show that while the communication time during private inference remains nearly the same, GPU-based computation achieves approximately a 10-fold speedup compared to its CPU counterpart. \autoref{fig:gpu_time} compares the GPU computation times of \sysname and the simple fine-tuning method across various batch sizes. The results reveal a clear advantage for \sysname, primarily attributed to its smaller number of learnable parameters. In contrast, the simple fine-tuning method requires a substantially larger set of learnable parameters and is unable to support private inference on GPUs with larger batch sizes due to limitations in GPU memory capacity.
\wenhao{we may not need the GPU result}}

\subsection{Ablation Study and Sensitivity Analysis}
\label{subsec:ablation}

\para{Sensitivity to network conditions}
%除仿真的WAN和LAN网络环境外，我们额外报告了一组\sysname在不同延迟和带宽下的隐私推理时间，我们以CIFAR-100数据集为例，与Performance breakdown这一节相同，我们选取SFT的Last layer作为baseline，评估结果如\autoref{tab:sensitivity_net_condition}所示。从结果可以看出，我们的方案在各个网络条件下均具有显著优势，特别的，如\autoref{tab:breakdown}所示，\sysname需要更小的通信量和通信轮数，，在较差的网络条件下表现更加突出，对比SFT，\sysname在最好的网络条件和最差的网络条件下，分别加速xx-xx倍。
%除仿真的WAN和LAN网络环境外，我们额外报告了一组\sysname在不同带宽下的隐私推理时间，我们以CIFAR-100数据集为例，固定网络延迟为4ms，与Performance breakdown这一节相同，我们选取SFT的Last layer作为baseline，评估结果如\autoref{tab:bandwidth}所示。从结果可以看出，我们的方案在各个网络带宽下均具有显著优势，特别的，如\autoref{tab:breakdown}所示，\sysname需要的通信量仅为SFT的1.94\%，在较差的网络条件下表现更加突出，对比SFT，\sysname在最好的网络条件和最差的网络条件下，分别加速$7.80\times$和$35.38\times$。基于MPC技术的隐私推理，需要频繁的通信，对通信环境十分敏感，这是其难以实际应用的原因之一，而\sysname更能容忍较差的网络环境，对实际部署将更加有利。
In addition to the simulated LAN and WAN environments, we further evaluated the private inference efficiency of \sysname under a range of network bandwidths using the CIFAR-100 dataset. The network latency was fixed at 4 ms, and the SFT configuration was adopted as the baseline. As shown in \autoref{tab:bandwidth}, \sysname consistently outperforms the baseline across all bandwidth settings. Notably, \sysname reduces the communication volume to only 3.87\% of that required by SFT (\autoref{tab:breakdown}). This efficiency gain becomes increasingly significant under limited bandwidth conditions. Specifically, \sysname achieves speedups of $9.56\times$ under optimal network conditions and up to $21.72\times$ under the most constrained setting. Given that MPC-based private inference is inherently communication-intensive and highly sensitive to network conditions, \sysname exhibits strong robustness to degraded network environments, thereby enhancing its viability for real-world applications.

%\wenhao{需要说一下用的哪个数据集，然后应该不能自己跟自己比，需要跟baseline对比，突出1. 带宽越低，优势越大 2. 延迟越高，优势越大。并解释这两个现象的原因。为了对比需要，我们可能要固定数据集、测试带宽的时候就固定延时，测试延时的时候就固定带宽。重点评估带宽吧，带宽对我们更有利，延迟变化的影响可以不做，参考comet}\xss{目前我们这个结构，对baseline可以说是两方面都占优势，通信量低，轮数也低，所以在更差的网络环境下应该更有优势}\wenhao{这部分我没有问题}\hl{请看我的comment}

\begin{table}[t!]
\centering
\caption{Total private inference latency (unit: second) under varying network bandwidths.}
\label{tab:bandwidth}
{
\begin{tabular}{c|ccc}
\toprule
 \textbf{Bandwidths}  & \makecell{\textbf{SFT} (baseline)} & \textbf{\sysname} & \textbf{Improvements} \\
\midrule
 100 Mbps        &  145.77     &    \cellcolor{lightgray}6.71    & \cellcolor{lightgray}$21.72\times$     \\
 500 Mbps       &  38.65       &       \cellcolor{lightgray}2.11   & \cellcolor{lightgray}$18.32\times$   \\
 1 Gbps         &     23.71     &      \cellcolor{lightgray}1.71       &    \cellcolor{lightgray}$13.87\times$          \\
 5 Gbps          &    12.95      &      \cellcolor{lightgray}1.35        &   \cellcolor{lightgray}$9.56\times$         \\
\bottomrule
\end{tabular}
}
\end{table}

% \begin{table}[htbp]
% \centering
% \caption{Private inference efficiency under different network conditions.\xss{不同延迟下的数据有些奇怪，固定延迟似乎更能突出我们的优势}}
% \label{tab:sensitivity_net_condition}
% \renewcommand{\arraystretch}{1.4}

% \begin{tabular}{>{\bfseries}lc|cccc}
% \toprule
%  & \multirow{2}{*}{\textbf{Latency}} & \multicolumn{4}{c}{\textbf{Time (s)}}  \\
% \cmidrule(lr){3-6}
%  & & 100Mbps & 500Mbps & 1Gbps & 5Gbps \\
% \midrule
% \multirow{4}{*}{\rotatebox{90}{SFT}} &
% 0.5ms  & 144.69 & 37.72 & 23.73 & 12.84 \\
% & 4ms    & 145.77 & 38.65 & 23.71 & 12.95    \\
% & 10ms   & 146.90 &  39.71 & 25.23 & 15.84   \\
% & 50ms   & 154.31 & 53.76 & 44.64 & 40.16  \\
% \midrule
% \multirow{4}{*}{\rotatebox{90}{\sysname}} &
% 0.5ms  & 3.77  & 1.29  & 1.02  & 0.99  \\
% & 4ms    & 4.12  & 2.03  & 1.82  & 1.66  \\
% & 10ms   & 5.12  & 3.30  & 3.23  & 3.10   \\
% & 50ms   & 13.44 & 12.28 & 12.06 & 11.93  \\
% \bottomrule
% \end{tabular}
% \end{table}

\begin{figure}
    \centering
    \includegraphics[width=\columnwidth]{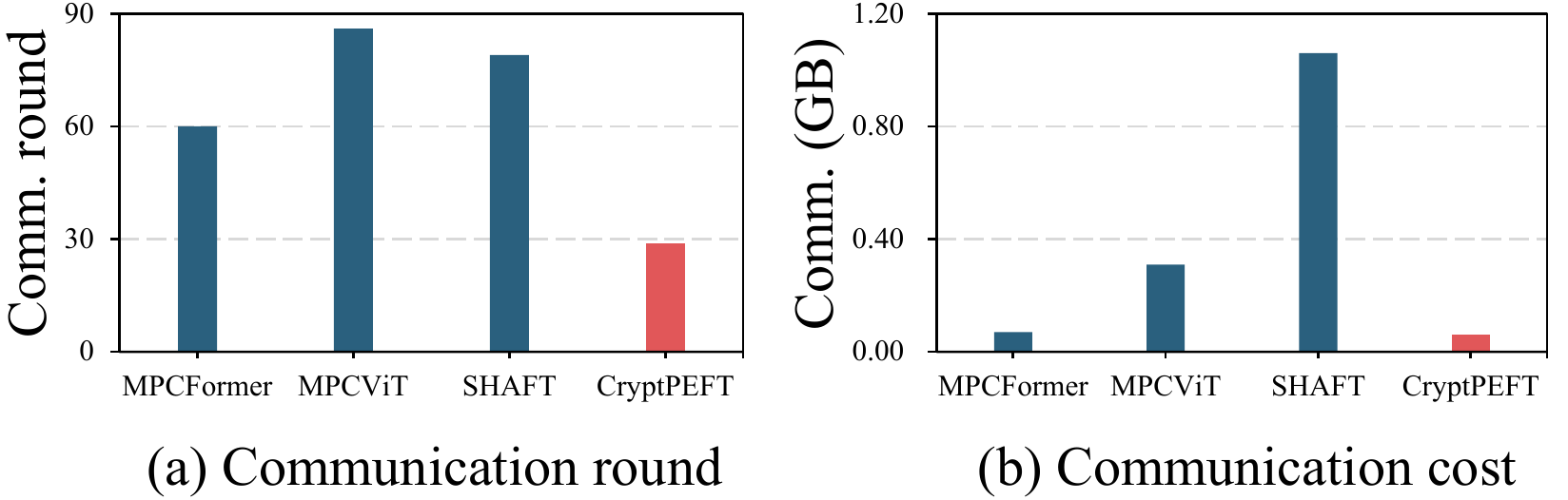}
    \caption{\sysname outperforms existing attention mechanisms in terms of both communication round and cost.}
    \label{fig:ablation}

    \phantomsubcaption
    \label{fig:ablation(a)}
    \phantomsubcaption
    \label{fig:ablation(b)}
\end{figure}

\para{Effectiveness of \atten}
%\wenhao{首先展示通信轮数、通信量的对比，正文描述一下数据：mpcformer的通信量小一点，除此之外，通信量和通信轮数都有优势}
%\wenhao{然后，用一个固定的数据集，我们在不同网络带宽和延迟下，评估了不同的attention替换后的性能，和\atten的实际latency对比}
%如Q2所述，目前存在的一些对MPC友好的attention mechanisms在通信上不高效，我们提出了对MPC更加友好的\atten和MPC-friendly MLP，在后续的评估中，统一称为\atten，
%我们进一步将\atten与目前多种对MPC友好的attention mechanisms方案进行比较，我们以CIFAR-100数据集为例，将MPCformer、SHAFT和MPCViT三种方案应用到\sysname中，在不同的网络环境下进行端到端的隐私推理，\autoref{fig:ablation}展示了不同方案下的communication round和communication cost，对比MPCViT和SHAFT，\sysname在communication round 和communication cost上都有显著优势，对比MPCFormer，\sysname的communication cost比MPCFormer高出17.86\%，但\sysname的communication round比MPCFormer少52.99\%，\sysname的隐私推理效率优于MPCFormer。
We evaluated \atten by comparing it with several attention mechanisms specifically designed for MPC-friendly computation. Using the CIFAR-100 dataset, we integrated three representative approaches---MPCFormer, SHAFT, and MPCViT---into the \sysname framework and conducted end-to-end private inference under various network conditions. As shown in \autoref{fig:ablation}, we report both communication rounds and communication cost for each method. 
Compared to existing attention mechanisms, \sysname achieves significant reductions in both metrics.
%\xss{\sysname reduces communication rounds by 51.67\% and communication cost by 14.29\%}, 
%MPCViT and SHAFT, 
% While \sysname incurs a 17.86\% increase in communication cost relative to MPCFormer, it reduces communication rounds by 52.99\%, 
%resulting in improved overall inference efficiency, as illustrated in \autoref{fig:CryptPEFT_efficiency}. 
This performance gain is attributed to the design of \atten, which provides a more communication-efficient attention mechanism. Specifically, \atten relies predominantly on MPC-friendly operations (additions and multiplications), thereby reducing both communication and computation overheads (see \autoref{fig:adapter}).

\begin{figure}
    \centering
    \includegraphics[width=\columnwidth]{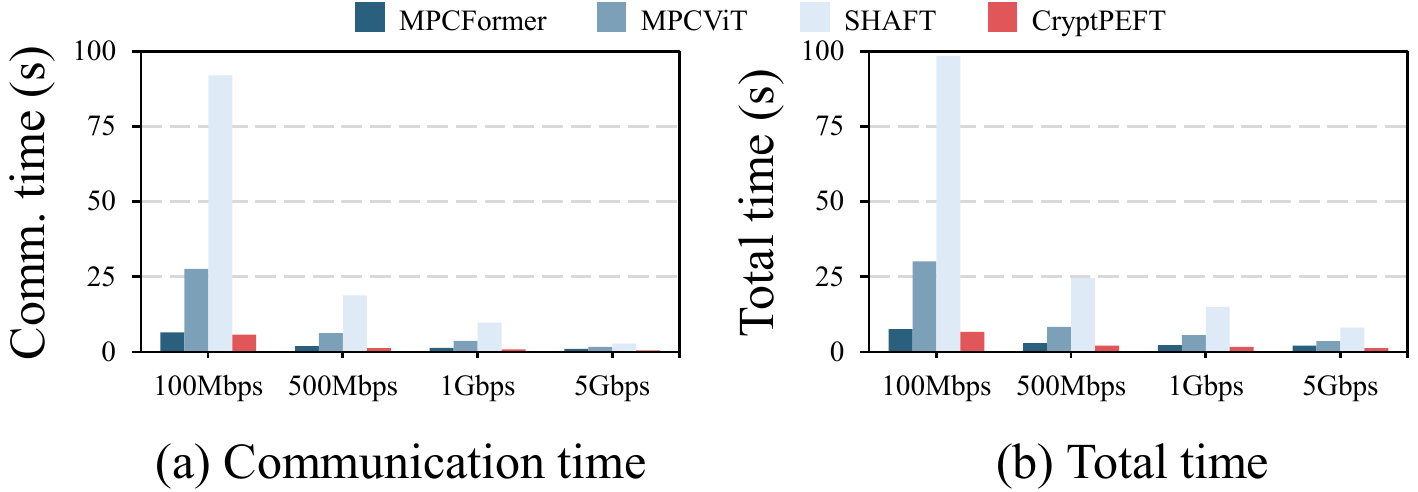}
    \caption{Inference efficiency across attention mechanisms.}
    \label{fig:CryptPEFT_efficiency}

    \phantomsubcaption
    \label{fig:CryptPEFT_efficiency_comm}
    \phantomsubcaption
    \label{fig:CryptPEFT_efficiency_total}
\end{figure}

Under different network conditions, \autoref{fig:CryptPEFT_efficiency_total} presents the total inference latency, where \sysname consistently outperforms all baselines. In particular, compared to MPCFormer, MPCViT, and SHAFT, \sysname achieves speedups of $1.13\times$ to $1.56\times$, $2.73\times$ to $4.49\times$, and $6.02\times$ to $14.70\times$, respectively, highlighting the effectiveness of \atten in improving private inference efficiency.

%\sysname在communication time and total time上显著降低，受益于我们提出的\atten，\sysname显著提升了隐私推理的效率。\wenhao{需要解释实验数据}\xss{参考comet，这里应该会放2张图，1*2放置，对应的指标分别是communication time and total time，单张图横坐标表示不同带宽，相同带宽下的一簇数据表示不同方案的通信时间或总时间}

%\wenhao{这里的图有点多，没有突出我们方案的优势，我建议画一个inferency latency的breakdown的图，分别在LAN和WAN环境下， 说明我们提出的attention显著降低了communication time，从而降低了整体的延迟（也就是不单独体现communication round/cost，直接展示communication/computation time）。另外，应该不能用一个简单的神经网络，还是要用efficiency-first下的某个真实的、典型的dataset做end-to-end展示。也可能只做WAN的实验，多放几个数据集}
%\xss{对比MPCFormer，优势应该不是特别明显，大概也就是加速1.3倍左右，主要是mpcformer这个方案确实挺快的，但是我们这个作为贡献的一个点，比他快30\%似乎也说得过去，mpcformer的缺点是通信轮数高，我们在通信轮数上才是它的一半，但是这个在低延迟的场景下，显得不是很关键，即使用WAN，延迟4ms也挺低的}\wenhao{也就是说其实mpcformer里的attention的近似效率在低延迟下影响不大，那是不是应该增大wan的延迟}

%在实际推理场景下，往往对数据做批量处理或利用GPU的并行计算能力来增加推理效率，基于此，我们设置了不同batch size来展示\sysname在不同batch size下的隐私推理效率，并且分别使用CPU和GPU两种计算单元，特别的，我们的网络环境设置为WAN，数据集设置为CIFAR-100，统计其amortized overhead。结果如\autoref{fig:batch_size}所示，我们观察到,提高batch size能进一步改善隐私推理效率，以使用CPU作为计算单元为例，batch size从1升高至128时，计算时间加速了$2.68\times$,通信时间加速了$2.28\times$，但隐私推理效率的增速会因硬件水平和通信环境而放缓，使用GPU作为计算单元，可以进一步加速计算，使用GPU作为计算单元时，batch size从1升高至128时，计算时间加速了$82.00\times$。
\begin{figure}[h!]
    \centering
    \includegraphics[width=\columnwidth]{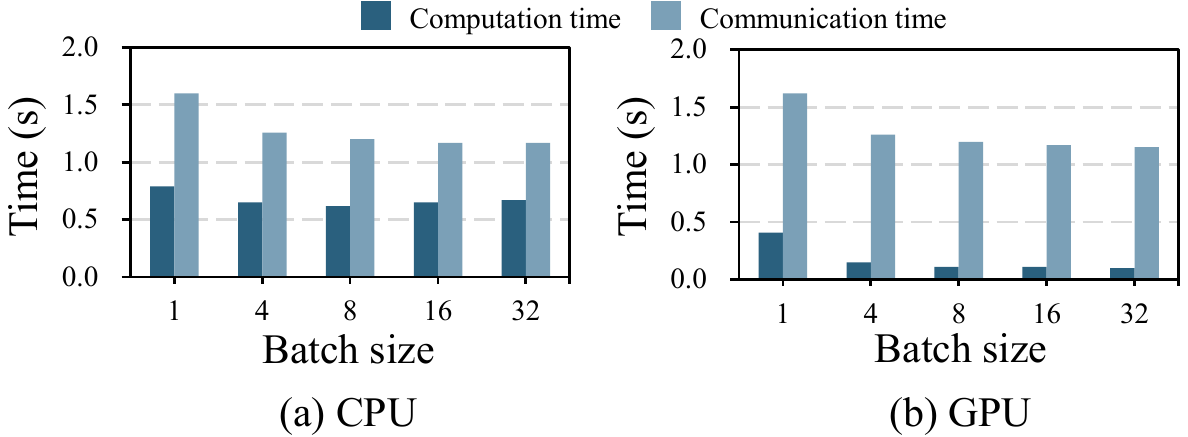}
    \caption{Private inference efficiency under different batch sizes.}
    \label{fig:batch_size}

    \phantomsubcaption
    \label{fig:batch_size(CPU)}
    \phantomsubcaption
    \label{fig:batch_size(GPU)}
\end{figure}

\para{Sensitivity to batch sizes}
In real-world inference scenarios, batching and GPU parallelism are commonly employed to improve inference efficiency. To evaluate the impact of batch size on private inference efficiency, we evaluated \sysname across a range of batch sizes using both CPU and GPU backends. All experiments were conducted on the CIFAR-100 dataset under a WAN network setting, and we report the amortized overhead per sample.
As illustrated in \autoref{fig:batch_size}, increasing the batch size leads to substantial improvements in private inference efficiency. For example, when using a CPU, scaling the batch size from 1 to 32 results in a $1.30\times$ speedup in computation time and a $1.37\times$ speedup in communication time per sample. However, the marginal gains decrease with larger batch sizes due to hardware and network bandwidth limitations. GPU-based computation yields even greater benefits: increasing the batch size from 1 to 32 achieves an $4.10\times$ speedup in computation time.

\ignore{
\xss{rewrite ablation study, only scaler on model utility}We further investigate the impact of scaler on \sysname. In this evaluation, we constrain the search space to a fixed configuration.

% \para{Impact of the scaler values on model utility}
In previous evaluations, the scaler value was fixed at 4.0. To further investigate the optimal scaler value for \sysname, we conducted additional experiments starting with a scaler value of 0.1 and exploring a range of alternative values. The downstream tasks evaluated included CIFAR-10, CIFAR-100, Caltech-101, Food-101, SVHN and Flowers-102, with all other settings kept consistent with the prior evaluations. As shown in \autoref{fig:ablation_adapter}, \sysname generally performs better with larger scaler values, such as 1.0 or 2.0.
%    \item \hl{\textit{Impact of batch size on private inference efficiency.}}

\para{Impact of approximation precision on model utility and efficiency}
In previous evaluations, the approximation component employed a polynomial of degree 6. In this evaluation, we investigate the impact of varying degrees of approximation on both model utility and inference efficiency. 
%Beginning with a degree of 2, we evaluated a range of values to determine their influence on 
%a range of degree values were evaluated to determine their influence on performance metrics. 
The results, illustrated in \autoref{fig:ablation(b)}, show that model utility remains relatively stable across different degrees of approximation, suggesting that \sysname exhibits a strong tolerance for low-precision approximation components. 

Subsequently, we evaluated inference efficiency in terms of inference time, as shown in \autoref{fig:ablation(c)}. As the approximation degree increases, the inference times also rises, indicating a decline in overall inference efficiency. Given that \sysname is robust to low-precision approximation components, we recommend employing lower-precision approximations to further enhance inference efficiency.

\para{Effectiveness of NAS}
In this evaluation, we kept the overall structure of \sysname the same as in previous evaluations but replaced its adapter with traditional PEFT adapters. We conducted two sets of experiments, substituting the \sysname adapter with AdaptFormer and LoRA, respectively. The results, shown in \autoref{fig:ablation(d)}, indicate that on the CIFAR-100 and Food-101 datasets, both LoRA and AdaptFormer achieve notably lower accuracy compared to \sysname's adapter. On the SVHN dataset, LoRA and AdaptFormer reach an average accuracy of only 60.88\%, much lower than the 89.27\% accuracy achieved by \sysname. These results demonstrate that traditional PEFT adapters do not integrate well with \sysname, underscoring the effectiveness of the NAS, which consistently delivers state-of-the-art utility across a range of tasks.
}

\begin{figure}
    \centering
    \includegraphics[width=0.9\columnwidth]{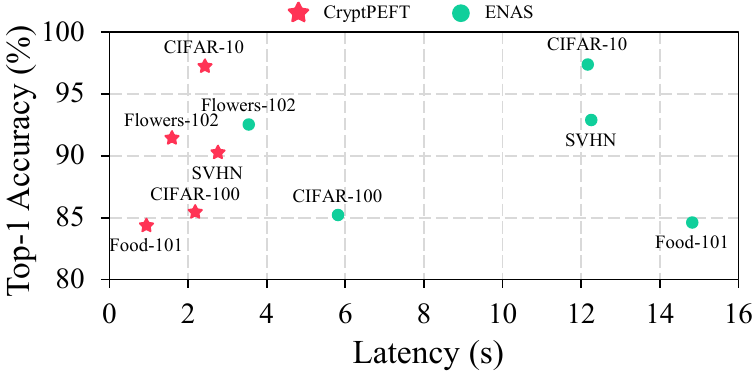}
    \caption{Comparison of accuracy and private inference latency for models discovered by \sysname and ENAS.}
    \label{fig:NAS_abalation}
\end{figure}

\para{\hl{Effectiveness of NAS}}
\hl{Our NAS approach incorporates a cost model tailored for private inference, prioritizing architectures with higher efficiency in this setting. To evaluate its effectiveness, we adopted ENAS~\cite{ENAS}, a representative reinforcement learning based NAS method, as the baseline. We conducted experiments under the WAN network environment across 5 downstream tasks. The end-to-end private inference latency was evaluated using the adapters discovered by both methods. As shown in \autoref{fig:NAS_abalation}, our method achieved a $4.07\times$ average speedup in private inference latency, with only a marginal drop in classification accuracy (89.73\% vs. 90.51\%). In addition, it identified target models $1.85\times$ faster on average compared to ENAS (3.03 hours vs. 5.60 hours on average)}.
%由于我们的搜索策略是在面向隐私推理的cost model的指导下搜索的，因此优先搜索隐私推理效率更高的模型。为此，我们选取一个典型的基于强化学习的ENAS~\cite{ENAS}作为baseline。
%针对一系列下游任务，如CIFAR-10，CIFAR-100，Food-101，SVHN和Flowers-102，\sysname执行搜索需要的时间分别为249，216，237，284和37分钟，ENAS方案为301，340，519，482和37分钟，\sysname比ENAS平均快了64.13\%。此外，\sysname提出的搜索算法能够快速且有效的搜索到目标adapter，\autoref{fig:NAS_abalation}展示了不同下游任务使用不同搜索算法得到的adapter的utility和其隐私推理延迟之间的关系，平均来看，我们的搜索策略得到的adapter结构，其隐私推理相比ENAS加速了$4.47\times$，而模型的分类准确率仅降低0.86\% (89.73\% v.s. 90.51\%)，

%作为对比，我们选择基于强化学习并带有参数共享的神经网络搜索ENAS~\cite{ENAS}作为baseline，

%% file: tex/discussion.tex
\section{Limitations and Discussions}
\label{sec:discussion}

\para{Backbone model overhead}
Under the OWC policy adopted by \sysname, a substantial portion of the computation on the backbone model is offloaded to the \muu and executed in plaintext. This design assumes the \muu has moderate computational capability---a reasonable assumption given that, in traditional MPC-based approaches, the \muu already engages in costly encrypted computations. Since encrypted computation typically dominates the \muu's workload, reducing its reliance through OWC offers a significant advantage. 

%\wenhao{Adjust cost analysis as requested by reviewer C}
%在\sysname系统下，\ms和\muu都需要从公开渠道(e.g., Hugging Face)下载Backbone，这在初始化时，需要消耗一定的时间，但这一行为对\ms和\muu是离线操作，且仅需要执行一次，不会影响隐私推理的效率，如\autoref{discussion:1}所述，\muu需要在Backbone上独立执行明文推理(无需与\ms交互)，我们在测试中包含了这一操作。此外，\ms和\muu需要独立存储Backbone，这引来部分存储开销，但\sysname和目前的PEFT方案具有相同的优势，即对多个下游任务，仅需存储一个Backbone，大大减少了存储开销
%\hl{
\hl{In \sysname, both \ms and \muu must download the backbone model (e.g., from public repositories like Hugging Face), incurring a one-time offline cost that does not impact private inference efficiency. Both parties also store a local copy of the backbone, introducing certain storage overhead. However, similar to existing PEFT methods, \sysname benefits from model reuse across multiple downstream tasks, significantly reducing overall storage demands.}
%In \sysname, both \ms and \muu are required to download the backbone model (e.g., from public repositories such as Hugging Face). This initialization step incurs a one-time offline cost for both parties, but does not affect the efficiency of private inference. As discussed in \autoref{sec:discussion}, \muu performs independent plaintext inference on the backbone without any interaction with \ms, and we include this operation in our evaluation. Furthermore, both \ms and \muu need to store a local copy of the backbone model, resulting in some additional storage overhead. However, \sysname shares the same advantage as existing PEFT approaches: a single backbone can be reused across multiple downstream tasks, substantially reducing overall storage requirements.

\para{Scalability to LLMs}
A promising avenue is to extend \sysname's applicability to a wider range of architectures and application domains. While our current focus is on vision-based tasks, adapting the proposed approach to large language models (LLMs) or multi-modal networks could yield novel insights and present new challenges. Specifically, addressing the unique functional blocks and data flow patterns inherent in language transformers may necessitate specialized adapter designs, enhanced OWC strategies, and innovative approximation methods tailored to natural language processing. 

\para{\hl{Trusted Execution Environment (TEE)}}
\hl{While our primary focus is on the MPC setting, our proposed method is compatible with TEE-based deployments with minimal modification. Specifically, the OWC constraint and the resulting architecture search procedure are agnostic to the underlying secure execution mechanism and can be readily applied in TEE environments.
TEEs offer advantages in terms of lower latency and simpler deployment, especially in single-device or edge scenarios. However, they rely on trusting the hardware vendor and are potentially susceptible to side-channel attacks, whereas MPC provides stronger security guarantees under standard cryptographic assumptions.}

%% file: tex/related.tex
\section{Related Works}
\label{sec:related}
\para{Privacy-preserving computation friendly neural network architectures}
Traditional neural networks were not originally designed to accommodate private inference. To bridge this gap, recent research has explored the optimization of neural architectures for privacy-preserving settings. 
Early works primarily focused on adapting CNNs for private inference under HE and MPC frameworks~\cite{kundu2023learning, cho2022sphynx, cho2022selective, kundu2023making}. 
Recently, Zeng et al. introduced MPCViT~\cite{zeng2023mpcvit}, a ViT architecture tailored for MPC settings. By reducing the scale of the ViT and implementing heterogeneous attention mechanisms, MPCViT achieves efficient and accurate inference under MPC frameworks.

%This approach demonstrates a significant advancement in designing transformer-based neural networks that are efficient for MPC. 

%传统的神经网络在设计之初，并没有考虑隐私推理的效率问题，因此，为了更高效的完成隐私推理，部分工作~\cite{kundu2023learning,cho2022sphynx,cho2022selective,kundu2023making,zeng2023mpcvit}开始尝试设计对隐私推理友好的神经网络，先前大部分工作~\cite{kundu2023learning,cho2022sphynx,cho2022selective,kundu2023making}以设计对HE或者MPC友好的卷积神经网络为主，MPCViT~\cite{zeng2023mpcvit}以ViT为基础，通过缩小ViT的规模，设计了一系列对MPC友好的transformer架构的神经网络

%\para{模型高效近似}
\para{Efficient approximation of neural networks}
To accelerate private inference, researchers have explored polynomial approximations of nonlinear functions~\cite{yu2022nn,li2022mpcformer,ao2024autofhe,ghodsi2020cryptonas,chen2022x,zhang2023sal}. While low-precision approximations can enhance computational efficiency, they often lead to utility degradation. 
MPCFormer employs knowledge distillation to improve model utility compromised by low-precision approximations~\cite{li2022mpcformer}. %By training a student model to emulate a high-performing teacher model, MPCFormer achieves efficient and private Transformer inference with minimal accuracy loss.
AutoFHE introduces layerwise mixed-degree polynomial approximations, assigning different polynomial degrees to various layers based on their sensitivity to approximation errors~\cite{ao2024autofhe}. 
Recent studies have proposed more efficient approximation methods to mitigate utility degradation caused by low-precision approximations~\cite{zimerman2024power,park2024powerformer}.
%为了加速深度神经网络隐私推理，很多工作~\cite{yu2022nn,li2022mpcformer,ao2024autofhe,ghodsi2020cryptonas,chen2022x,zhang2023sal}提出对非线性函数进行多项式近似，低精度近似能加速隐私推理，但会带来性能上的损失，为了解决这个问题，MPCFormer~\cite{li2022mpcformer}提出使用知识蒸馏改善模型性能，AutoFHE~\cite{ao2024autofhe}提出模型的不同位置使用不同精度的多项式来保证效率和性能，通过使用多目标优化搜索算法，自动调整神经网络模型来适配不同下游任务。同时一些更高效的近似方案被提出~\cite{zimerman2024power, park2024powerformer}来解决低精度近似导致的模型性能下降的问题。

%\para{高效隐私推理系统}
\para{Private transformer inference systems}
The widespread adoption of Transformer architectures has introduced notable challenges for private inference. In response, several studies have developed efficient private inference systems tailored for Transformers~\cite{hao2022iron,pang2024bolt,lu2023bumblebee,hou2023ciphergpt,ndss/KeiC25_SHAFT}.
Specifically, Iron addresses the intensive computational demands of large-scale matrix multiplications and complex nonlinear functions like Softmax and GELU inherent in Transformer models~\cite{hao2022iron}. 
BumbleBee~\cite{lu2023bumblebee} and Bolt~\cite{pang2024bolt} implement advanced protocols for matrix multiplication and activation functions. CipherGPT extends these advancements to generative large language models~\cite{hou2023ciphergpt}. SHAFT~\cite{ndss/KeiC25_SHAFT} extends \textsc{CrypTen}~\cite{crypten-github} by introducing efficient protocols for constant-round Softmax and GELU computations.
%, thereby enabling a more efficient private inference system that outperforms BumbleBee, Bolt, and \textsc{CrypTen} in both computation and communication.
%SHAFT扩展了\textsc{CrypTen}~\cite{crypten-github}，提出固定轮数的Softmax和GELU高效计算协议，进而实现更加高效的隐私推理系统，在计算和通信方面超越BumbleBee、Bolt和\textsc{CrypTen}.
%(LLMs). It introduces a private inference system specifically designed to handle the substantial communication and computation costs associated with generative models, enabling efficient and private inference for models like GPT~\cite{hou2023ciphergpt}.
%针对transformer架构的广泛应用，而transformer隐私推理开销大的问题，部分工作开始开发针对transformer的高效隐私推理系统~\cite{hao2022iron,pang2024bolt,lu2023bumblebee,hou2023ciphergpt},Iron~\cite{hao2022iron}针对transformer架构中大规模矩阵乘法、高计算复杂度的非线性函数Softmax、GELU等，提出一系列安全计算协议优化通信开销和计算开销，~\cite{lu2023bumblebee,pang2024bolt}又进一步降低安全推理过程中的通信开销，~\cite{hou2023ciphergpt}提出了针对生成式大语言模型的隐私推理系统，解决目前的协议对生成式大模型通信和计算开销过大的问题
In comparison, \sysname introduces a novel architecture tailored for PEFT methods, which can be seamlessly integrated into existing backbone models.
%to improve privacy-preserving inference. 
%Furthermore, \sysname can also be deployed in high-performance privacy-preserving inference systems to achieve further acceleration of private inference.
%Furthermore, CryptPEFT is implemented as a PyTorch module, facilitating efficient approximation and enabling the development of high-performance private inference systems.

%和上面的工作不同，我们从模型架构角度加速隐私推理，CryptPEFT用一种全新的架构，既支持参数高效微调，又可以应用到任何backbone上，进行隐私推理加速，同时，CryptPEFT以pytorch模块实现，支持高效近似和高效隐私推理系统

%% file: tex/conclusion.tex
\section{Conclusions}
\label{sec:conclusion}
%In this paper, we introduced \sysname, a novel solution that integrates one-way communication and parameter-efficient fine-tuning within private inference settings. By focusing on LoRA, selective layer adaptation, and automated search, \sysname effectively balances accuracy with privacy-preserving requirements, achieving substantial gains in both runtime efficiency and security.

In this paper, we present \sysname, the first PEFT architecture designed specifically for private inference. \sysname introduces an OWC paradigm that confines encrypted computation to the adapter, thereby eliminating the need for expensive two-way interactions with the backbone. To ensure strong model utility under the OWC constraint, we explore the design space of OWC-compliant adapters and incorporate an automated search mechanism to identify optimal configurations. Our evaluations demonstrate that \sysname achieves significant improvements in private inference efficiency. 

%% file: tex/ack.tex
\section*{Acknowledgments}  
We would like to express our sincere gratitude to the anonymous reviewers and our shepherd for their insightful and valuable feedback. The work of the authors from the Institute of Information Engineering was supported in part by the Strategic Priority Research Program of the Chinese Academy of Sciences (Grant No. XDB0690100), the National Natural Science Foundation of China (Grant Nos. 92270204 and 62272452), and a research grant from Huawei Technologies.

%% file: tex/ndss_ae_appendix_template_v1.tex
% Artifact Appendix template for the NDSS Artifact Evaluation
% version 1.1 (20250525)

% remove the following block when merging the appendix with the camera-ready full paper
%%%
% \documentclass[conference]{IEEEtran}
% \pagestyle{plain}
% \usepackage{url}
% \begin{document}
%%%

\section{Artifact Appendix}

% The artifact appendix is meant to be a self-contained document presenting a roadmap for setting up and evaluating your artifact. It should provide the following elements:
% \begin{enumerate}
% \item a list of the hardware, software, and configuration \textbf{requirements} for running the artifact;
% \item a clear description of how, and in what respects, the artifact \textbf{supports} the research presented in the paper;
% \item a guide for how others can \textbf{execute and validate} the artifact for its functional and usability aspects;
% \item the \textbf{major claims} made by your paper and a clear description of how to obtain data for each claim through your supplied artifact.
% \end{enumerate}

% Points 3-4 are not required if only the \textit{Available} badge is requested. Linking the paper claims to the artifact is a necessary step that allows artifact evaluators to functionally test and reproduce your results. Towards that end, explicitly list down items (e.g., results, plots, tables) from the paper and cross-reference them with the experiments to be reproduced.
 
% Unless otherwise stated, filling every section below with the requested contents is mandatory for participating in the artifact evaluation process. Section titles cannot be changed.

\subsection{Description \& Requirements}

\sysname is a parameter-efficient fine-tuning (PEFT) solution specifically designed for private inference scenarios. It minimizes the cost of encrypted computation by introducing a novel one-way communication (OWC) architecture, significantly reducing both computational and communication overhead. To maintain strong model utility under these constraints, \sysname incorporates OWC-compatible adapters and employs an automated neural architecture search (NAS) algorithm.
Our artifact includes the code and scripts necessary to evaluate both the model utility and private inference efficiency of the proposed method. %Additionally, it contains the dataset used for the evaluation.

% This section should list all the information necessary to recreate the experimental setup you used to run your artifact.

% Provide also a link to an archival repository where all the artifact's main components (software, datasets, documentation, etc.) can be accessed and, where this applies, the minimal hardware and software requirements to run your artifact.

% It is also very good practice to list and describe in this section benchmarks where those are part of your artifacts or simply have been used to produce results with it.

\subsubsection{How to access}
\label{subsec:access}
% Describe here how to access your artifact. During the artifact evaluation, in case of a private repository, you should provide instructions on how to access it. For the camera-ready version of the appendix, you must provide a DOI link to the AEC-approved artifact version that you must upload by then to permanent storage.
The source code of \sysname is publicly available at {\url{https://doi.org/10.5281/zenodo.17036866}}.

\subsubsection{Hardware dependencies}
Our artifact can be executed on a standard Linux server without requiring specialized hardware. The original experiments were conducted on a machine equipped with two NVIDIA GeForce RTX 4090 GPUs, used primarily for neural architecture search (NAS) and for assessing the benefits of GPU acceleration in batched task processing. To reduce hardware dependency and enhance accessibility, GPU-related components have been excluded from the artifact. This modification does not impact the paper’s core claims: \sysname achieves high model utility and efficient private inference through the introduction of the OWC policy and an OWC-compliant adapter design.
%Simply write ``None." where this does not apply to your artifact.
%Our experimental platform consists of an Intel Xeon Silver 4310 CPU, 64 GB of RAM, and two NVIDIA GeForce RTX 4090 GPUs.

\subsubsection{Software dependencies}
%Simply write ``None." where this does not apply to your artifact.
We implemented \sysname using Python 3.10.14. The dependencies for \sysname (and the following experiments) are detailed in \texttt{requirements.txt}.

\subsubsection{Benchmarks}
%Describe here any data (e.g., datasets, models, workloads, etc.) required by the experiments with this artifact reported in your paper. Simply write ``None." where this does not apply to your artifact.
(1) Datasets: CIFAR-10~\cite{cifar10}, CIFAR-100~\cite{CIFAR100}, Food-101~\cite{food101}, SVHN~\cite{SVHN} and Flowers-102~\cite{flowers102}.
(2) Models: ViT-B-16, a set of adapters searched using \autoref{alg:gridsearch}.

\subsection{Artifact Installation \& Configuration}

%This section should include all the high-level installation and configuration steps required to prepare the environment to be used for the evaluation of your artifact.
Download the source code locally following \autoref{subsec:access}. In the AE folder under the project root directory (default: \sysname), you will find a \texttt{README.md} file with detailed installation and configuration instructions.

% \subsection{Experiment Workflow}

% This section should provide a high-level view of the experimental workflow and how it is implemented, invoked, and (if needed) customized. The section is optional if the experiment workflow can be easily embedded in the Evaluation section.

% \textit{Detailed information for submission, reviewing, and badging process followed for the evaluation of NDSS 2026 artifacts can be found at:} \url{https://secartifacts.github.io/ndss2026/}.

\subsection{Major Claims}
% Enumerate here the major claims (Cx) made in your paper. For this purpose, we ask you to use the this and the following section and cross-reference the items therein, as explained next.

% Follows an example:

% \begin{itemize}
%     \item (C1): \textsc{System} achieves the same accuracy of the state-of-the-art systems for a task X while saving 2x storage resources. This is proven by the experiment (E1) whose results are illustrated/reported in [refer to your paper's plots, tables, sections, etc.].
%     \item (C2): \textsc{System} has been used to uncover new bugs in the Y software. This is proven by the experiments (E2) and (E3) in [ibid].
% \end{itemize}
\begin{itemize}
    \item (C1): \sysname achieves classification accuracy comparable to the state-of-the-art methods, while significantly reducing the number of parameters involved in encrypted computation. This is proven by the experiment (E1), with results reported in \autoref{tab:performance}.
    \item (C2): \sysname significantly reduces private inference latency in both LAN and WAN environments. This is proven by the experiment (E2), with results reported in \autoref{tab:LAN_batch64} and \autoref{tab:breakdown}.
    \item (C3): \sysname significantly outperforms MPCViT~\cite{zeng2023mpcvit} by providing higher utility and achieving $9.19\times$ to $13.14\times$ speedup in private inference on CIFAR-10 and CIFAR-100. This is proven by the experiment (E1), experiment (E2) and experiment (E3), with results reported in \autoref{tab:mpcvit_vs_CryptPEFT}.
    \item (C4): The \atten proposed in \sysname significantly outperforms the attention mechanisms used in MPCFormer~\cite{li2022mpcformer}, MPCViT~\cite{zeng2023mpcvit}, and SHAFT~\cite{ndss/KeiC25_SHAFT} in terms of private inference efficiency. This is proven by the experiment (E4), with results reported in \autoref{fig:CryptPEFT_efficiency}.
    
\end{itemize}

\subsection{Evaluation}

% This section include all the operational steps and experiments which must be performed to evaluate if your artifact is functional and validate the proposed/presented experiments. For this purpose, we ask you to use the this and the preceding section and cross-reference the items therein.

% If the \textit{Reproduced} badge is requested, the section should reference your paper's key results: describe the expected numbers and, where applicable, the maximum variation expected (particularly important for performance numbers). If an experiment is a scaled-down version of the one run for the paper's evaluation, provide a clear indication of it with a justification of the scaling-down being meaningful.

% We also highly encourage you to provide your estimates of human- and compute-time for each of the listed experiments.

% Follows an exemplary structure for one experiment (Ey):

\subsubsection{Experiment (E1)}
[\sysname utility] [5 human-minutes + 20 compute-minutes]: This experiment is designed to evaluate the utility of \sysname under both Utility-first and Efficiency-first strategies.

\textit{[Preparation]}
Go to the project root directory (default name: \sysname).

\textit{[Execution]}
Run the command below.
\begin{tcolorbox}[colback=gray!5!white, colframe=gray!80!black, left=1mm]
\begin{verbatim}
$: bash AE/eval_model_utility.sh
\end{verbatim}
\end{tcolorbox}

\textit{[Results]}
All results are saved in the directory \path{AE/eval_result}. Open the file named like \path{WAN_CRYPTPEFT_Efficiency_first_cifar100} to find results similar to the example below (note that results may vary by about 1\% depending on the experimental environment and hardware).
\begin{tcolorbox}[colback=gray!5!white, colframe=gray!80!black, left=1mm]
\begin{verbatim}
========= eval result ...... =========
acc:85.47
n_param:1.12141
\end{verbatim}
\end{tcolorbox}

\subsubsection{Experiment (E2)}
[\sysname private inference latency] [5 human-minutes + 90 compute-minutes]: This experiment is designed to evaluate the private inference latency of \sysname under LAN and WAN environments.

\textit{[Preparation]}
Go to the project root directory (default name: \sysname).

\textit{[Execution]}

(1): Execute the following command in the Linux terminal to simulate a LAN network environment (avoid using VSCode, as it may cause the execution to fail).
\begin{tcolorbox}[colback=gray!5!white, colframe=gray!80!black, left=1mm]
\begin{verbatim}
$: sudo tc qdisc add dev lo root \
netem rate 1gbit delay 0.5ms
\end{verbatim}
\end{tcolorbox}

(2): Open two terminals and run the following commands separately (you can use either VSCode or a Linux terminal).
\begin{tcolorbox}[colback=gray!5!white, colframe=gray!80!black, left=1mm]
\begin{verbatim}
$0: bash AE/eval_CRYPTPEFT_PI.sh 0 LAN
$1: bash AE/eval_CRYPTPEFT_PI.sh 1 LAN

$0: bash AE/eval_SFT_Last_PI.sh 0 LAN
$1: bash AE/eval_SFT_Last_PI.sh 1 LAN
\end{verbatim}
\end{tcolorbox}

(3): Execute the following command in the Linux terminal to simulate a WAN network environment (avoid using VSCode, as it may cause the execution to fail).
\begin{tcolorbox}[colback=gray!5!white, colframe=gray!80!black, left=1mm]
\begin{verbatim}
$: sudo tc qdisc del dev lo root
$: sudo tc qdisc add dev lo root \
netem rate 400mbit delay 4ms
\end{verbatim}
\end{tcolorbox}

(4): Open two terminals and run the following commands separately (you can use either VSCode or a Linux terminal).
\begin{tcolorbox}[colback=gray!5!white, colframe=gray!80!black, left=1mm]
\begin{verbatim}
$0: bash AE/eval_CRYPTPEFT_PI.sh 0 WAN
$1: bash AE/eval_CRYPTPEFT_PI.sh 1 WAN

$0: bash AE/eval_SFT_Last_PI.sh 0 WAN
$1: bash AE/eval_SFT_Last_PI.sh 1 WAN
\end{verbatim}
\end{tcolorbox}

(5): Restore the default network environment (avoid using VSCode, as it may cause the execution to fail).
\begin{tcolorbox}[colback=gray!5!white, colframe=gray!80!black, left=1mm]
\begin{verbatim}
$: sudo tc qdisc del dev lo root
\end{verbatim}
\end{tcolorbox}

\textit{[Results]}
All results are saved in the directory \path{AE/eval_private_inference_result}. Open the file named like \{e.g., \texttt{eval\_CryptPEFT\_LAN\_cifar100}\} to find results similar to the example below (the total\_time and comm\_time may vary due to differences in CPU specifications; however, the improvements compared to the baseline, comm\_cost and comm\_round remain consistent).
\begin{tcolorbox}[colback=gray!5!white, colframe=gray!80!black, left=1mm]
\begin{verbatim}
total_time: 1.1336361408233642
comm_round: 29.0
comm_cost: 0.052045270800590515
comm_time: 0.49003771375864746
\end{verbatim}
\end{tcolorbox}

\subsubsection{Experiment (E3)}
[Comparation with MPCViT] [5 human-minutes + 20 compute-minutes]: This experiment is designed to compare the model utility and private inference latency between \sysname and MPCViT (the utility results of MPCViT are taken from the results reported in their paper~\cite{zeng2023mpcvit}).

\textit{[Preparation]}

(1): Go to the project root directory (default name: \sysname).

(2): Simulate WAN and LAN network environments following similar steps as in Experiment (E2).

\textit{[Execution]}

(1): Make sure you are in the simulated LAN network environment. Open two terminals and run the following commands separately (you can use either VSCode or a Linux terminal).
\begin{tcolorbox}[colback=gray!5!white, colframe=gray!80!black, left=1mm]
\begin{verbatim}
$0: bash AE/eval_MPCViT_PI.sh 0 LAN
$1: bash AE/eval_MPCViT_PI.sh 1 LAN
\end{verbatim}
\end{tcolorbox}

(2): Make sure you are in the simulated WAN network environment. Open two terminals and run the following commands separately (you can use either VSCode or a Linux terminal).
\begin{tcolorbox}[colback=gray!5!white, colframe=gray!80!black, left=1mm]
\begin{verbatim}
$0: bash AE/eval_MPCViT_PI.sh 0 WAN
$1: bash AE/eval_MPCViT_PI.sh 1 WAN
\end{verbatim}
\end{tcolorbox}

\textit{[Results]}
All results are saved in the directory \path{AE/eval_private_inference_result}. Open the file named like  \texttt{eval\_MPCViT\_WAN\_cifar100} to find results similar to the example below.
\begin{tcolorbox}[colback=gray!5!white, colframe=gray!80!black, left=1mm]
\begin{verbatim}
total_time: 22.374131655693056
comm_round: 513.0
comm_cost: 0.5615268349647522
comm_time: 18.25921990380448
\end{verbatim}
\end{tcolorbox}

\subsubsection{Experiment (E4)}
[The private inference efficiency of \atten] [5 human-minutes + 50 compute-minutes]: This experiment is designed to evaluate the efficiency of private inference with various attention mechanisms.

\textit{[Preparation]}

(1): Go to the project root directory (default name: \sysname).

(2): Simulate different network bandwidths following similar steps as in Experiment (E2). Note that the network latency should be fixed at 4ms, and the network environment must be reset to default before changing to a different bandwidth.

\textit{[Execution]}
We assume you are in a network environment with 1Gbps bandwidth and 4ms latency. Open two terminals and run the following commands separately.
\begin{tcolorbox}[colback=gray!5!white, colframe=gray!80!black, left=1mm]
\begin{verbatim}
$0: bash AE/ablation_LinAtten_PI.sh 0 1G
$1: bash AE/ablation_LinAtten_PI.sh 1 1G
\end{verbatim}
\end{tcolorbox}

\textit{[Results]}
All results are saved in the directory \path{AE/eval_private_inference_result}. Open the file named like \path{ablation_LinAtten_1G_CryptPEFT_cifar100} to find results similar to the example below.
\begin{tcolorbox}[colback=gray!5!white, colframe=gray!80!black, left=1mm]
\begin{verbatim}
total_time: 1.5270877838134767
comm_round: 29.0
comm_cost: 0.06424663960933685
comm_time: 0.8196889258921146
\end{verbatim}
\end{tcolorbox}

% \subsection{Customization}
% Provide here notes on how to customize your experiments, when applicable. The section is optional.

% \subsection{Notes}
% This section is meant to allow authors to include any further important notes that may not fall within any of the previous categories. We kindly encourage you to remove this section where this sort of content may not be strictly needed (rather than filling it with unnecessary or redundant information). 

%\subsection{Necessary Changes to Fulfill Major Revision Requirements}

%In response to the reviewers' feedback, we plan to include an ablation study to highlight the effectiveness of our proposed NAS strategy, which was not part of the original submission. However, due to time constraints, we may not be able to complete these additional experiments before the AE deadline. As such, we understand that this component may not be considered during the AE process. %We will update the code repository with the new results once the study is completed, so the AE reviewer can access and evaluate them at that time.

% remove the following block when merging the appendix with the camera-ready full paper
%%%
% \end{document}
%%%